\documentclass[twocolumn]{aastex62}
\usepackage{graphicx}
\pdfoutput=1

\usepackage{amsmath}
\def\approxprop{%
  \def\p{%
    \setbox0=\vbox{\hbox{$\propto$}}%
    \ht0=0.6ex \box0 }%
  \def\s{%
    \vbox{\hbox{$\sim$}}%
  }%
  \mathrel{\raisebox{0.7ex}{%
      \mbox{$\underset{\s}{\p}$}%
    }}%
}

\begin{document}

\title{Constraining Theories of Polarized SiO Maser Transport:\\ Multi\-Epoch Analysis of a $\pi/2$ Electric Vector Rotation Feature}
\shorttitle{SiO Maser Polarization of EVPA Rotation}

\author{T. L. Tobin}
\affil{Department of Astronomy, University of Illinois at Urbana-Champaign\\1002 W. Green Street, Champaign, IL 61801, USA}
\email{tltobin2@illinois.edu}

\author{A. J. Kemball}
\affil{Department of Astronomy, University of Illinois at Urbana-Champaign\\1002 W. Green Street, Champaign, IL 61801, USA}
 
\author{M. D. Gray}
\affil{Jodrell Bank Centre for Astrophysics, Alan Turing Building, University of Manchester\\Manchester M13 9PL, UK}

\received{2018 May 01} \accepted{2018 Dec 20 }

\begin{abstract}
The detailed polarization mechanisms of SiO masers originating from the near circum\-stellar environment of Asymptotic Giant Branch stars are not yet definitively known. Prevailing theories are broadly classified as either Zeeman or non-Zeeman in origin, the latter including effects such as anisotropic pumping or anisotropic resonant scattering. The predicted behavior of the linear and circular polarization fractions and electric vector position angle vary by theory. In particular, individual maser features that exhibit a rotation in linear polarization of $\sim \pi/2$ as a function of frequency over their extent can be utilized as a test of several maser polarization transport theories. In this paper, we analyze one SiO ($\nu=1$, $J=1-0$) maser feature toward the Mira variable, TX Cam that exhibits this internal polarization rotation and persists across five epochs (spanning $\sim3$ months). We compare our results to the predictions by several maser polarization theories and find that the linear polarization across the feature is consistent with a geometric effect for a saturated maser originating when the angle between the projected magnetic field and the line of sight ($\theta$) crosses the Van Vleck angle $\theta_F \sim 55^{\circ}$. However, the electric vector position angle (EVPA) exhibits a smooth rotation across the spatial extent of the feature rather than the expected abrupt $\pi/2$ flip. We discuss possible explanations for this discrepancy and alternative theories. Circular polarization across the feature is also analyzed and it is the most accurately described by Zeeman effects giving rise to a circular polarization fraction of the form $m_c \approxprop \cos \theta$.
\end{abstract}


\section{Introduction}\label{Intro}

By nature of their high intrinsic brightness and spatial compactness, individual astrophysical maser components serve as unique high-resolution probes of their host molecular gas populations. Sub-milliarcsecond angular resolution imaging of SiO masers in the near-circumstellar environment (CSE) of large-amplitude, long-period variable (LPV) stars \citep{hab96, eli80, cku96, wu17} is possible using Very Long Baseline Interferometry (VLBI) \citep{moran79, miyoshi94, diamond94, greenhill95, kemb02, cotton06, assaf13}. The near-CSE in large-amplitude LPV stars has both complex kinematics and dynamics \citep{witt08, diakem03,hink84,khouri16,wong16,vlem17}\explain{Replaced reference to Castro-Carrizo et al. (2010) with references to Khouri et al. (2016), Wong et al. (2016), and Vlemmings et al.(2017)}; it is dominated by significant mass loss through the stellar wind, shocks driven by stellar pulsation, and associated outflow and infall velocity gradients \citep{humph02,hink84,hink97,ireland08, groen16, maer16,doan17}. These factors may be further influenced by the magnetic field, but the magnitude, origin, and dynamical influence remains uncertain \citep{denpin07, black01, sokerzoabi, nord07, nordblack06, fabas11, lealf13}.Both magnetic fields and binary companions have been proposed as dynamically significant in the distribution and kinematics of the expelled mass \citep{matt00, black00, black01, nord07, nordblack06, duthu17}, and could play a role in shaping the resulting planetary nebula and the return of processed material to the ISM \citep{black01}.

Polarization measurements of sufficiently compact regions within the near-CSE may help trace the magnitude and morphology of the magnetic fields around these stars. SiO $\nu = 1, J=1-0$ maser components are highly linearly polarized \citep{trol79, herp06} and thus have significant potential as compact probes of the magnitude and morphology of the magnetic field in the near-CSE \citep{vlem05}. However, these millimeter-wavelength SiO masers have low levels of intrinsic circular polarization \citep{herp06} and can only be accurately interferometrically calibrated and imaged using data reduction techniques that address systematic sources of error in Stokes $V$  at commensurately low levels \citep{kembrich11}. Furthermore, the extent to which the observed maser polarization arises from an underlying magnetic field directly from the Zeeman effect, as opposed to other mechanisms, is still the subject of theoretical debate \citep{watson09,eli96}. Anisotropic pumping \citep{west83, deguchi76, buj94}, non-Zeeman \citep{watson09} or Hanle effects \citep{aram05}, and anisotropic resonant scattering \citep{houde14} have all been proposed as significant contributors to the polarization signature. To study the magnetic field in this environment, one must first develop greater certainty about which theoretical model of polarized maser transport prevails in this complex environment.

Each theory of SiO maser polarization can be considered a different forward data model connecting the underlying key physical properties of the maser region - such as the incident or seed radiation field, the magnetic field distribution, line excitation conditions, maser geometry, beaming angles, electron densities, and local turbulence - to the measured Stokes parameters. In the idealized limit of high signal-to-noise imaging at high fidelity with high angular and frequency resolution, and where the key physical parameters in the maser region are also able to be measured, it would be possible to use the observed Stokes parameters over the ensemble of observed maser components to constrain the theory of polarization maser transport with precision as an inverse problem. However, this idealization is not realizable in practice, primarily due to incomplete knowledge of physical conditions in the maser region, but significant progress can be made by  constructing tests of as high a theoretical discriminant value as possible using accurate component-level maser polarization measurements within achievable sensitivity limits \citep{richter16}.

Analyses of maser features exhibiting an abrupt $\sim90^\circ$ rotation in electric vector position angle (EVPA) can provide constraints on maser polarization theory as they can allow inference of the angle between the magnetic field and the line of sight across the maser feature, as described in detail below. \citet{kdrgx11} analyzed such a feature in a single epoch of VLBA polarimetric observations from a prior monitoring campaign of the $\nu=1, J=1-0$ 43 GHz SiO maser emission toward the Mira variable TX Cam \citep{diakem03, kemb09, GDK10, GDK13}. This behavior has also been observed in H$_2$O masers \citep{vlemdiam06}.

 The theoretical explanations for the EVPA rotation were discussed by \citet{kdrgx11} and are revisited here in summary. The prevailing explanations include (i) a projection effect caused by the global magnetic field passing through the Van Vleck angle with respect to our line-of-sight, (ii) a sharp change in the local magnetic field orientation, or (iii) a change in the orientation of radiation anisotropy, independent of the magnetic field.

In the case of (i), a number of asymptotic limit solutions were originally derived by \citet{GKK} (hereafter GKK) for linear, $J=1-0$ masers undergoing isotropic pumping and assuming zero circular polarization. For the case of saturated SiO $\nu=1, J=1-0$ masers, an EVPA reversal of $90^\circ$ will arise if the angle $\theta$ between the magnetic field and the line of sight passes through the Van Vleck angle, $\theta_F \equiv \sin^{-1}\sqrt{2/3} \approx 55 ^\circ$.  A later analysis by \citet{eli96} resulted in the same expected polarization profiles across the EVPA reversal in masers but under less extreme levels of saturation. Conversely, numerical simulations studies by \citet{watwyld01} were more strongly supportive of these polarization profiles only in the limit of strong saturation, as originally considered by GKK. 

In case (ii), the EVPA reversal would be caused by the magnetic field rapidly changing orientation. \citet{sokerclayton} argued that cool magnetic spots may protrude from the photosphere of the AGB star, causing a dramatic change in the direction of the magnetic field just above the spot \citep{soker02}. In this case, rather than traversal of the critical Van Vleck angle between the magnetic field and the line of sight, the EVPA would be tracing an abrupt rotation of the magnetic field from tangential to radial in the image plane.

Finally, in the case of (iii), the linear polarization is driven instead by anisotropic pumping of the masing SiO molecules \citep{aram05, west83}. An EVPA flip of the magnitude studied here could be a result of the source  function or optical depth varying across the feature.

In addition, the profile of the circular polarization fraction, $m_c$, across the EVPA reversal profile, and thereby the inferred $m_c(\theta)$ can provide constraints on the mechanism responsible for circular polarization in SiO masers. Although GKK assumed zero $m_c$, subsequent thoeretical studies have attempted to include Zeeman circular polarization in their calculations. For masers with overlapping Zeeman components such as SiO, a non-paramagnetic molecule, \citet{eli96} found that $m_c$ is inversely proportional to $\cos\theta$ for saturated masers. Work by \citet{gray12} found that $m_c \approxprop \cos \theta$. Numerical simulations by \citet{watwyld01} predicted that circular polarization would be proportional to $\cos\theta$ for highly unsaturated masers and would peak around $\cos \theta \sim 0.2$ for more saturated masers, as shown in Figure 1 of their paper. 

A Zeeman interpretation of SiO maser circular polarization may produce strong magnetic field estimates ranging from a few to a hundred Gauss \citep{barv87,kdmd97,amiri12}. If assumed global, this could lead to a significant disequilibrium between the magnetic and thermal gas pressure \citep{watson09}. In comparison to other observational magnetic field estimates, such a high magnetic field at the SiO maser radius may be consistent with extrapolated estimates from H$_2$O maser polarimetry \citep{vlem05, lealf13} and some optical polarimetry measurements \citep{konst10, konst14, lebre14}. It has also been proposed that the Zeeman magnetic field strengths inferred from SiO masers sample local enhancement of a sparse global field \citet{kemb09}. Similarly, if the EVPA reversal is due to a change in magnetic field orientation as described above in case (ii), the local magnetic field near the cool spot would only need to be $\sim1-10$ G \citep{soker02} and could reside within a weaker global field. 

However, non-Zeeman circular polarization could be produced by several mechanisms, including resonant foreground scattering \citep{houde14}.  In addition, linear polarization can be converted to elliptically polarized radiation when the maser traverses an anisotropic medium \citep{ned94, wiewat98, watson09}, and would display a correlation between $m_l$ and $m_c$, with $m_c \leq m_l^2/4 $ \citep{wiewat98}. Circular polarization of non-Zeeman origin would imply a significantly lower magnetic field strength than required for Zeeman circular polarization of observed SiO masers. We note that the resonant scattering mechanism proposed by \citet{houde14} requires a magnetic field strength in the foreground lower-density gas that is not inconsistent with that derived from H$_2$O maser observations \citep{vlem05, lealf13}.

In this paper, we extend the single-epoch analysis begun by \citet{kdrgx11} to additional adjacent epochs in the observing series \citep{diakem03, kemb09, GDK10, GDK13} that show similar  $\sim90^\circ$ EVPA rotation. This expands the sample of polarization profiles across such features. We find that the fractional linear polarization profiles $m_l(\theta)$ are broadly consistent with GKK; however the EVPA profiles $\chi(\theta)$ show greater discrepancies. Possible origins for such discrepancies are discussed below.

The remainder of this paper is organized as follows: The observations are described in Section 2. Data reduction and results are described in Section 3. Section 4 contains a discussion of the data as they constrain theories of maser polarization, and a summary of our conclusions are presented in Section 5.

\section{Observations\label{obs}}
The data analyzed here consist of a subset of epochs from a long-term monitoring campaign of circumstellar $\nu=1, J=1-0$ SiO maser emission toward the Mira variable TX Cam. In this campaign, TX Cam was observed biweekly or monthly in full polarization using the 43 GHz band of the Very Long Baseline Array (VLBA)\footnote{http://vlba.aoc.nrao.edu} and one antenna of the Very Large Array (VLA)\footnote{http://vla.aoc.nrao.edu}, both of which were operated by the NRAO\footnote{The National Radio Astronomy Observatory is a facility of the National Science Foundation operated under cooperative agreement by Associated Universities, Inc.}. Analyses of the total intensity from this monitoring campaign were presented in \citet{diakem03,GDK10,GDK13}, while the linear polarization was described in \citet{kemb09}.

\begin{deluxetable*}{lllllll}
\tabletypesize{\scriptsize}
\tablecaption{Epochs with EVPA Reversal and Corresponding Absolute EVPA of J0359+509\label{tbl-1}}
\tablewidth{0pt}
\tablehead{\colhead{Epoch Code} & \colhead{VLBA Observing Date} & \colhead{Optical Phase\tablenotemark{a} ($\phi$)} & \colhead{Date of VLA Observation} & \colhead{$\chi$ (deg)\tablenotemark{c}} & \colhead{$\epsilon_\chi$ (deg)\tablenotemark{d}} & \colhead{VLA Config.}}
\startdata
BD46AM\tablenotemark{b} & 1998 Dec 6 & $1.68\pm0.01$  & ...\tablenotemark{e} & 90.1 & ... & ...\\
BD46AN & 1998 Dec 23 & $1.71\pm0.01$ & 1998 Dec 22 & 75.6 & 1.5 & C \\
BD46AO & 1999 Jan 5 & $1.74\pm0.01$ & ...\tablenotemark{e} & 85.0 & ... & ... \\
BD46AP & 1999 Jan 23 & $1.77\pm0.01$ & 1999 Jan 22 & 60.1 & 3.7 & C\\
BD46AQ & 1999 Feb 6 & $1.79\pm0.01$ & 1999 Feb 7 & 73.8 & 4.7 & CD \\
BD46AR & 1999 Feb 19 & $1.82\pm0.01$ & 1999 Mar 9 & 75.2 & 6.5 & D \\
\enddata
\tablecomments{Epoch VLBA observing date and optical phase from \citet{diakem03}. Absolute EVPA of J0359+509 and related data for epochs with associated VLA calibration observations from \citet{kemb09}, Table 2.}
\tablenotetext{a}{Optical phase was computed using the optical maximum at MJD$=50773$ cited by \citet{gray99} and assuming their quoted uncertainty of $\Delta\phi\sim0.01$. A mean period of 557.4 days is adopted \citep{khol85}.}
\tablenotetext{b}{The electric vector position angle (EVPA) reversal was present in this epoch, but after data reduction, only three channels had $m_l$ with SNR $>3$, while no $m_c$ values reached SNR$=3$. As such, it was not used to test the aforementioned model predictions.}
\tablenotetext{c}{Absolute EVPA of linearly polarized emission from J0359+509, relative to an assumed 43 GHz EVPA for 3C138 of $\chi_{3C138}=-14^\circ$ \citep{pt03}.}
\tablenotetext{d}{Estimated standard error in $\chi$, derived from independent analyses of each of the two 50 MHz VLA continuum spectra windows, as $\frac{1}{\sqrt{2}} | \chi_1 - \chi_2 |$}
\tablenotetext{e}{Associated VLA observations for these epochs were not available. $\chi$ values determined by fitting observed $\chi$ values with a weighted cubic polynomial over a scrolling five-sample window, and extracting the resulting value for the date of the VLBA observations of that epoch.}
\end{deluxetable*} 

The dates and observing codes of the epochs reduced for their EVPA reversal, along with the optical phase of TX Cam, are given in Table \ref{tbl-1}. However, few linear or circular polarization data points in the epoch designated BD46AM were detected with sufficient signal-to-noise to allow a meaningful fit to the models, so only BD46AN through BD46AR were carried forward in the analysis.

The observational parameters of the larger monitoring campaign, from which this subset of EVPA reversal epochs are drawn for re-analysis, are described in detail by \citet{diakem03} and summarized here. Observations were taken with a 4 MHz bandwidth centered on the SiO $\nu=1, J=1-0$ rest frequency of $43.122027$ GHz shifted to TX Cam's LSR velocity of $+9$ km s$^{-1}$ at the mid-point of the observations and the center of the array. The baseband was spanned by 128 channels, each with a width of 31.25 kHz or $\sim 0.2$ km s$^{-1}$. The data were sampled with one-bit quantization over a correlator accumulation interval of 4.98 s.

Each epoch consists of 6.5 hrs of total observing time spread over an 8 hour period shared with another project. Of this observing block, 17 scans were scheduled on the primary target, TX Cam. Seven scans were scheduled for the primary continuum calibrator, J0359+509, which was used for polarization, group delay, and bandpass calibration. Secondary calibrators, 3C454.3 and J0609-157, were observed with one scan each, and were later both also used for group delay and bandpass calibration. Each scan lasted 13 minutes. For optimal \textit{u-v} coverage, observations of TX Cam and J0359+509 were spread as evenly as possible over the 8 hour time span.

\section{Results}
The data were reduced and calibrated with the Astronomical Image Processing System (AIPS) adapted to implement the accurate circular polarization calibration techniques described in \citet{kembrich11}. The data presented here were imaged using an image size of $4096 \times 4096$ pixels and a pixel size of $30 \mu as$. However, the fully reduced images presented in this paper were re-sampled to a size of $2048 \times 2048$ pixels with a pixel size of $50$ $\mu as$ to allow direct alignment with prior reduced image series. The results presented in this work are derived from analysis of the re-sampled images.

After reduction, each epoch resulted in separate Stokes \{I, Q, U, V\} images in the central 113 frequency channels. These inner channels span the SiO maser emission and also avoid bandpass edge effects. For conformity with \citet{kemb09} and \citet{kdrgx11}, a $540 \times 420$ $\mu$as elliptical Gaussian restoring beam at a position angle of $20^\circ$ was adopted across all epochs.

Calibration of the absolute EVPA of the observed linearly polarized emission follows the approach described by \citet{kemb09}. Associated VLA observations of 3C138 (the primary EVPA calibrator) and J0359+509 (the transfer calibrator) are used to establish the absolute EVPA of the VLBI observations. This method has a conservative estimated peak-to-peak error of $\sim 10^\circ - 20^\circ$ \citep{kemb09}. Absolute EVPA values for epochs BD46AM and BD46AO were estimated by fitting the existing calibration values with a weighted cubic polynomial over a scrolling five-sample window \citep{kemb09}. The resulting absolute EVPA of J0359+509 relative to the assumed 3C138 EVPA of $\chi_{3C138}=-14^\circ$ at 43 GHz \citep{pt03} is given in Table \ref{tbl-1}.

\begin{figure}[t!]
\figurenum{1}
\plotone{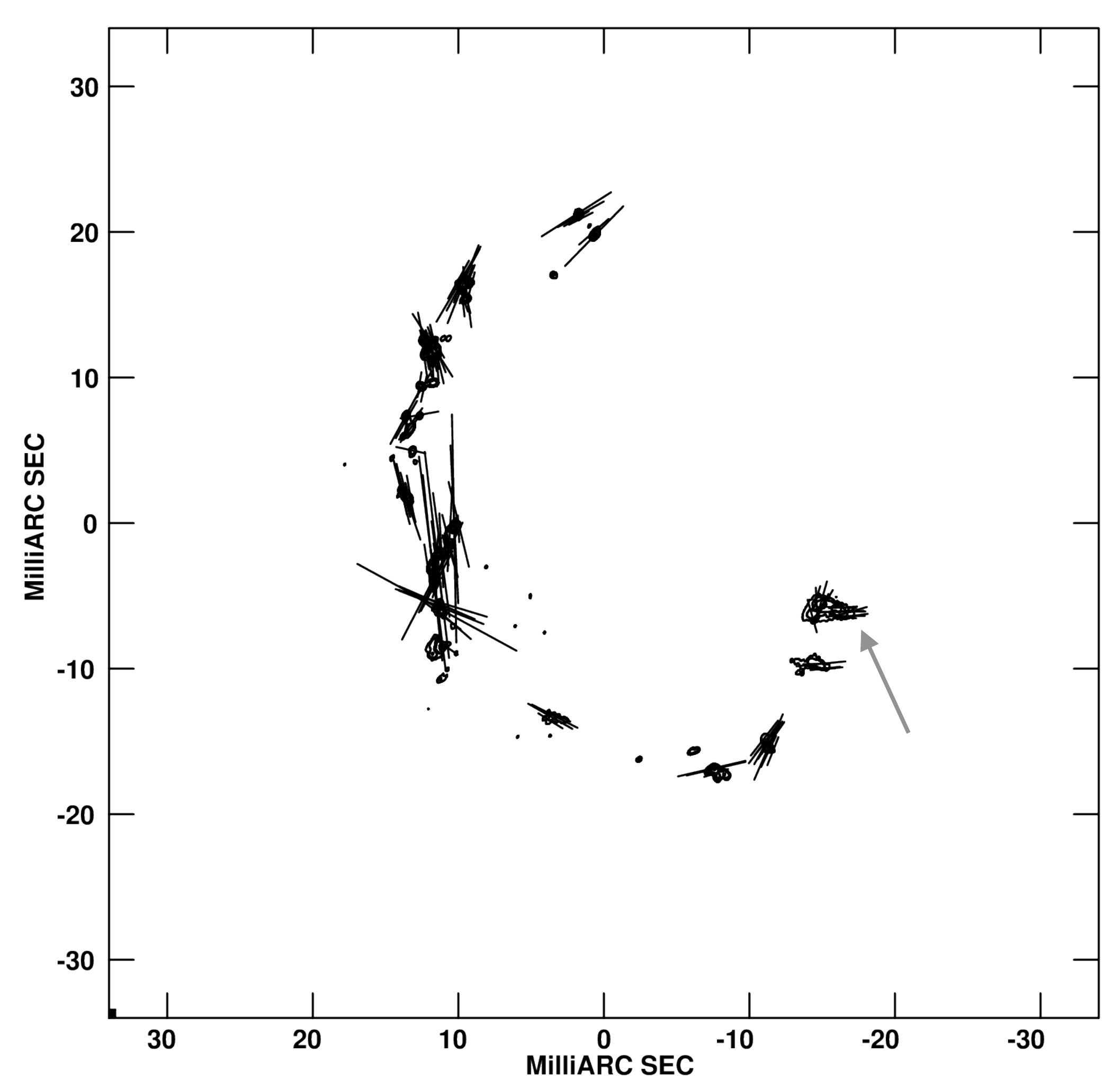}
\caption{Frequency--averaged Stokes I contour image with linear polarization contours of the $\nu=1,J=1-0$ SiO maser emission toward TX Cam from the epoch BD46AO. Contour levels are $\{-10,-5,5,10,20,40,80,160,320\} \times \sigma$, where $\sigma_{AO} = 1.6430$ mJy beam$^{-1}$. Vectors are at the angle of the EVPA and with a length proportional to the zeroth moment linearly polarized intensity such that 1 mas in length indicates $P=4$ mJy beam$^{-1}$. CLEAN beam size in this epoch is $ 443 \times 404$ $\mu$as, and too small to be visible at this scale. Spatial coordinates are with reference to the center of the aligned subimage. The grey arrow denotes the maser feature analyzed in this work. Corresponding images for all epochs are available in the online journal. \label{fig-iconpvec}}
\end{figure}

\begin{figure}[t!]
\figurenum{2}
\plotone{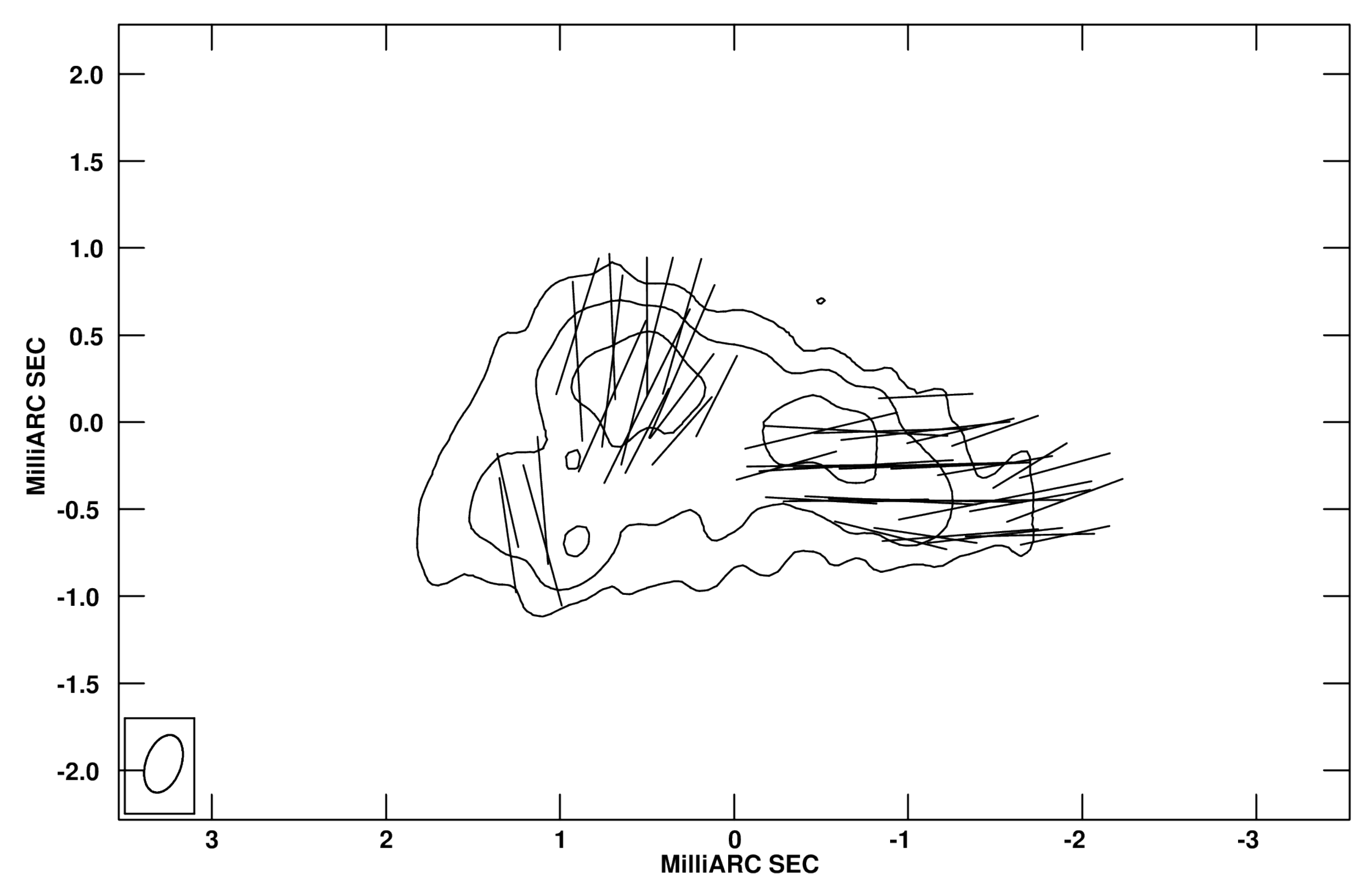}
\caption{Frequency--averaged Stokes I contour image with linear polarization contours of the $\nu=1,J=1-0$ SiO maser emission  for the EVPA rotation feature from the epoch BD46AO. Contour levels plotted are $\{-10,-5,5,10,20\} \times \sigma$, where $\sigma_{AO} = 1.6430$ mJy beam$^{-1}$. The lowest contour level visible  here is $5\sigma$. Vectors are at the angle of the EVPA and with a length proportional to the zeroth moment linearly polarized intensity such that 1 mas in length indicates $P=10$ mJy beam$^{-1}$. Spatial coordinates are with reference to the center of the aligned subimage. Corresponding images for all epochs are available in the online journal. \label{fig-2}}
\end{figure}

\begin{figure*}[t!]
\figurenum{3}
\centering
\includegraphics[width=0.9\textwidth]{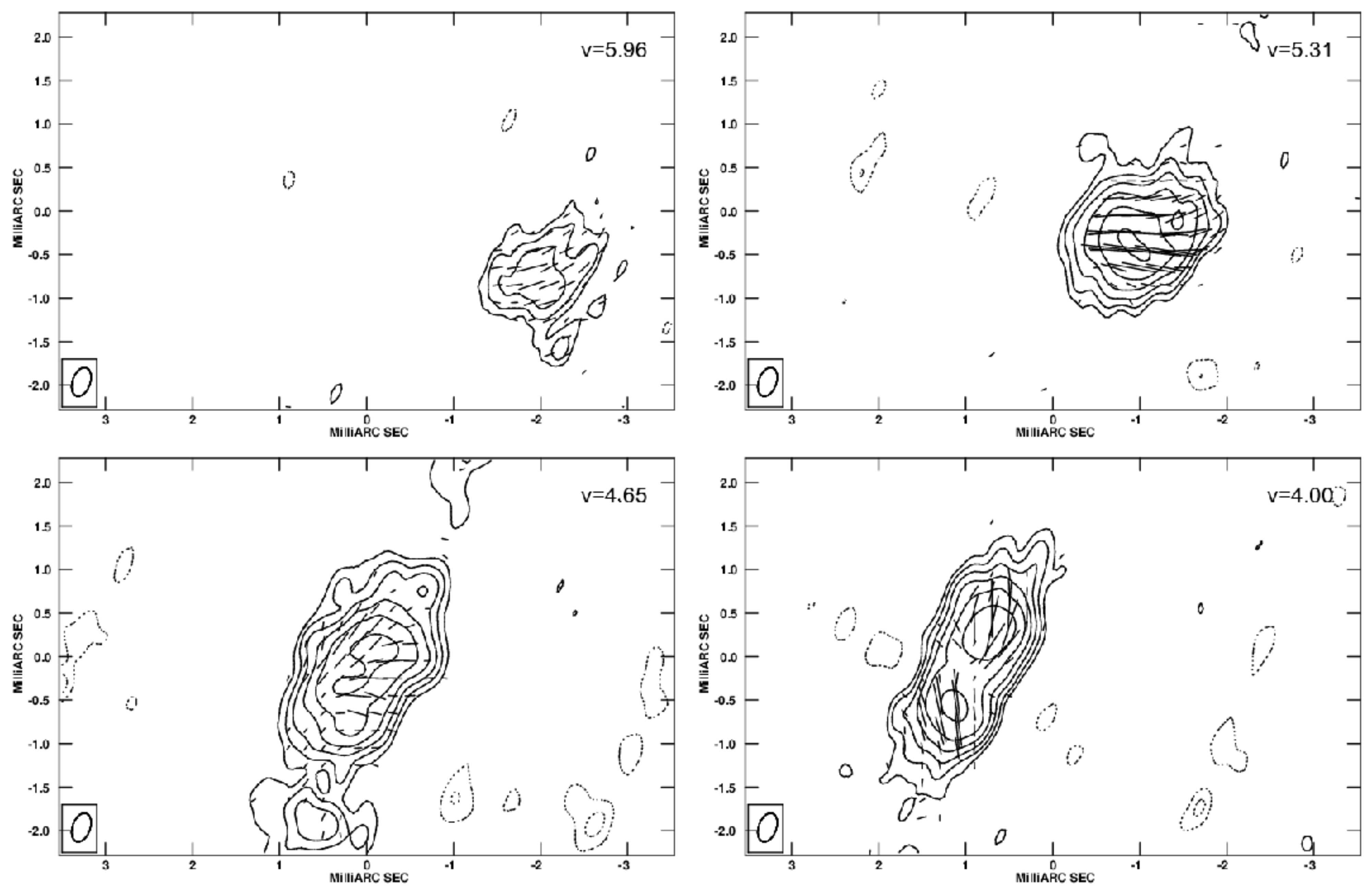}
\centering
\caption{Stokes I (contours) in a selection of equispaced channels from the epoch BD46AO, labeled with the LSR velocity (km s$^{-1}$). Contour levels are $\{ -12, -6, -3, 3, 6, 12, 24, 48, 96  \} \times \sigma$, where $\sigma_{AO} = 12.403$ mJy beam$^{-1}$. Vectors are at the angle of the EVPA and with a length proportional to the linearly polarized intensity such that 1 mas in length indicates $P=0.5$ Jy beam$^{-1}$. Spatial coordinates are with reference to the center of the aligned subimage. The plot with $v=4.65$ km s$^{-1}$ (lower left) is the channel with the minimum linear polarization as seen in Figure \ref{fig-6} and represents an approximate midpoint of the EVPA rotation as seen in Figure \ref{fig-7}. \label{fig-3}}
\end{figure*}

\begin{figure*}[t!]
\figurenum{4}
\centering
\includegraphics[width=0.9\textwidth]{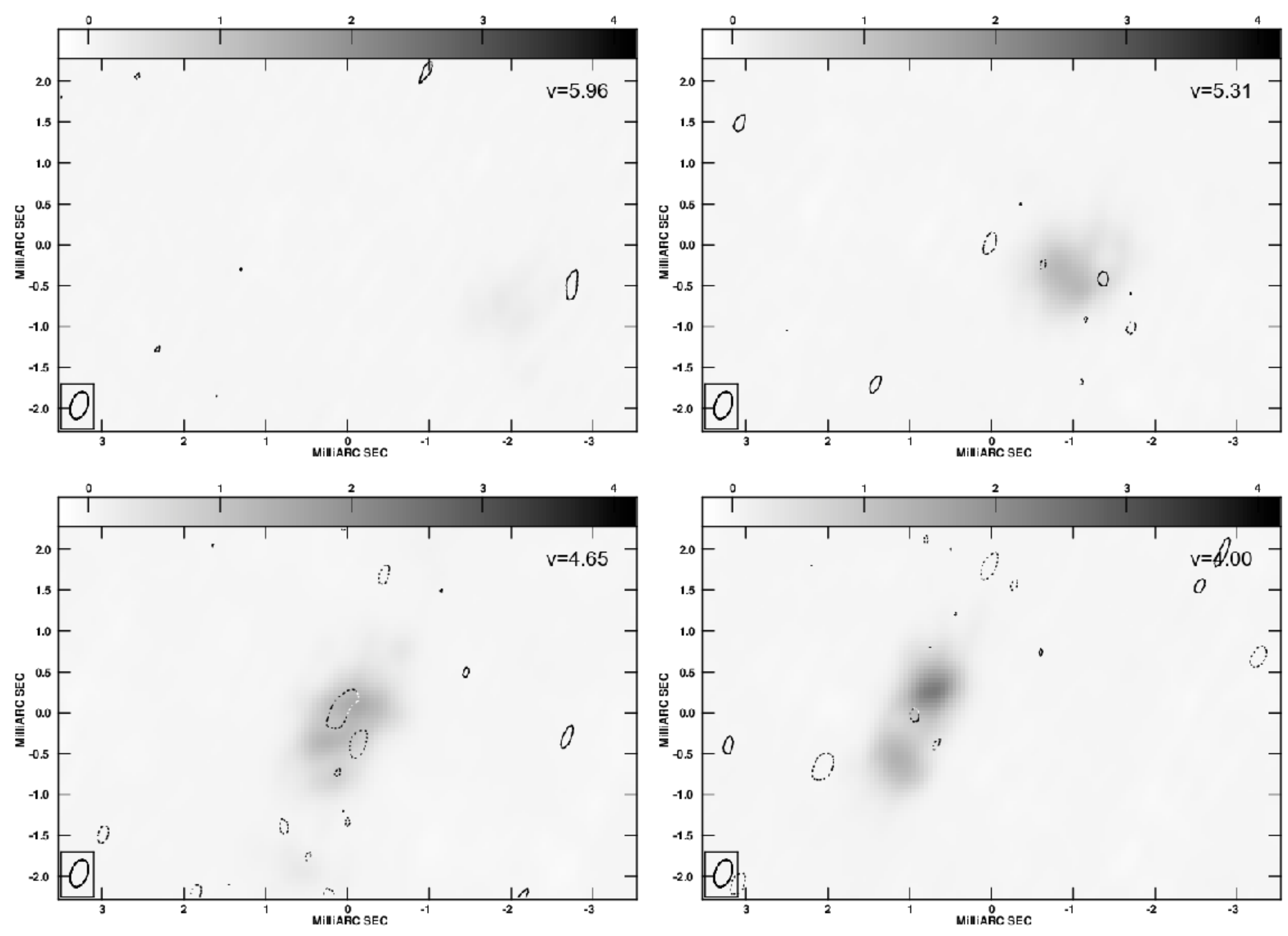}
\centering
\caption{Stokes V contours over Stokes I (greyscale) in a selection of equispaced channels from the epoch BD46AO, labeled with the LSR velocity (km s$^{-1}$) Contour levels are $\{ -6, -3, 3, 6 \} \times \sigma$, where $\sigma_{AO} = 12.403$ mJy beam$^{-1}$. Spatial coordinates are with reference to the center of the aligned subimage.  \label{fig-4}}
\end{figure*}

The frequency-averaged intensity and linear polarization for each of the six reduced epochs are shown in Figure Set \ref{fig-iconpvec}. The Stokes I flux density averaged over all velocity channels is represented by a contour map, with vectors representing the intensity and absolute EVPA of the average linearly polarized emission, calculated as $P=\sqrt{\left(Q^2+U^2\right)}$. The maser feature of interest for the EVPA reversal analysis is the one at approximately $\left(-16,-8\right)$ mas in Figure \ref{fig-iconpvec}, or the rightmost feature in each image. This feature is shown in in more detail in Figure \ref{fig-2}, again averaged over frequency. Figure \ref{fig-3} shows the feature's intensity and linear polarization as a function of frequency, while Figure \ref{fig-4} shows the feature's circular polarization across frequency.

\figsetstart
\figsetnum{1}
\figsettitle{Frequency--Averaged Stokes I Contours of $\nu=1,J=1-0$ SiO Maser Ring Towards TX Cam with Linear Polarization Vectors}

\figsetgrpstart
\figsetgrpnum{1.1}
\figsetgrptitle{BD46AN}
\figsetplot{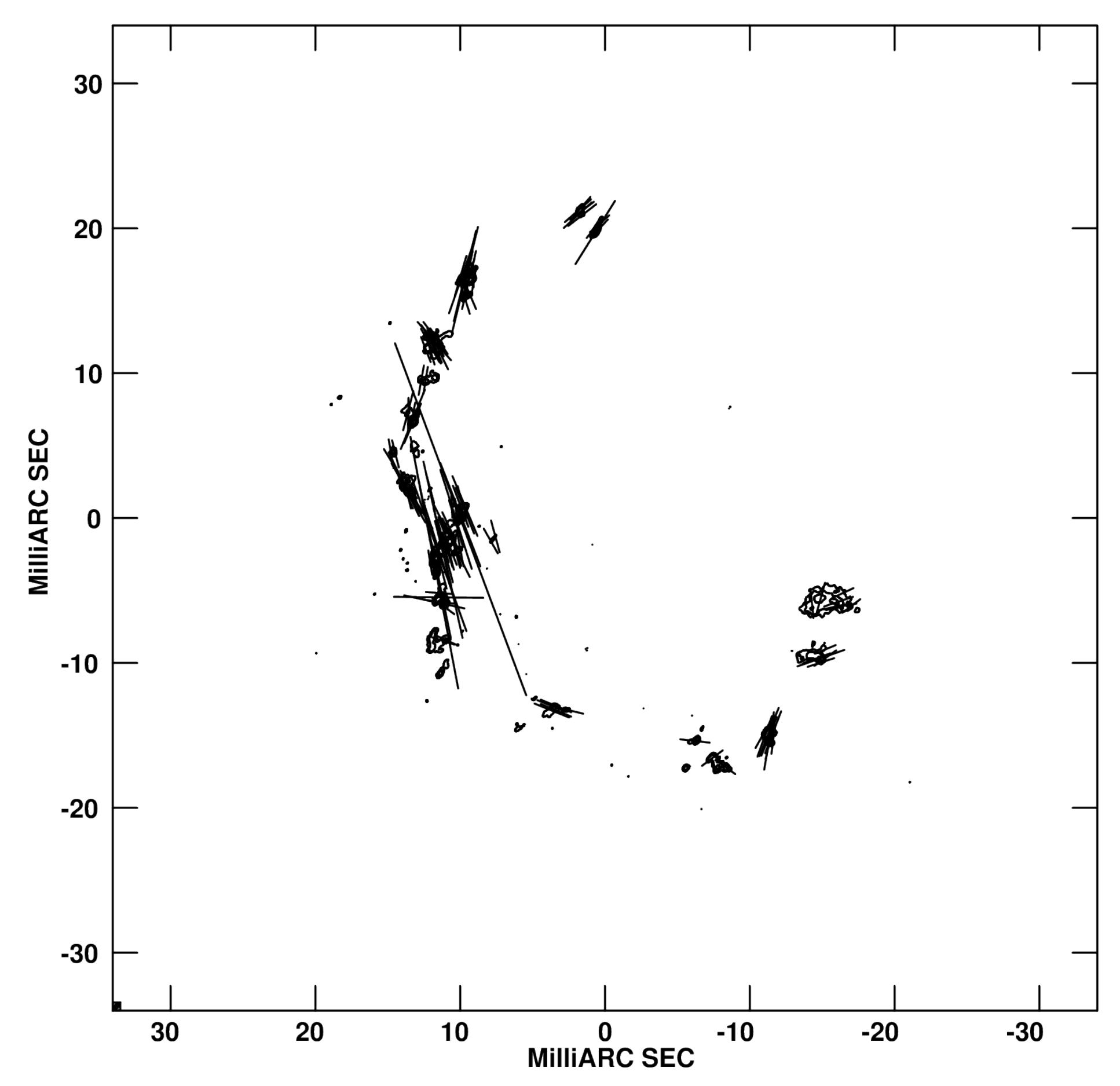}
\figsetgrpnote{Stokes I contours and linear polarization vectors for the full maser ring in epoch BD46AN. Contour levels are $\{-10,-5,5,10,20,40,80,160,320\} \times \sigma$, where $\sigma_{AN} = 1.5209$ mJy beam$^{-1}$. Vectors are at the angle of the EVPA and with a length proportional to the zeroth moment linearly polarized intensity such that 1 mas in length indicates $P=4$ mJy beam$^{-1}$. Spatial coordinates are with reference to the center of the aligned subimage.}
\figsetgrpend

\figsetgrpstart
\figsetgrpnum{1.2}
\figsetgrptitle{BD46AO}
\figsetplot{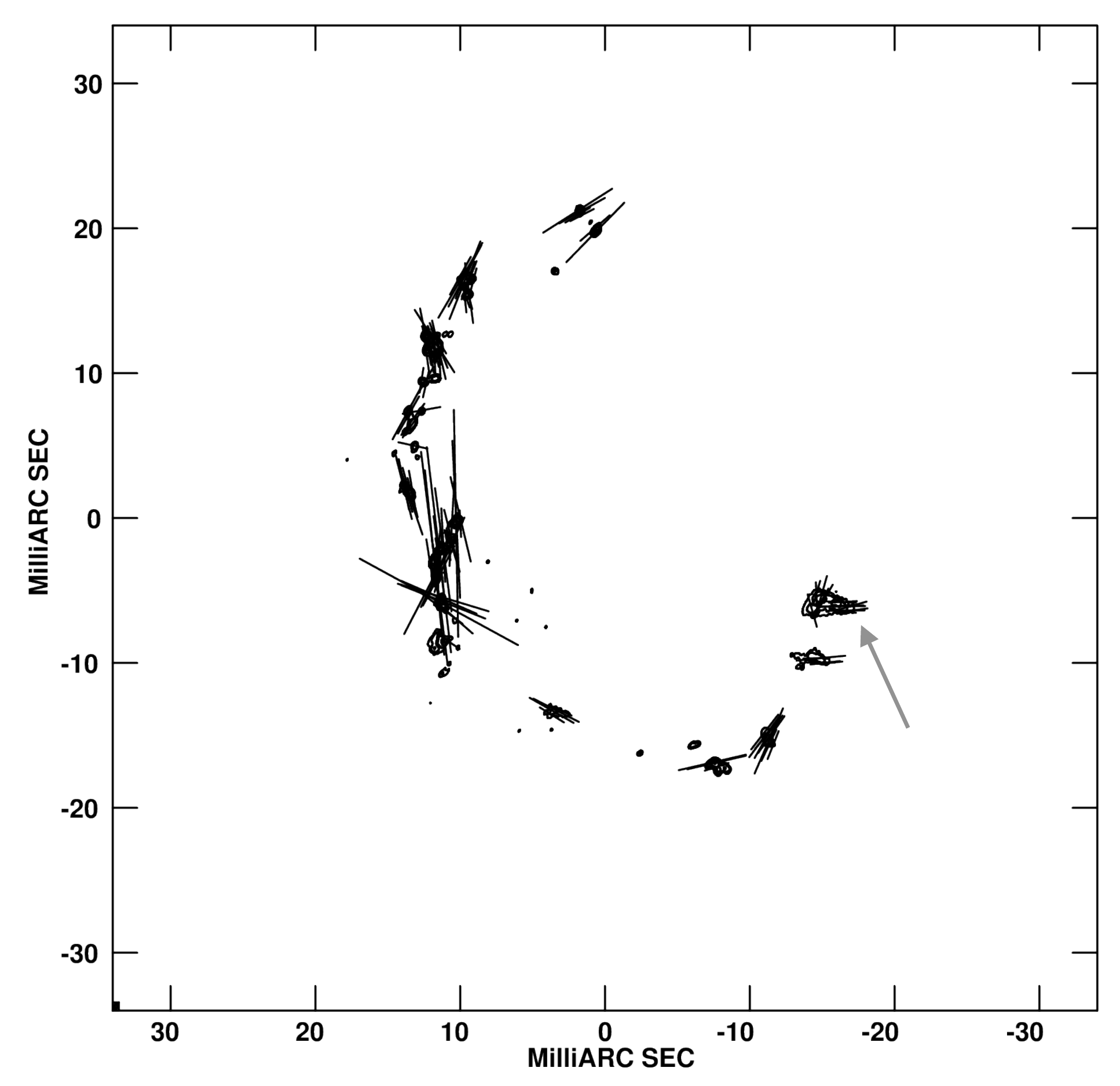}
\figsetgrpnote{Stokes I contours and linear polarization vectors for the full maser ring in epoch BD46AO. Contour levels and spatial scale are as in 1.1, with instead $sigma_{AO} = 1.6430$ mJy beam$^{-1}$.}
\figsetgrpend

\figsetgrpstart
\figsetgrpnum{1.3}
\figsetgrptitle{BD46AP}
\figsetplot{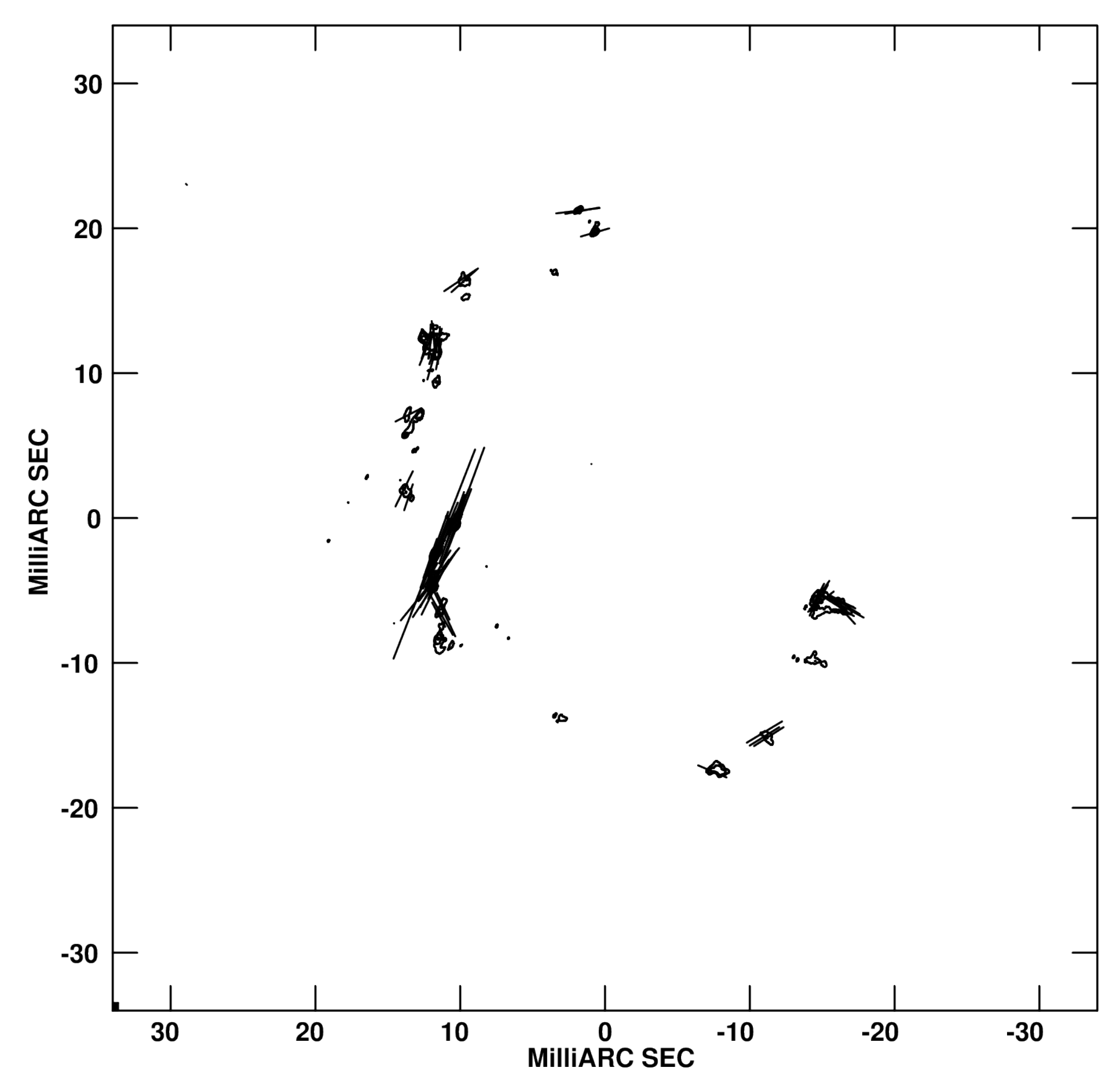}
\figsetgrpnote{Stokes I contours and linear polarization vectors for the full maser ring in epoch BD46AP. Contour levels and spatial scale are as in 1.1, with instead $sigma_{AP} = 1.8594$ mJy beam$^{-1}$.}
\figsetgrpend

\figsetgrpstart
\figsetgrpnum{1.4}
\figsetgrptitle{BD46AQ}
\figsetplot{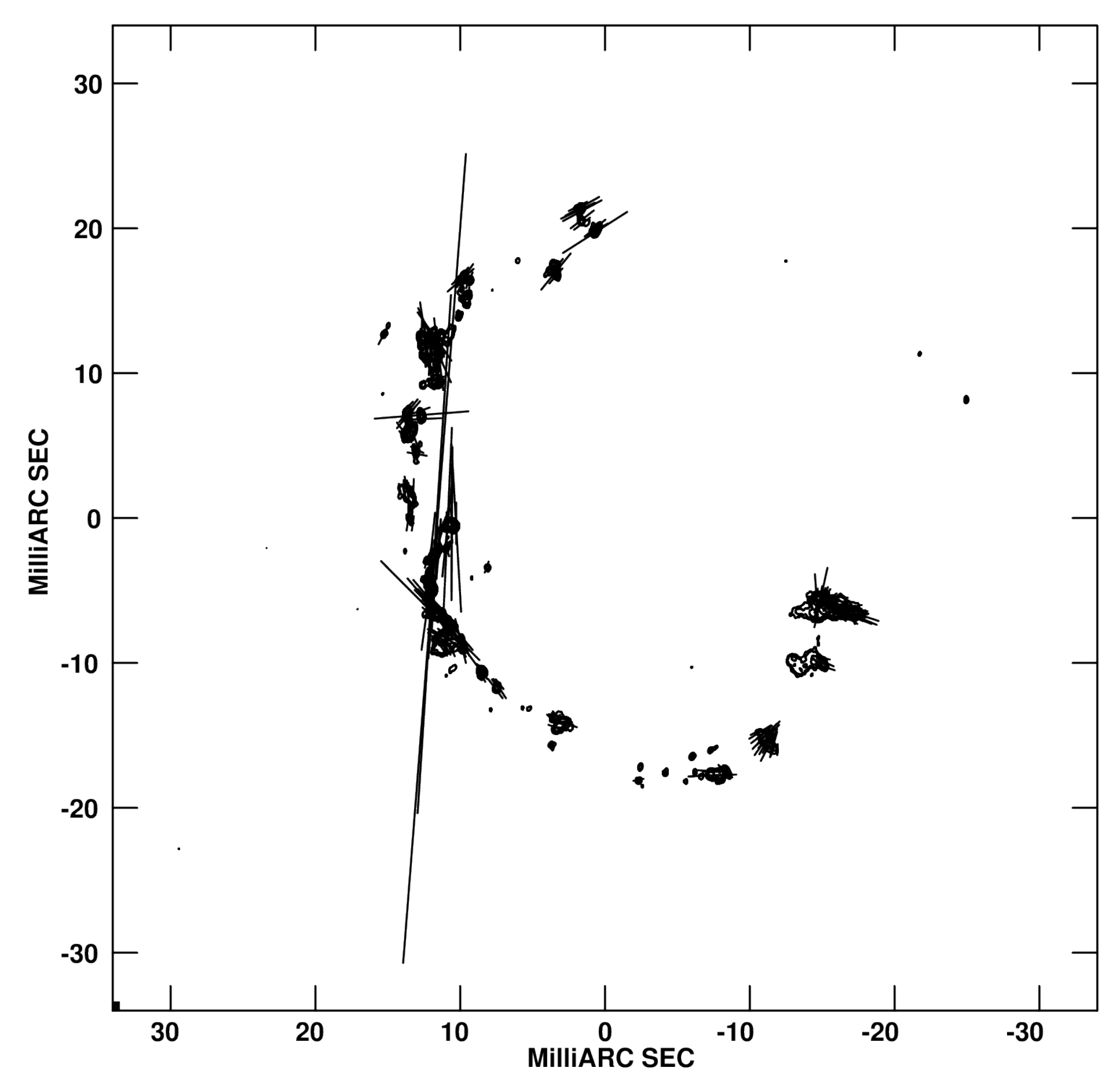}
\figsetgrpnote{Stokes I contours and linear polarization vectors for the full maser ring in epoch BD46AQ. Contour levels and spatial scale are as in 1.1, with instead $sigma_{AQ} = 0.64265$ mJy beam$^{-1}$.}
\figsetgrpend

\figsetgrpstart
\figsetgrpnum{1.5}
\figsetgrptitle{BD46AR}
\figsetplot{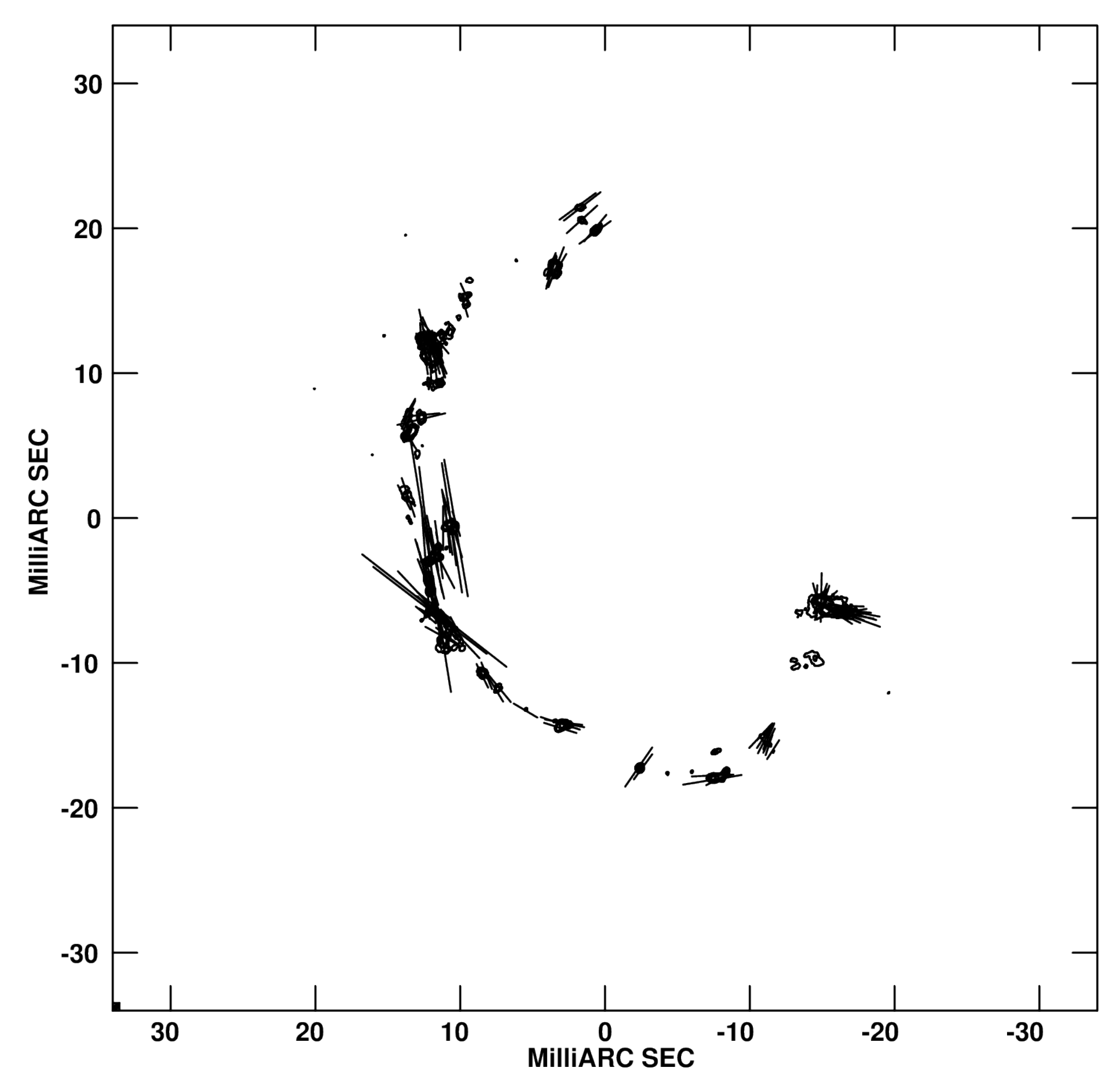}
\figsetgrpnote{Stokes I contours and linear polarization vectors for the full maser ring in epoch BD46AR. Contour levels and spatial scale are as in 1.1, with instead $sigma_{AR} = 1.4597$ mJy beam$^{-1}$.}
\figsetgrpend

\figsetend

\figsetstart
\figsetnum{2}
\figsettitle{Frequency--Averaged Stokes I Contours with Linear Polarization Vectors of $\nu=1,J=1-0$ SiO Maser Feature with $\pi/2$ EVPA Rotation}

\figsetgrpstart
\figsetgrpnum{2.1}
\figsetgrptitle{BD46AN}
\figsetplot{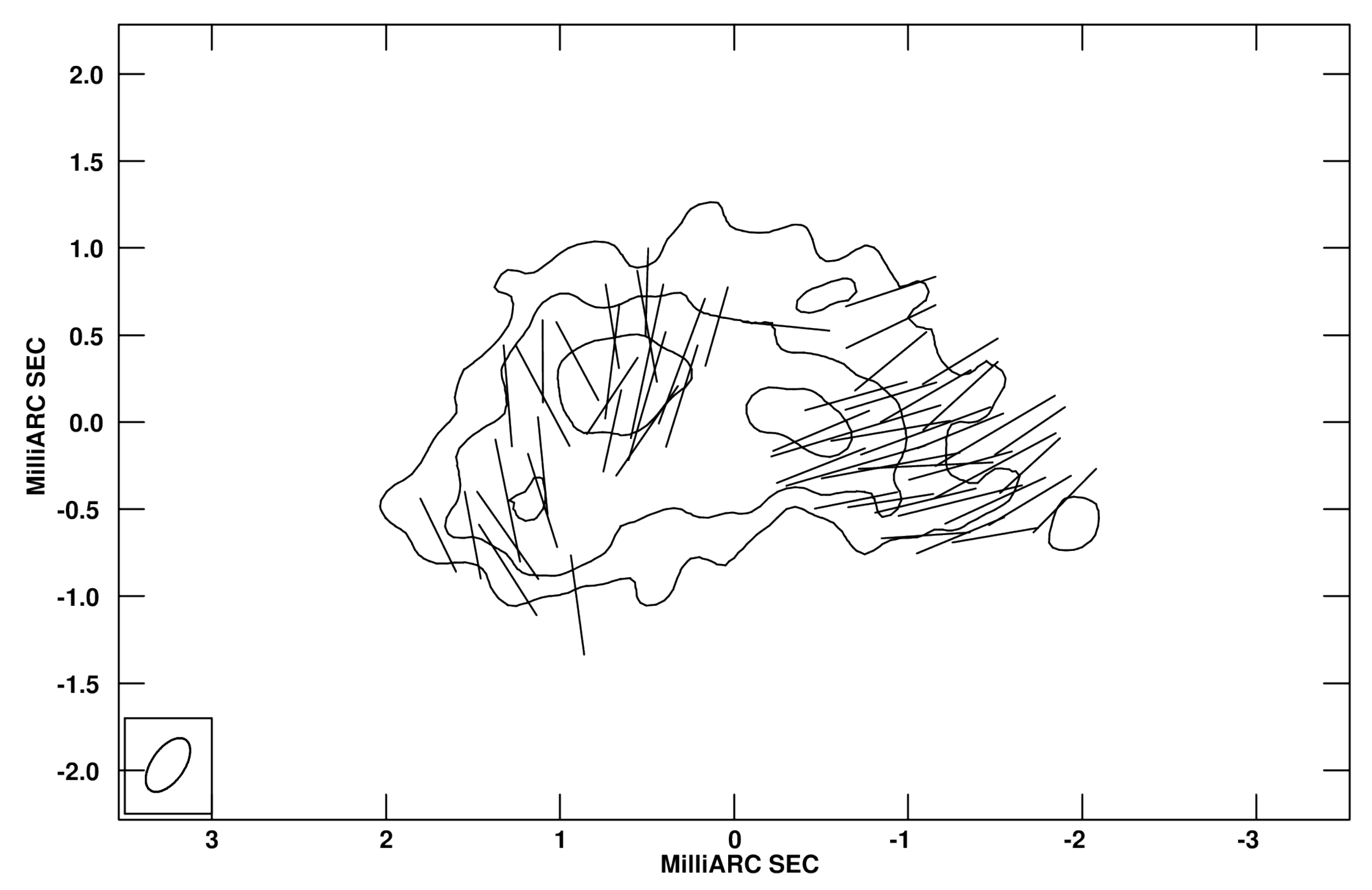}
\figsetgrpnote{Stokes I contours and linear polarization vectors for the target maser feature in epoch BD46AN. Contour levels are $\{-10,-5,5,10,20\} \times \sigma$, where $\sigma_{AN} = 1.5209$ mJy beam$^{-1}$. The lowest contour visible here is $5\sigma$. Vectors are at the angle of the EVPA and with a length proportional to the zeroth moment linearly polarized intensity such that 1 mas in length indicates $P=10$ mJy beam$^{-1}$. Spatial coordinates are with reference to the center of the aligned subimage.}
\figsetgrpend

\figsetgrpstart
\figsetgrpnum{2.2}
\figsetgrptitle{BD46AO}
\figsetplot{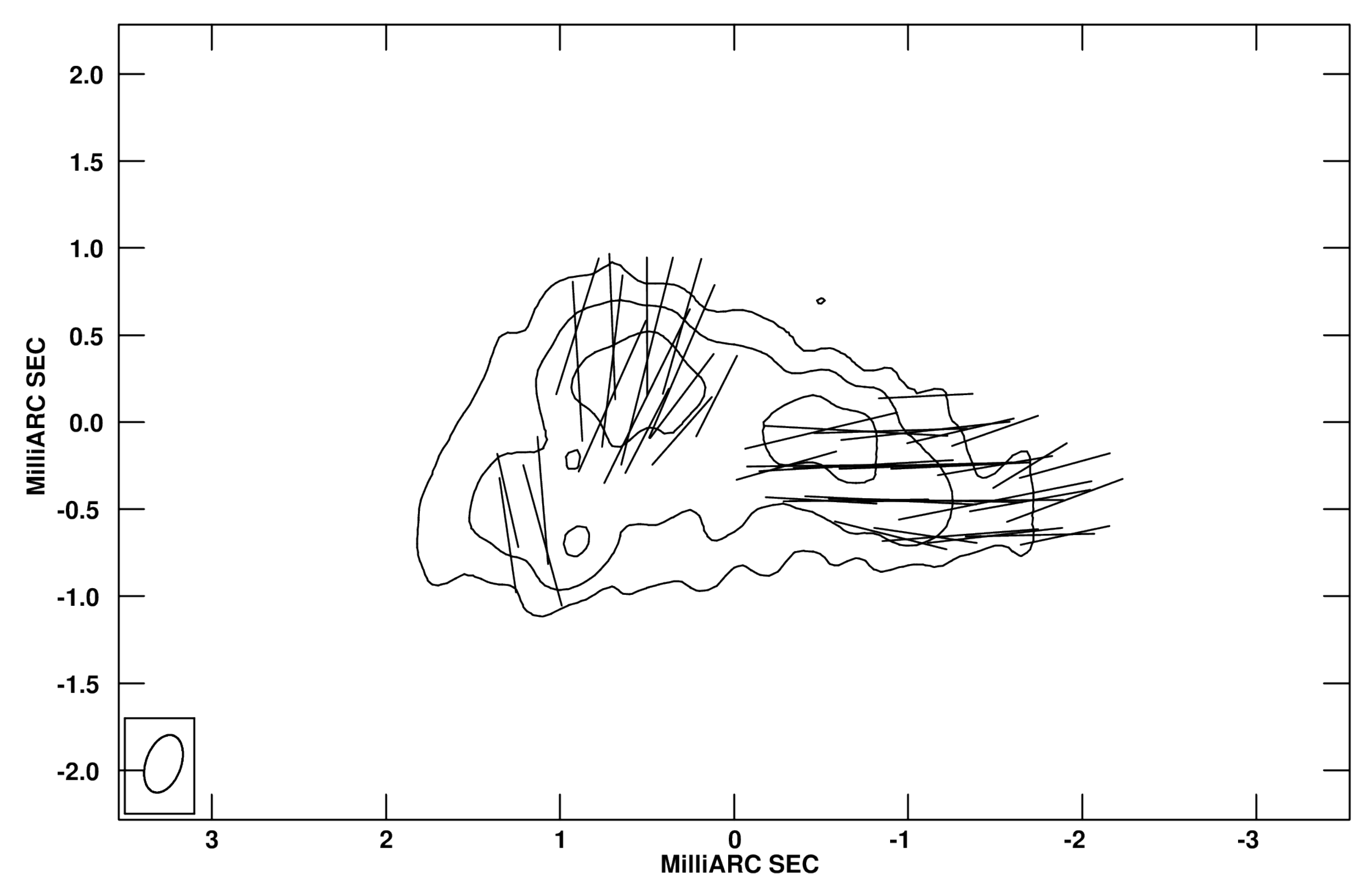}
\figsetgrpnote{Stokes I contours and linear polarization vectors for the target maser feature in epoch BD46AO. Spatial coordinates and polarization vectors are as in 2.1. Contour levels are $\{-10,-5,5,10,20\} \times \sigma$, where $sigma_{AO} = 1.6430$ mJy beam$^{-1}$. The lowest contour visible here is $5\sigma$.}
\figsetgrpend

\figsetgrpstart
\figsetgrpnum{2.3}
\figsetgrptitle{BD46AP}
\figsetplot{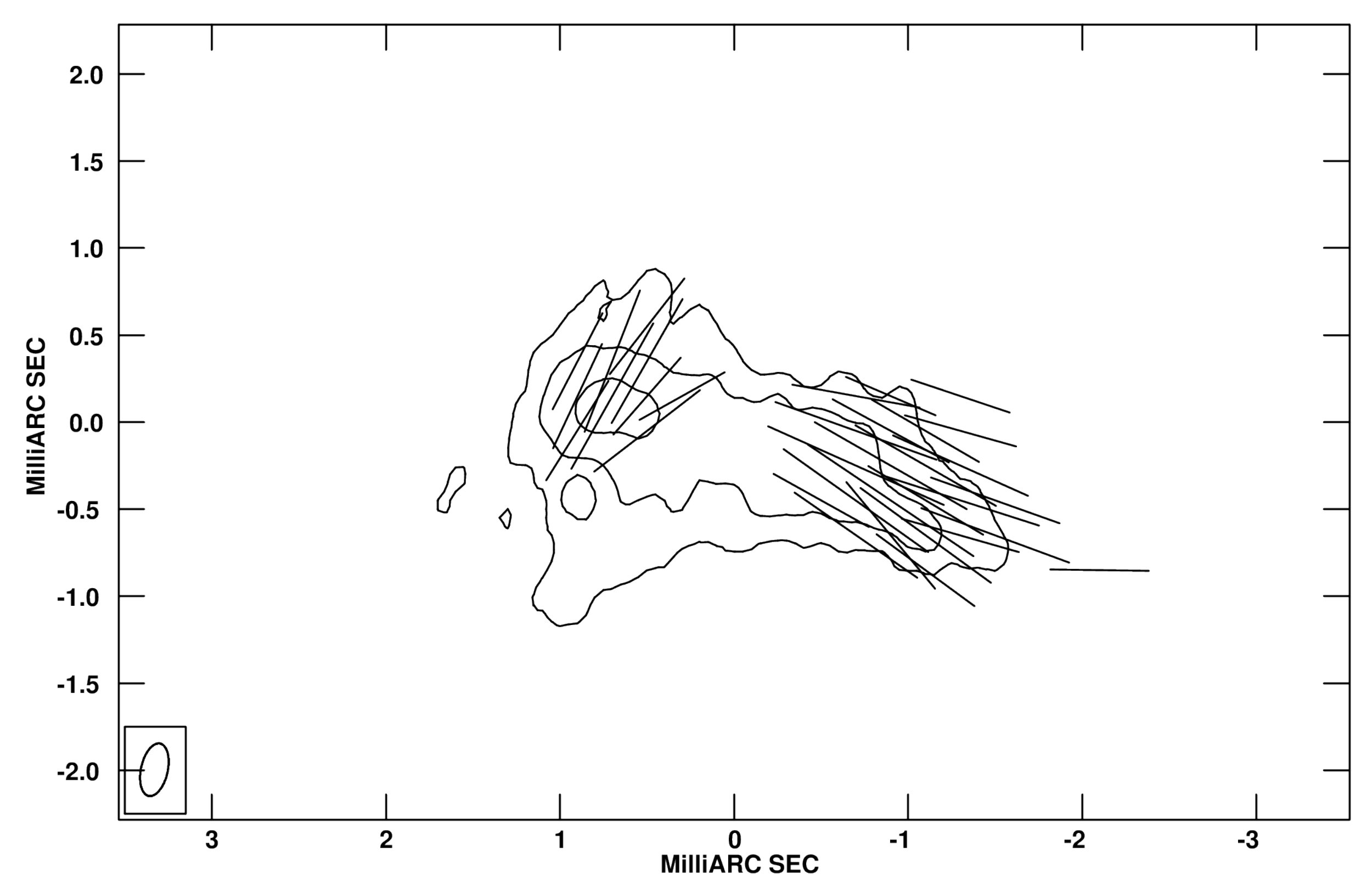}
\figsetgrpnote{Stokes I contours and linear polarization vectors for the target maser feature in epoch BD46AP. Spatial coordinates and polarization vectors are as in 2.1. Contour levels are $\{-10,-5,5,10,20\} \times \sigma$, where $sigma_{AP} = 1.8594$ mJy beam$^{-1}$. The lowest contour visible here is $5\sigma$.}
\figsetgrpend

\figsetgrpstart
\figsetgrpnum{2.4}
\figsetgrptitle{BD46AQ}
\figsetplot{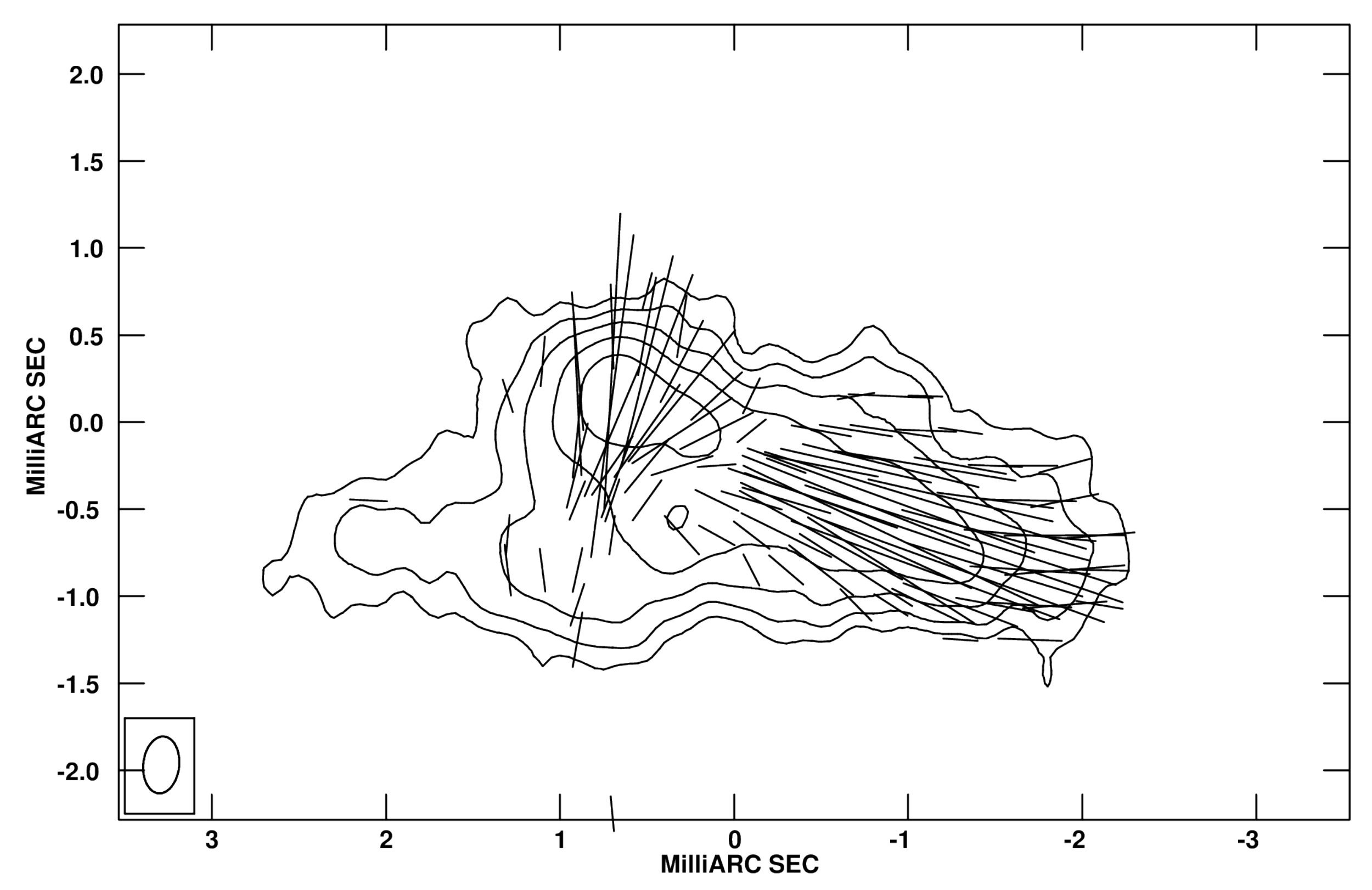}
\figsetgrpnote{Stokes I contours and linear polarization vectors for the target maser feature in epoch BD46AQ. Spatial coordinates and polarization vectors are as in 2.1. Contour levels are $\{-10,-5,5,10,20,40,80\} \times \sigma$, where $sigma_{AQ} = 0.64265$ mJy beam$^{-1}$. The lowest contour visible here is $5\sigma$.}
\figsetgrpend

\figsetgrpstart
\figsetgrpnum{2.5}
\figsetgrptitle{BD46AR}
\figsetplot{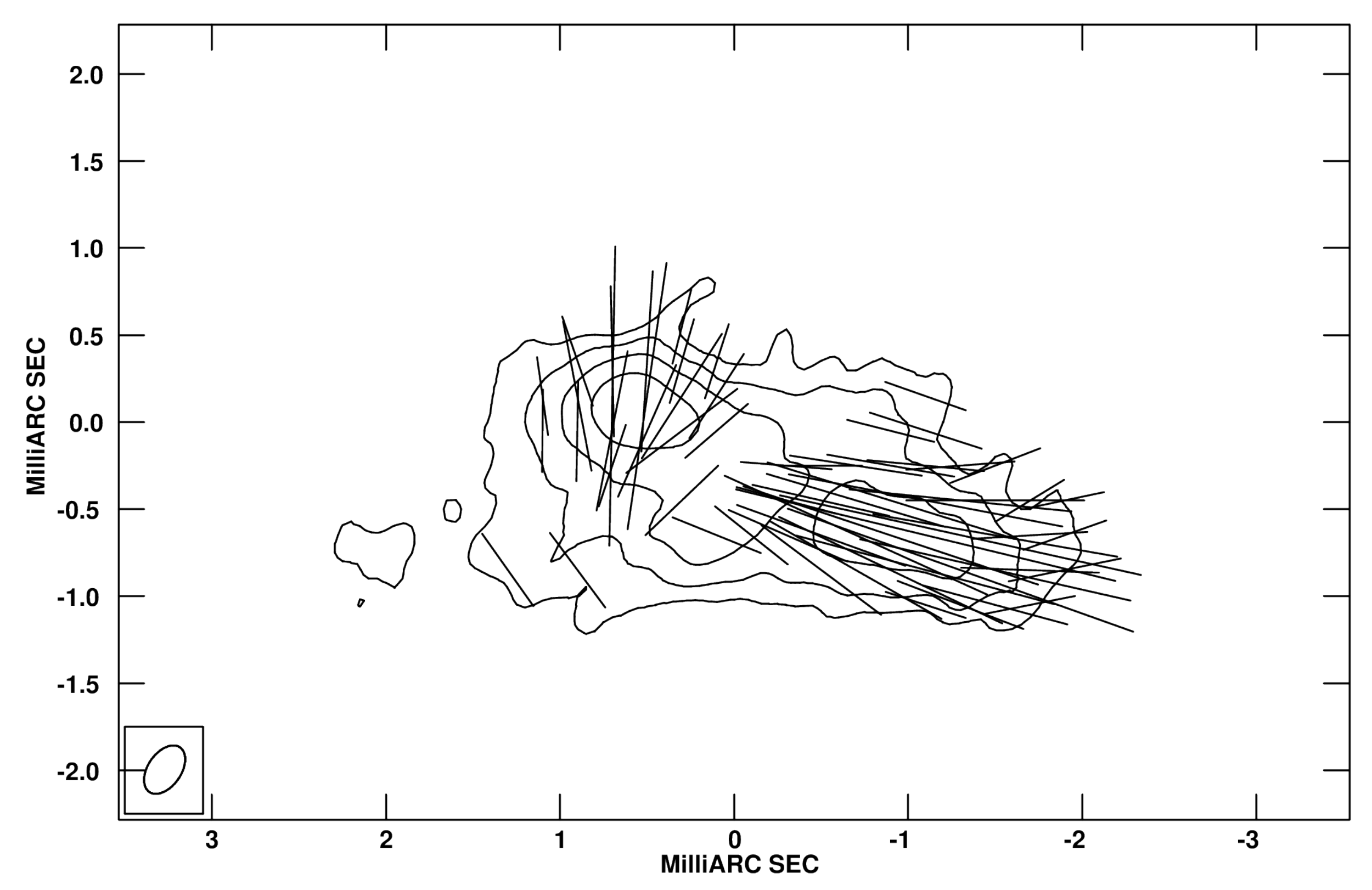}
\figsetgrpnote{Stokes I contours and linear polarization vectors for the target maser feature in epoch BD46AR. Spatial coordinates and polarization vectors are as in 2.1. Contour levels are $\{-10,05,5,10,20,40\} \times \sigma$, where $sigma_{AR} = 1.4597$ mJy beam$^{-1}$. The lowest contour visible here is $5\sigma$.}
\figsetgrpend

\figsetend

\figsetstart
\figsetnum{3}
\figsettitle{Channel-Level Stokes I Contours with Linear Polarization Vectors of $\nu=1,J=1-0$ SiO Maser Feature with $\pi/2$ EVPA Rotation}

\figsetgrpstart
\figsetgrpnum{3.1}
\figsetgrptitle{BD46AN}
\figsetplot{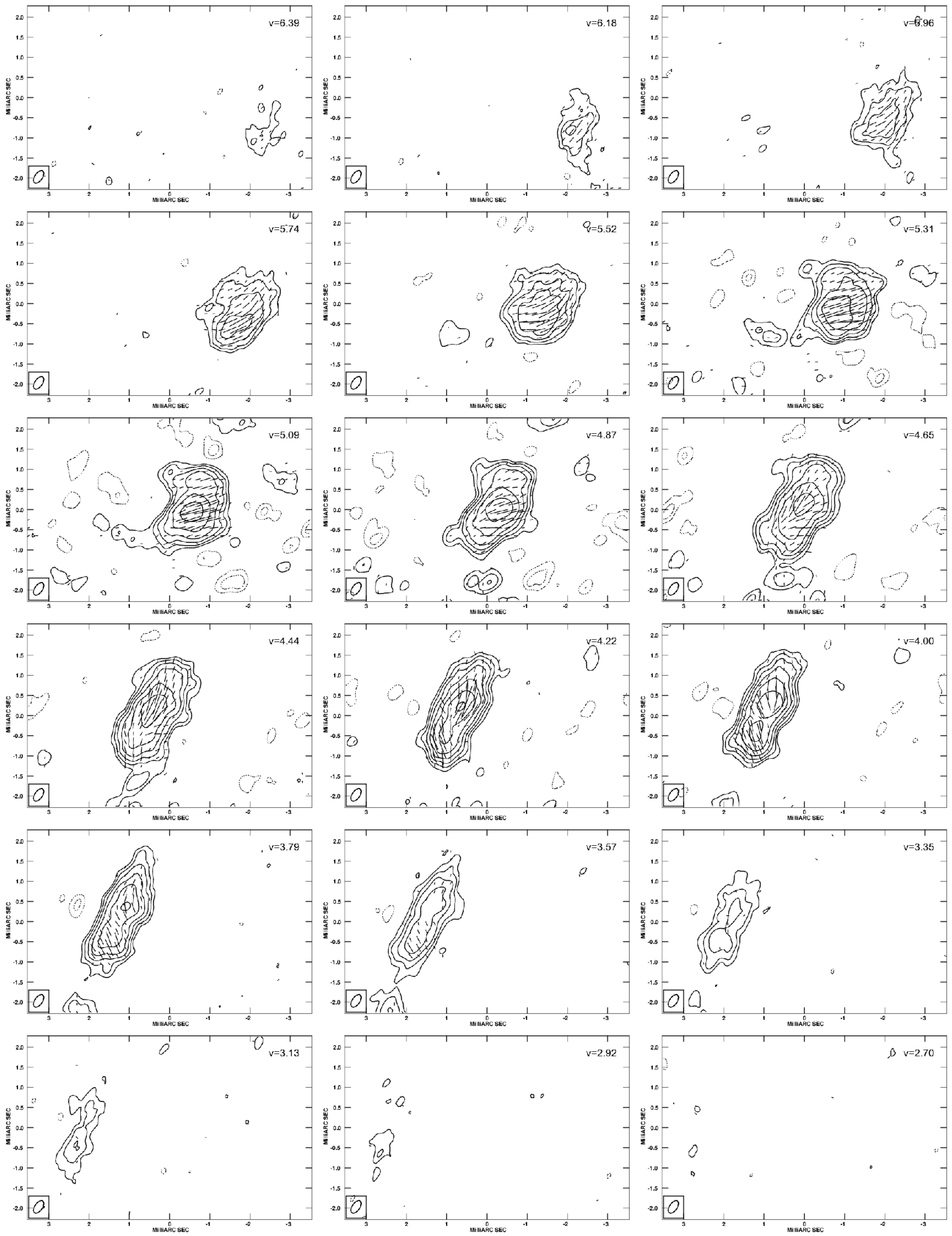}
\figsetgrpnote{Stokes I channel images, labeled with the LSR velocity (km s$^{-1}$), with linear polarization vectors for epoch BD46AN. Contour levels are $\{ -12, -6, -3, 3, 6, 12, 24, 48, 96, 192, 384, 768\} \times \sigma$,, where $\sigma_{AN} = 11.349$ mJy beam$^{-1}$. Vectors are at the angle of the EVPA and with a length proportional to the linearly polarized intensity such that 1 mas in length indicates $P=0.5$ Jy beam$^{-1}$. Spatial coordinates are with reference to the center of the aligned subimage.}
\figsetgrpend

\figsetgrpstart
\figsetgrpnum{3.2}
\figsetgrptitle{BD46AO}
\figsetplot{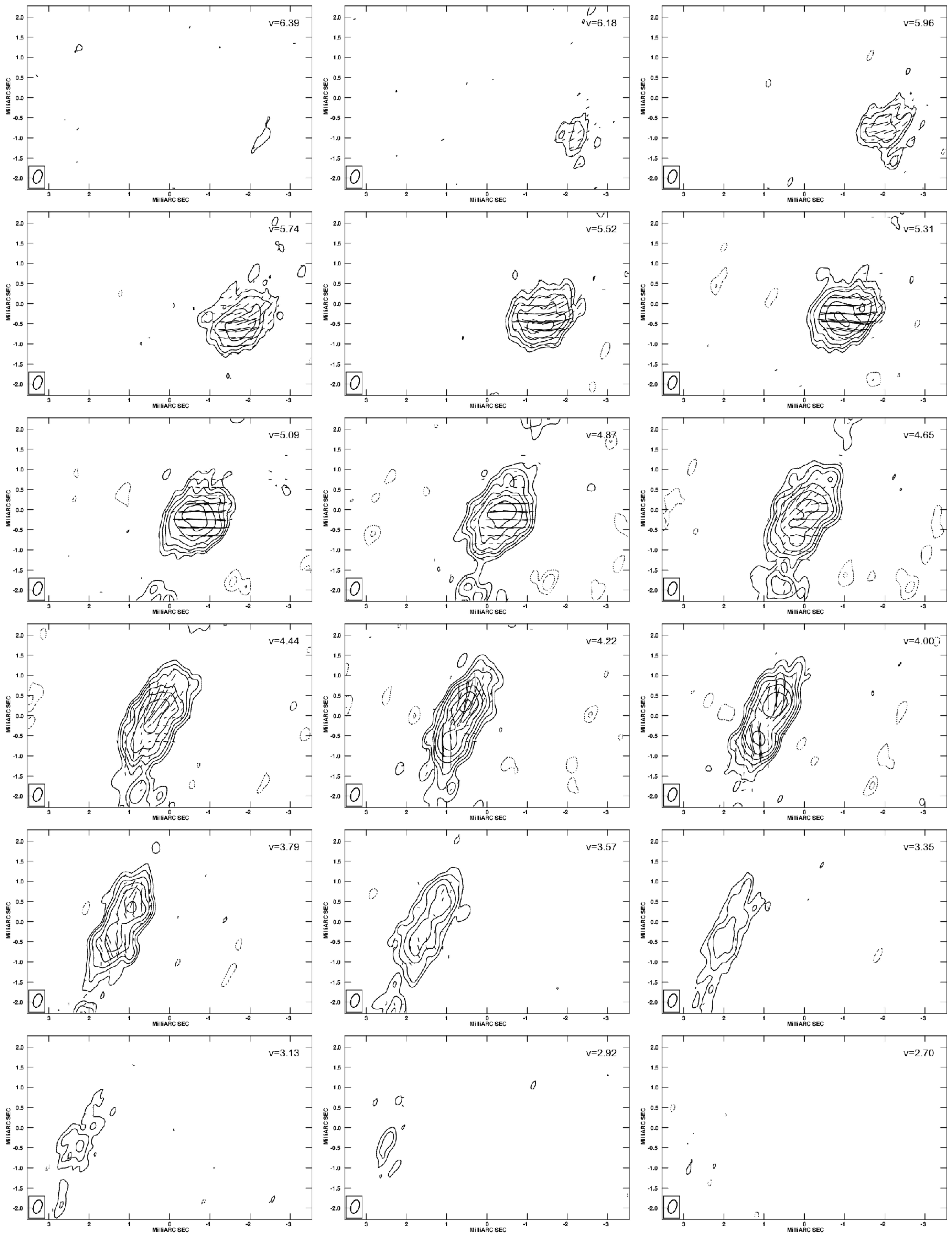}
\figsetgrpnote{Stokes I channel images with linear polarization vectors for epoch BD46AO. Contour levels and spatial scale are as in 3.1, with instead $sigma_{AO} = 12.403$ mJy beam$^{-1}$.}
\figsetgrpend

\figsetgrpstart
\figsetgrpnum{3.3}
\figsetgrptitle{BD46AP}
\figsetplot{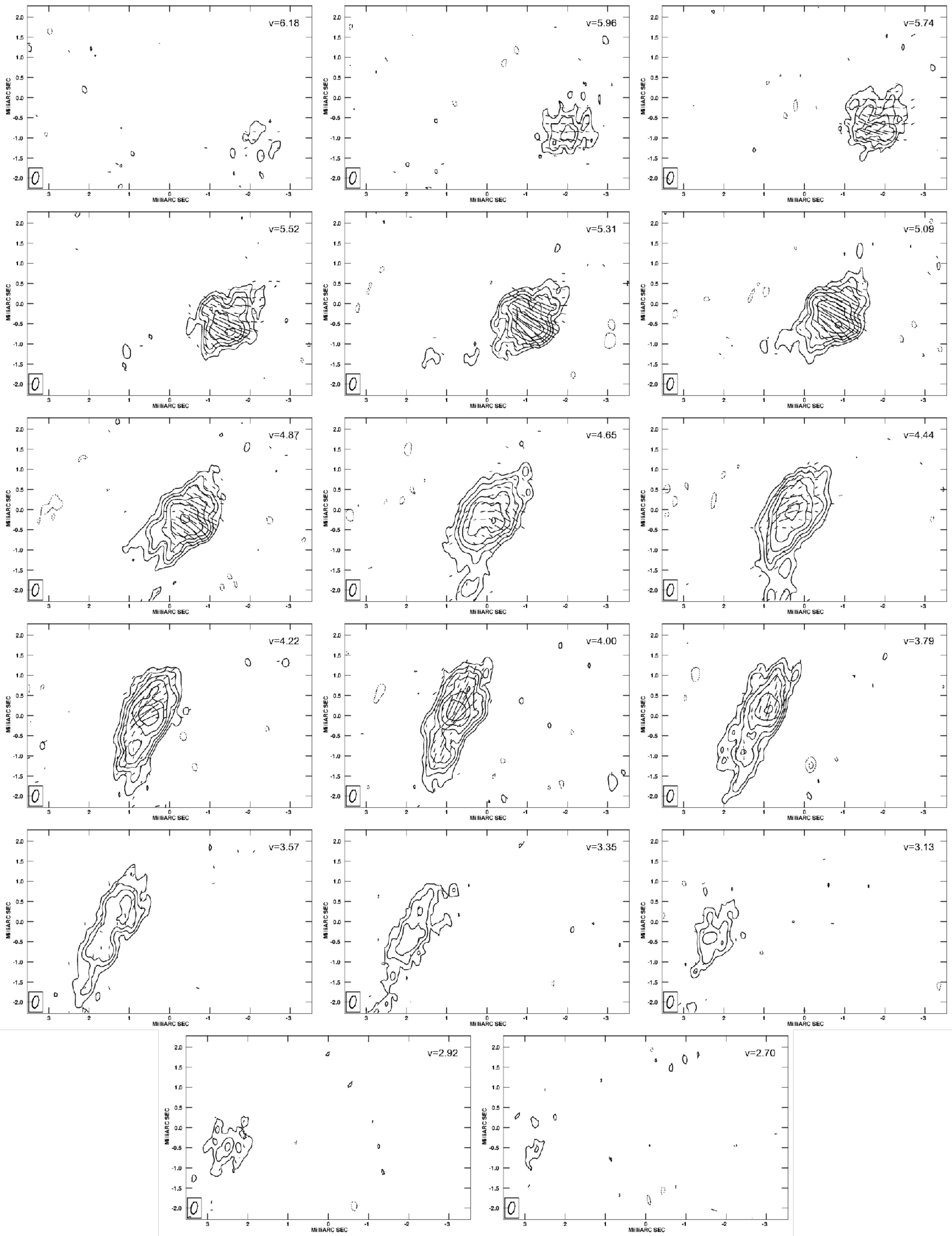}
\figsetgrpnote{Stokes I channel images with linear polarization vectors for epoch BD46AP. Contour levels and spatial scale are as in 3.1, with instead $sigma_{AP} = 13.293$ mJy beam$^{-1}$.}
\figsetgrpend

\figsetgrpstart
\figsetgrpnum{3.4}
\figsetgrptitle{BD46AQ}
\figsetplot{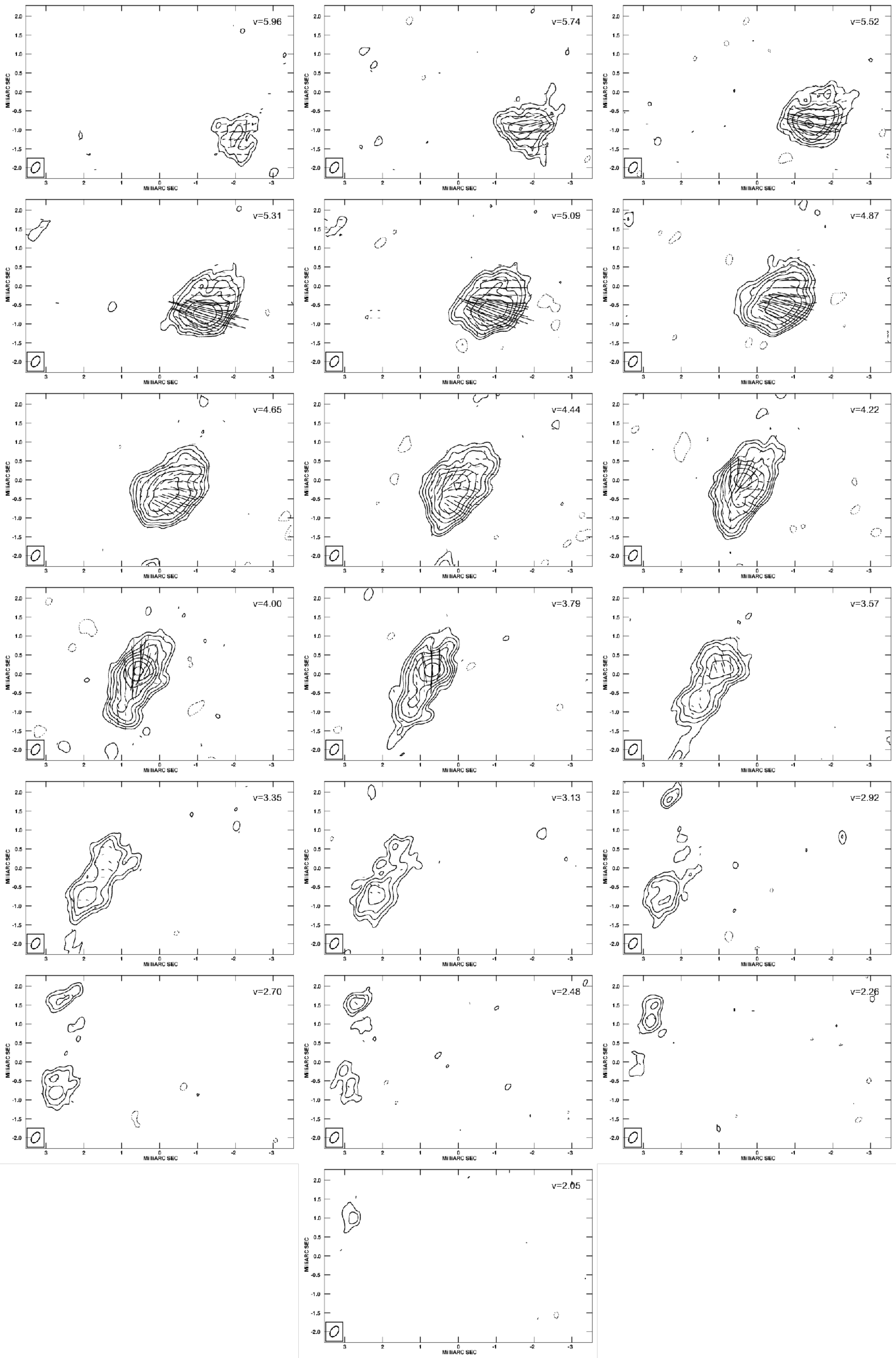}
\figsetgrpnote{Stokes I channel images with linear polarization vectors for epoch BD46AQ. Contour levels and spatial scale are as in 3.1, with instead $sigma_{AQ} = 5.5904$ mJy beam$^{-1}$.}
\figsetgrpend

\figsetgrpstart
\figsetgrpnum{3.5}
\figsetgrptitle{BD46AR}
\figsetplot{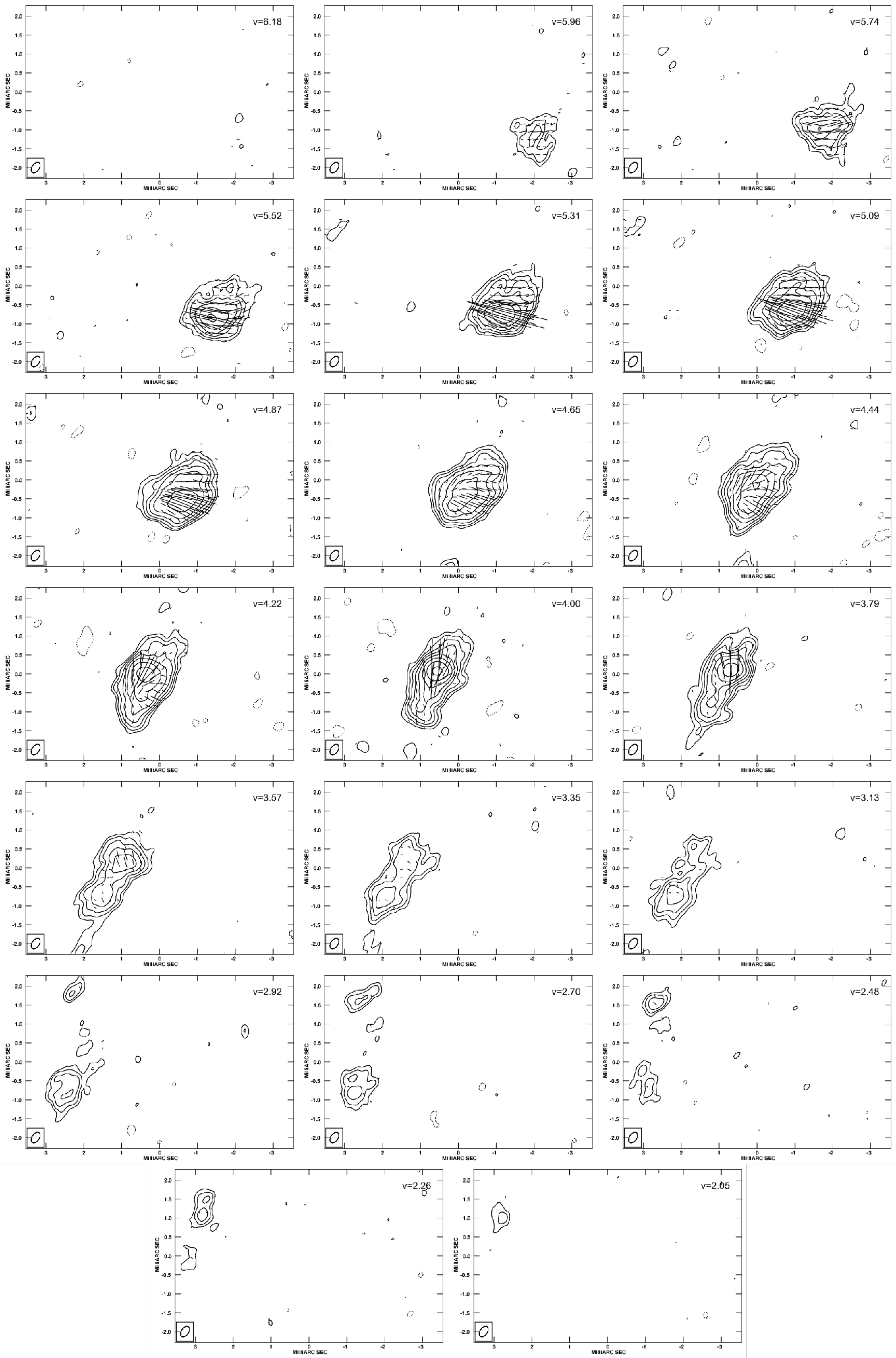}
\figsetgrpnote{Stokes I channel images with linear polarization vectors for epoch BD46AR. Contour levels and spatial scale are as in 3.1, with instead $sigma_{AR} = 11.334$ mJy beam$^{-1}$.}
\figsetgrpend

\figsetend

\figsetstart
\figsetnum{4}
\figsettitle{Channel-Level Stokes V Contours of $\nu=1,J=1-0$ SiO Maser Feature with $\pi/2$ EVPA Rotation}

\figsetgrpstart
\figsetgrpnum{4.1}
\figsetgrptitle{BD46AN}
\figsetplot{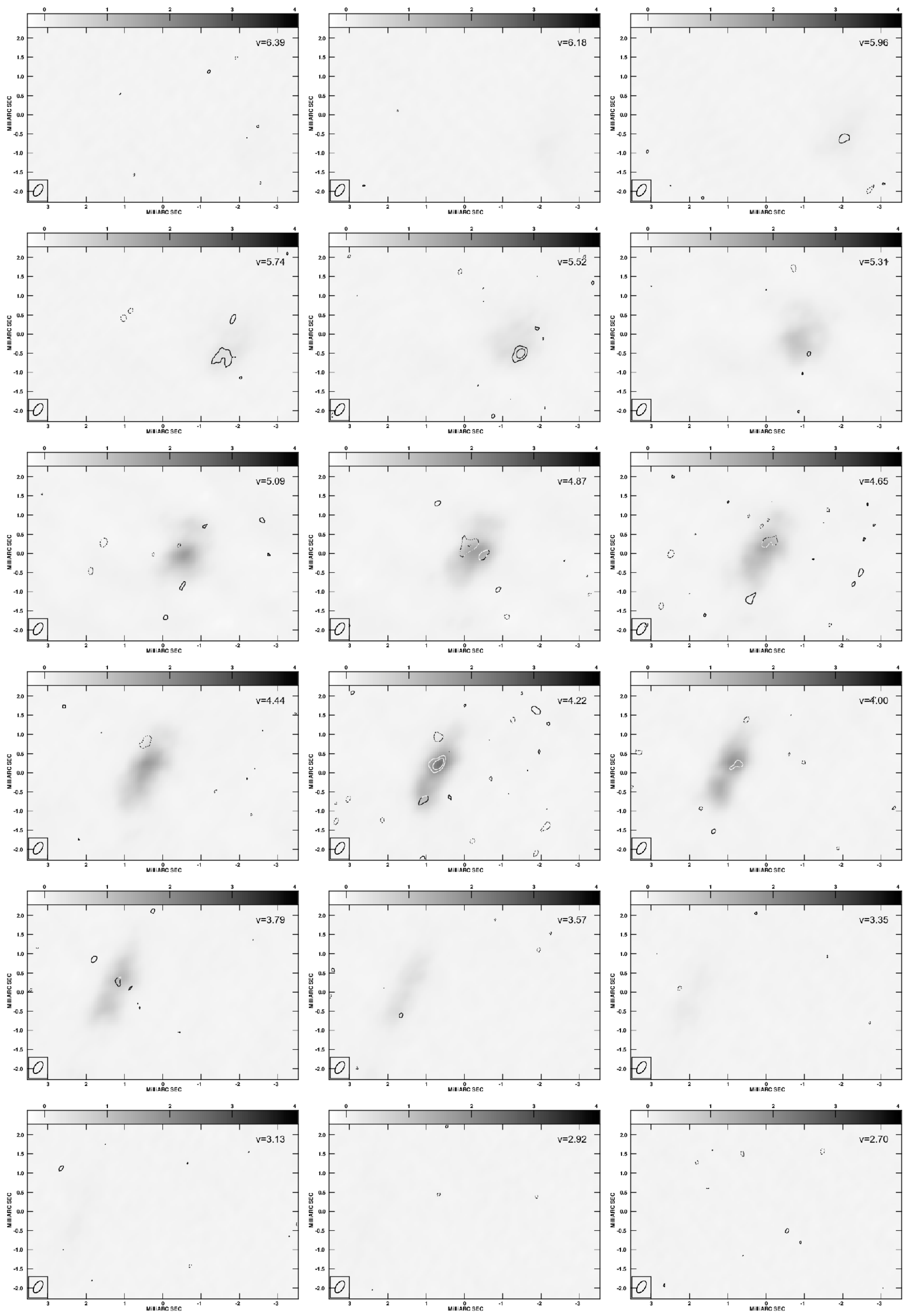}
\figsetgrpnote{Stokes V contours over Stokes I (greyscale) in each channel for epoch BD46AN, labeled with the LSR velocity (km s$^{-1}$) Contour levels are $\{-24, -12, -6, -3, 3, 6, 12, 24\} \times \sigma$, where $\sigma_{AN} = 11.349$ mJy beam$^{-1}$. Spatial coordinates are with reference to the center of the aligned subimage.}
\figsetgrpend

\figsetgrpstart
\figsetgrpnum{4.2}
\figsetgrptitle{BD46AO}
\figsetplot{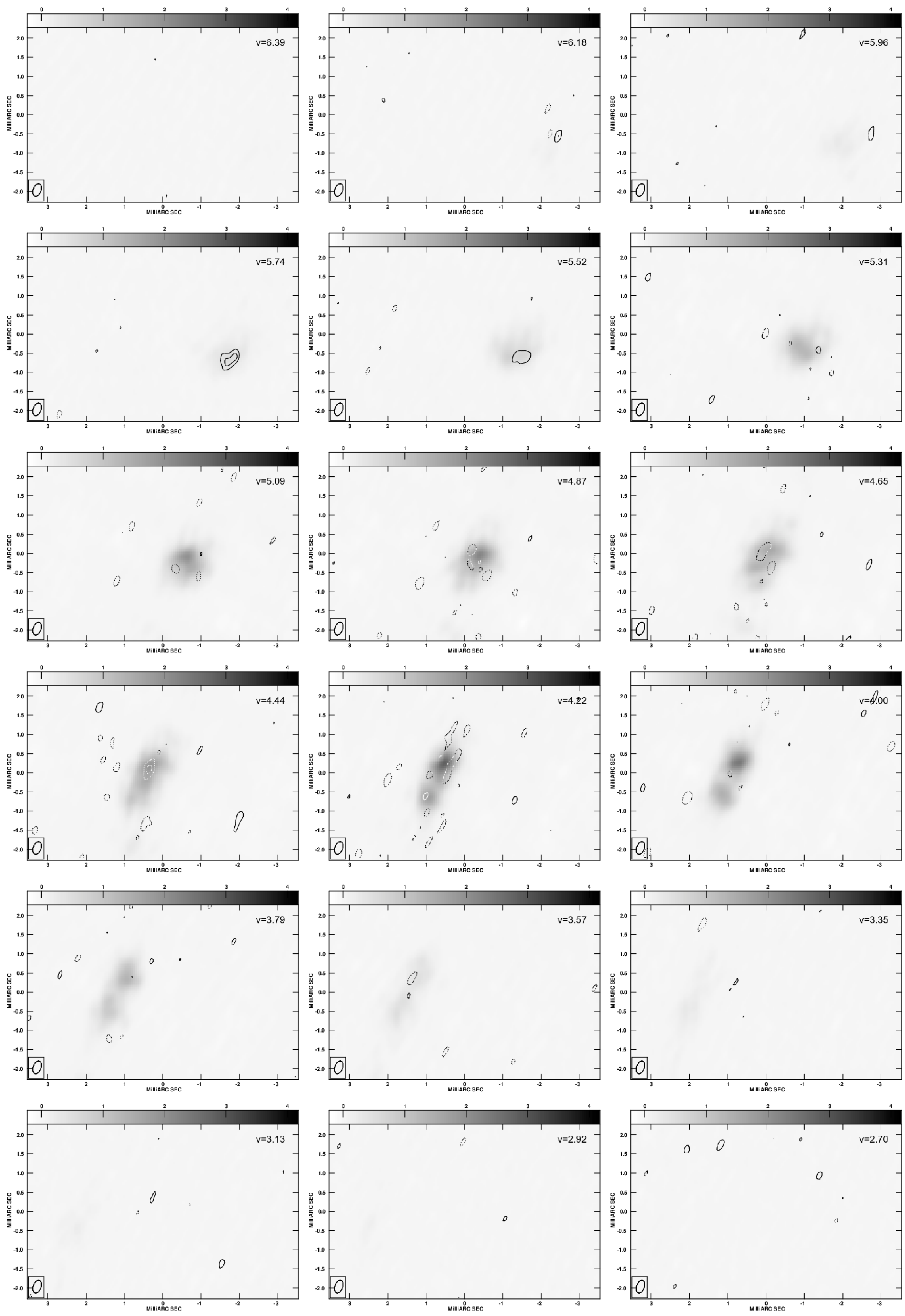}
\figsetgrpnote{Stokes V contours over Stokes I (greyscale) in each channel for epoch BD46AO. Contour levels and spatial scale are as in 4.1, with instead $sigma_{AO} = 12.403$ mJy beam$^{-1}$.}
\figsetgrpend

\figsetgrpstart
\figsetgrpnum{4.3}
\figsetgrptitle{BD46AP}
\figsetplot{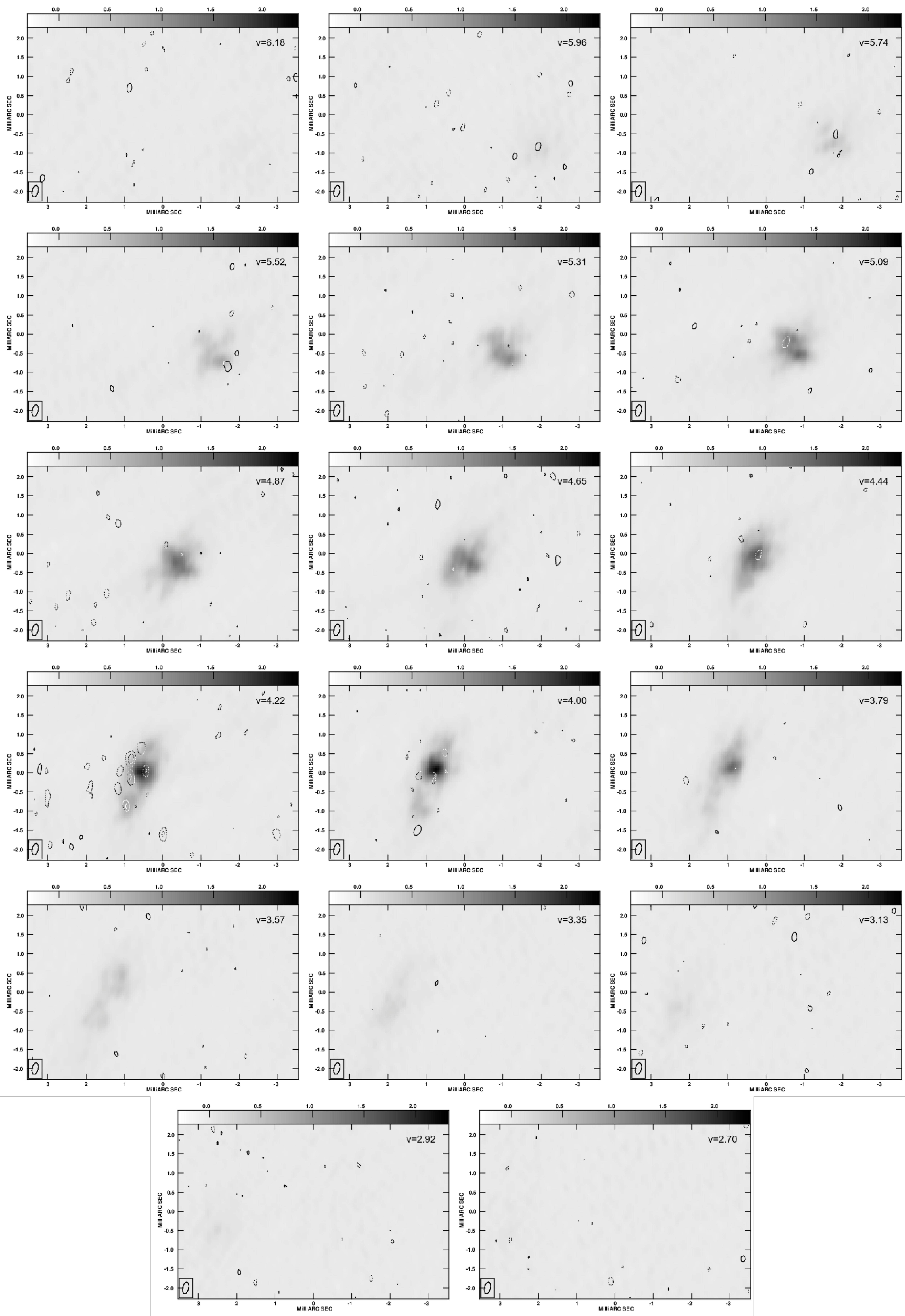}
\figsetgrpnote{Stokes V contours over Stokes I (greyscale) in each channel for epoch BD46AP. Contour levels and spatial scale are as in 4.1, with instead $sigma_{AP} = 13.293$ mJy beam$^{-1}$.}
\figsetgrpend

\figsetgrpstart
\figsetgrpnum{4.4}
\figsetgrptitle{BD46AQ}
\figsetplot{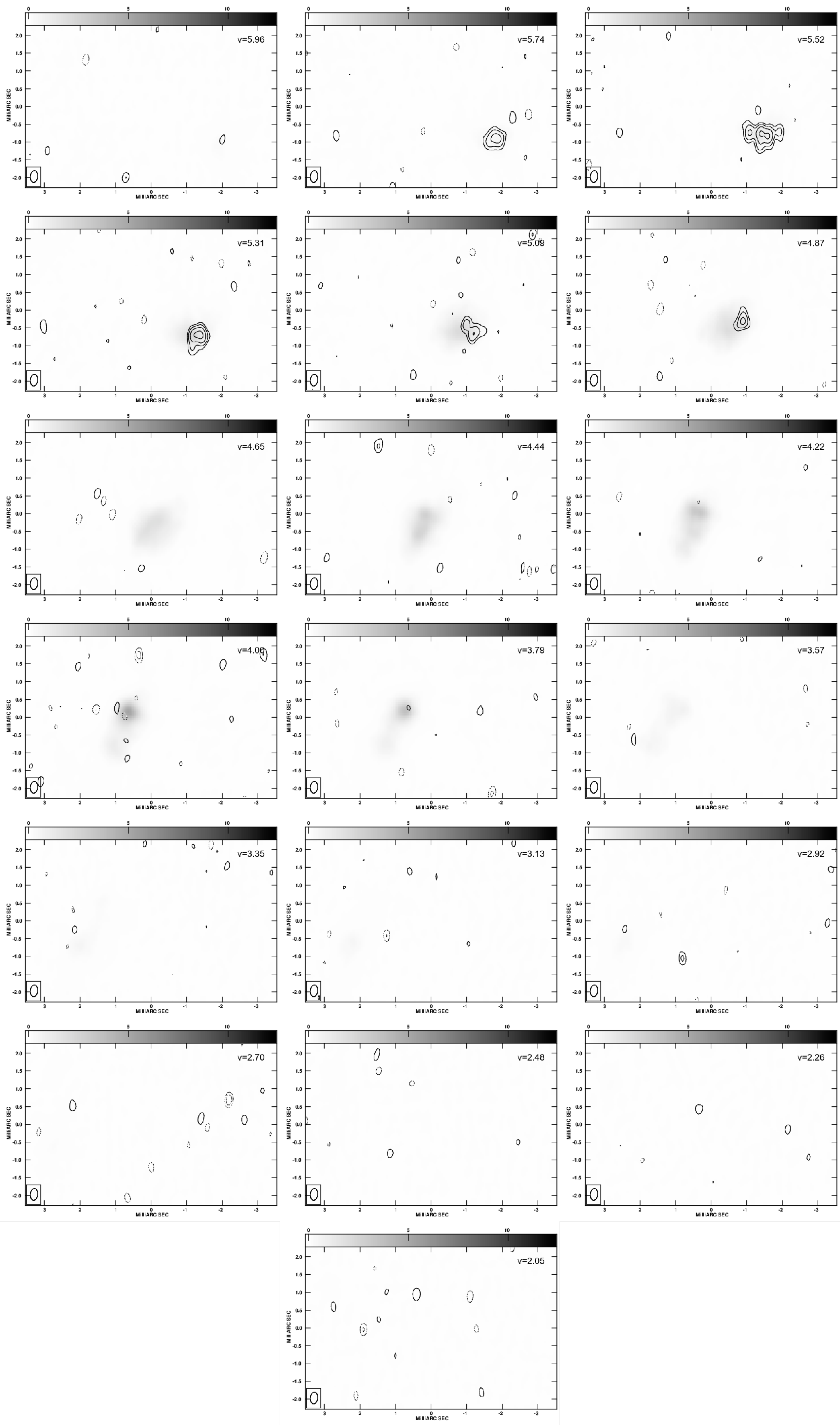}
\figsetgrpnote{Stokes V contours over Stokes I (greyscale) in each channel for epoch BD46AQ. Contour levels and spatial scale are as in 4.1, with instead $sigma_{AQ} = 5.5904$ mJy beam$^{-1}$.}
\figsetgrpend

\figsetgrpstart
\figsetgrpnum{4.5}
\figsetgrptitle{BD46AR}
\figsetplot{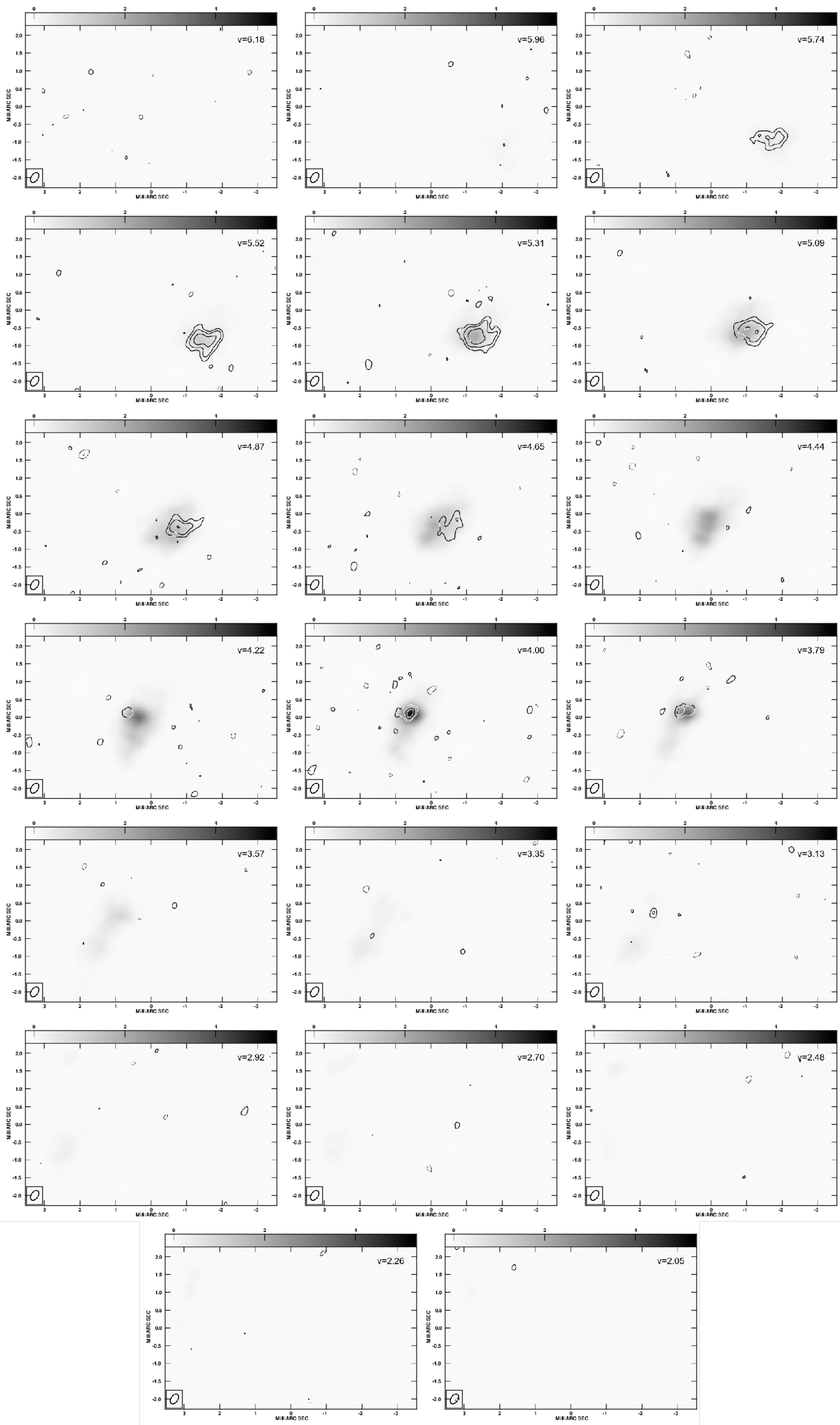}
\figsetgrpnote{Stokes V contours over Stokes I (greyscale) in each channel for epoch BD46AR. Contour levels and spatial scale are as in 4.1, with instead $sigma_{AR} = 11.334$ mJy beam$^{-1}$.}
\figsetgrpend

\figsetend

As mentioned earlier, the target feature in epoch BD46AM is detected with insufficient signal-to-noise in $m_l$ ($S/N \gtrsim 3$) to allow rigorous fitting to the predicted polarization profile shapes. The feature does not vary greatly in general shape or intensity distribution between BD46AN through BD46AR, but is only emerging in intensity at epoch BD46AM. Therefore, in the analysis that follows in Sections \ref{disc} and \ref{concl}, only epochs BD46AN through BD46AR are included.

As any absolute or relative positional information between the epochs is destroyed by phase self-calibration during image reduction, the final images need to be astrometrically aligned by other means, including cross-correlation and feature registration \citep{diakem03}. We adopt the resulting relative astrometric alignment of \citet{diakem03} as a reference here. The zeroth-moment Stokes $I$ images for the re-analyzed epochs forming part of the current work were aligned against the corresponding reference epochs in \citet{diakem03} using Fourier-based matched filtering as described in \citet{kemb09}.
The systemic error in this alignment was estimated by manually calculating the expected offsets for epochs BD46AO and BD46AQ against their corresponding reference epochs in \citet{diakem03} using three Stokes $I$ maser features as fiducial reference points. The three features were chosen because they were relatively isolated, did not vary significantly in morphology across epochs, and were approximately evenly distributed around the maser ring. By comparison with the offsets produced by the matched-filter alignment, the systematic alignment error is estimated to be 
$\sim \sqrt{2}$ pixels.

Due to the irregular morphology of the fine-scale intensity distribution of individual maser features, a combination of aperture fitting and intensity--weighted position averaging was used when determining the overall position of a feature, as needed for the alignment error estimation described above and also for proper motion calculations. For aperture fitting, a Gaussian filter is applied to a window containing the feature in each frequency channel of the Stokes $I$ image using SciPy's \textit{gaussian{\_}filter} function with $\sigma=5$ pixels. The location of maximum intensity is used as the center of the aperture in each channel. If that maximum intensity is greater than a specified background threshold, the aperture fit is performed for that channel. Finally, the results were examined by eye and the final channels containing the desired contiguous feature were stacked for the intensity--weighted average location.

\subsection{Proper Motion\label{pmspread}}
\begin{figure}[t!]
\figurenum{5}
\plotone{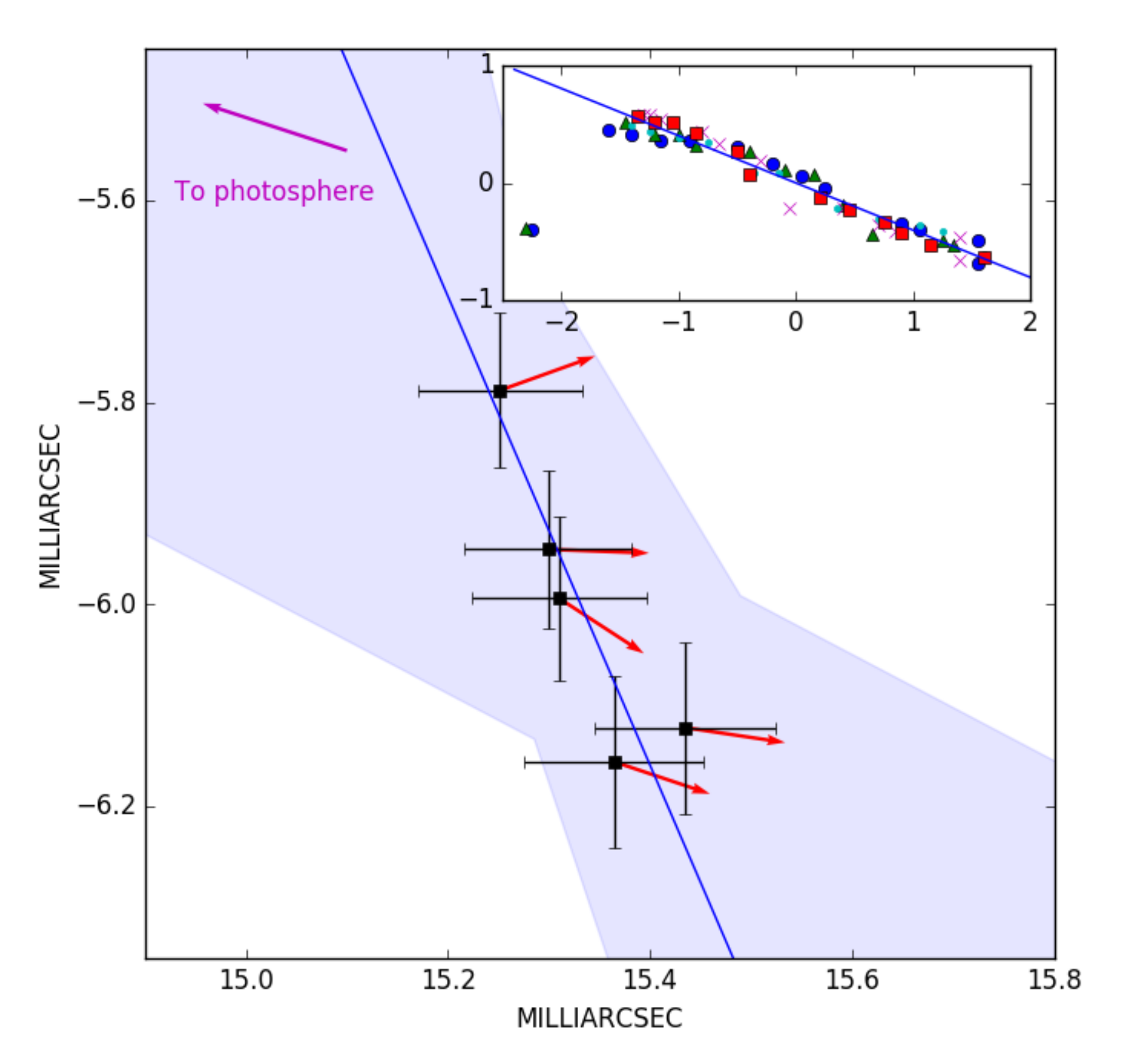}
\caption{The intensity--averaged position over time (black squares with error bars) was fit with a simple linear relation (blue line, fit error indicated by shaded region). X and Y axes are in mas from the center of the final aligned image, the approximate location of the star. The feature starts, in BD46AN, at the upper--leftmost position and moves to the southwest over time. Red vectors indicate the orientation of the magnetic field assuming GKK with $K=0$ in each epoch. The inset plot in the upper right shows the average direction in which the feature is spread as a function of velocity (blue line), as well as the relative positions of the sampled components presented in the analysis for epochs BD46AN (large blue circles), BD46AO (green triangles), BD46AP (small cyan circles), BD46AQ (magenta x's), and BD46AR (red squares). Axes of the inset plot are also in mas are with respect to $x=+16$ mas, and $y={-5.86, -5.92, -6.09, -6.14, -6.17}$ mas, respectively, in each of the aforementioned epochs. The location of the stellar photosphere is highly uncertain as no continuum emission from the star is visible in the observations, but the approximate direction of the star is indicated in the upper left. \explain{Added MILLIARCSEC labels to the x and y axes of the plot.}\label{fig-5}}
\end{figure}

For the aligned images, the average intensity--weighted position for the feature of interest was computed for each epoch using the method described above. A linear relation was then fitted to the feature's position over time using the least squares linear fitting code LSTSQ\footnote{LSTSQ Fortran package written and distributed by Benjamin Weiner.}. The average position in each epoch and the resulting fit is shown in Figure \ref{fig-5}. The direction of motion derived from this fit corresponds to a position angle of $203.33^\circ \pm 16.11$ (measured N through E).

Similarly, to determine the direction in which the feature is elongated over velocity in projection on the sky, the intensity--weighted average position was computed in each frequency channel and fitted with a linear relation. In BD46AN and BD46AO, this average position in the highest frequency channel deviated significantly from the general trend of the rest. As these fits are only used as a reference to the general orientation of the feature, both of these points were excluded in this particular instance. The mean position angle for the feature's spatial orientation over velocity was found to be $\sim248^\circ \pm 1$ (see Figure \ref{fig-5} for comparison).

\begin{figure*}[t!]
\figurenum{6}
\gridline{\fig{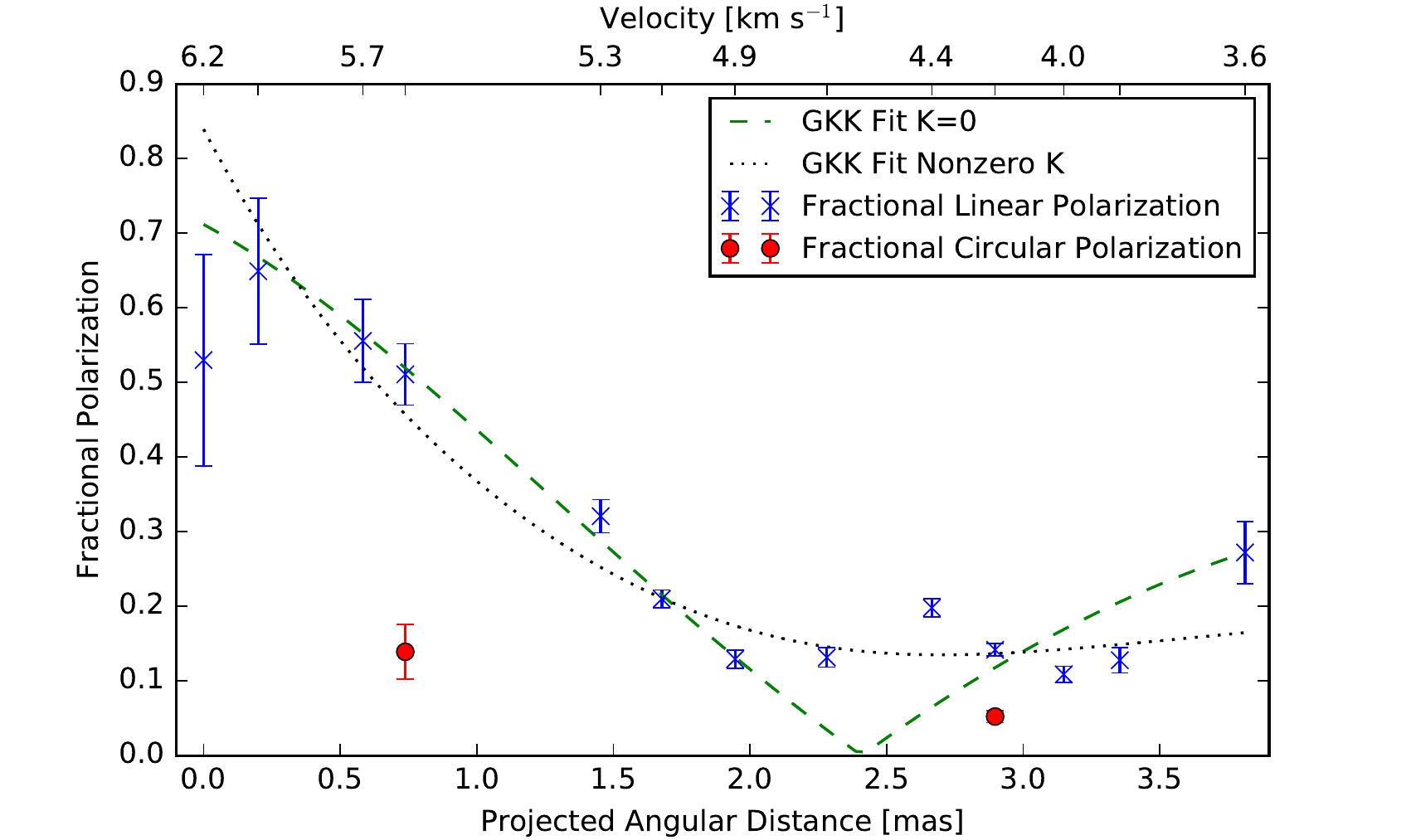}{0.49\textwidth}{(a)}
          \fig{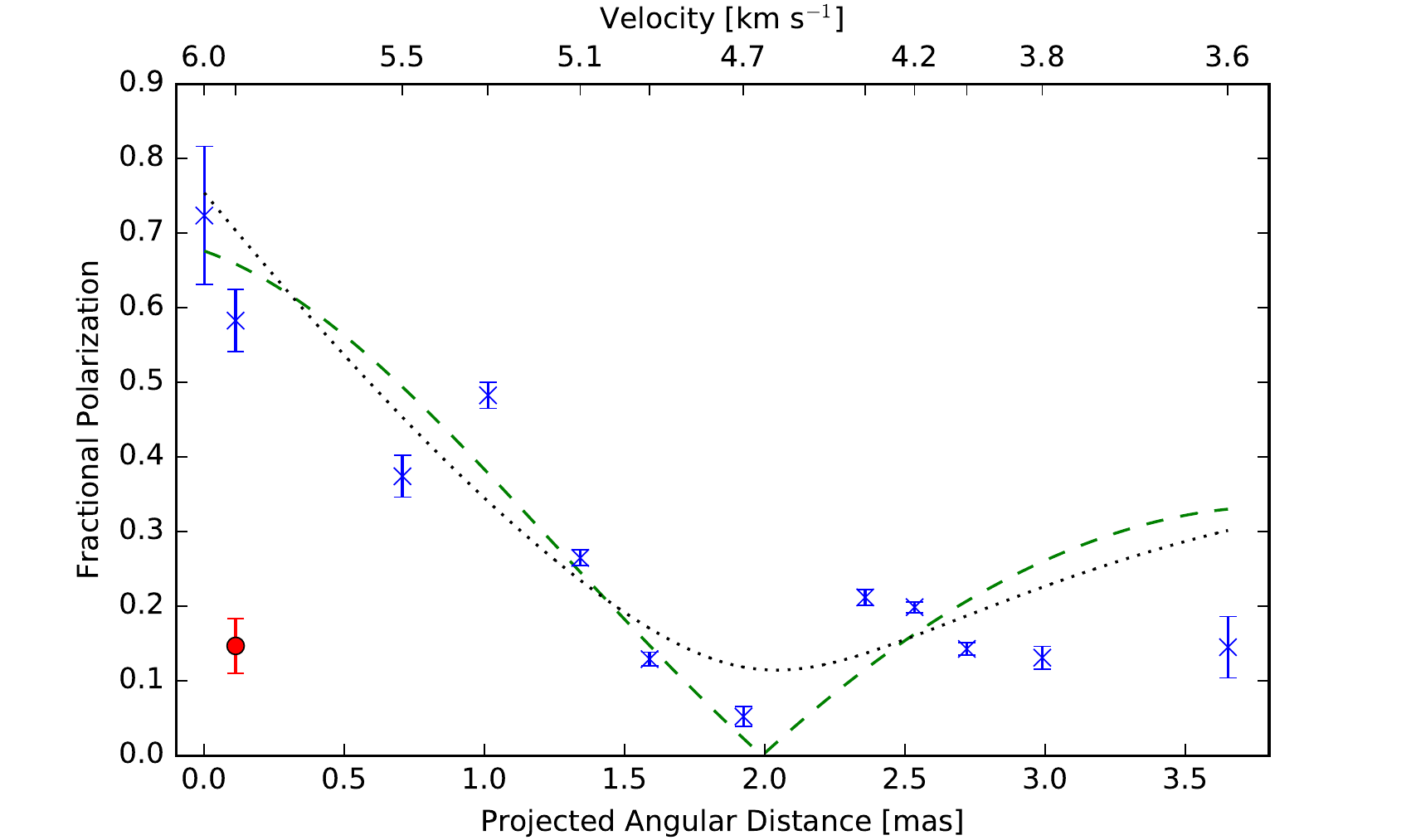}{0.49\textwidth}{(b)}}
\gridline{\fig{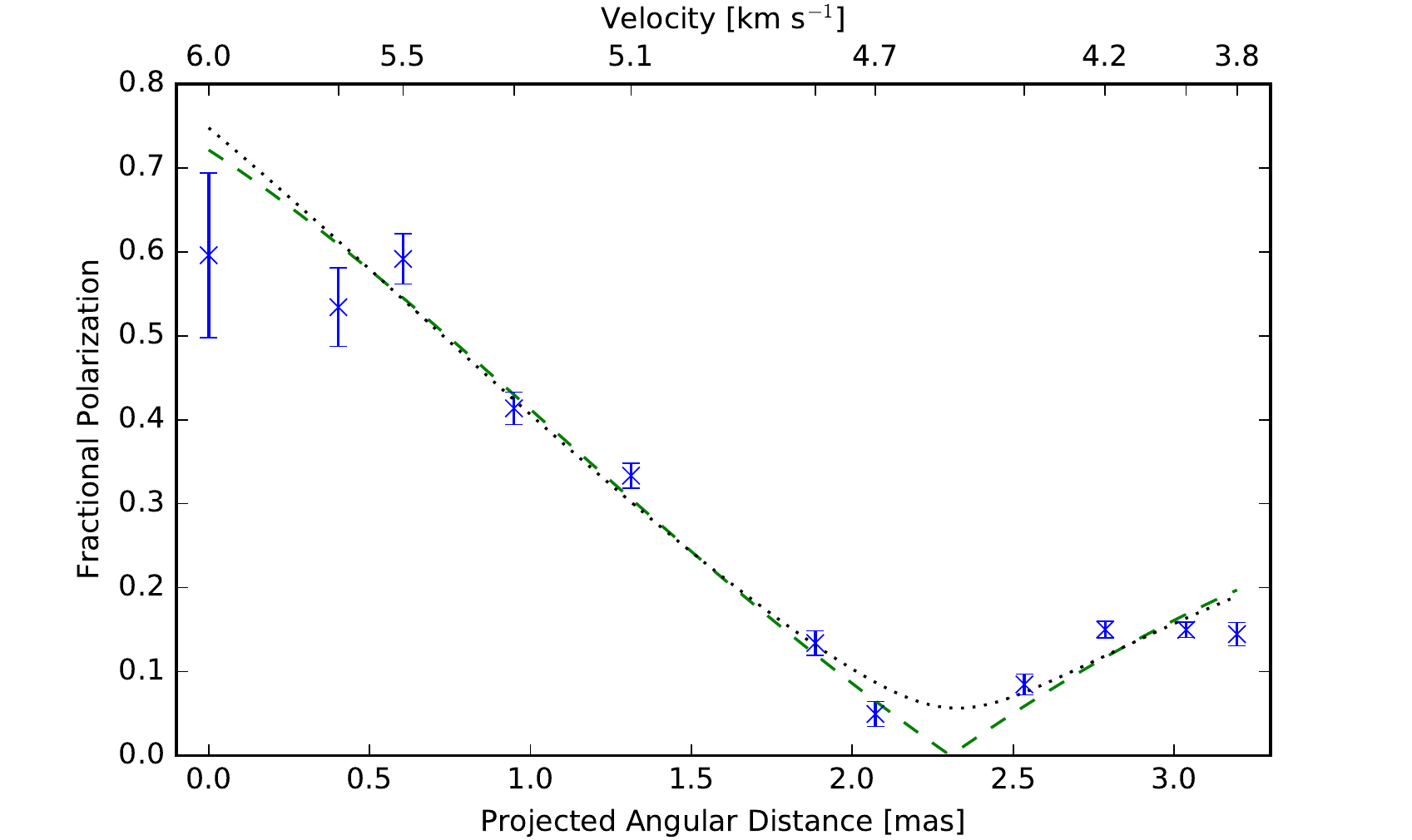}{0.49\textwidth}{(c)}
          \fig{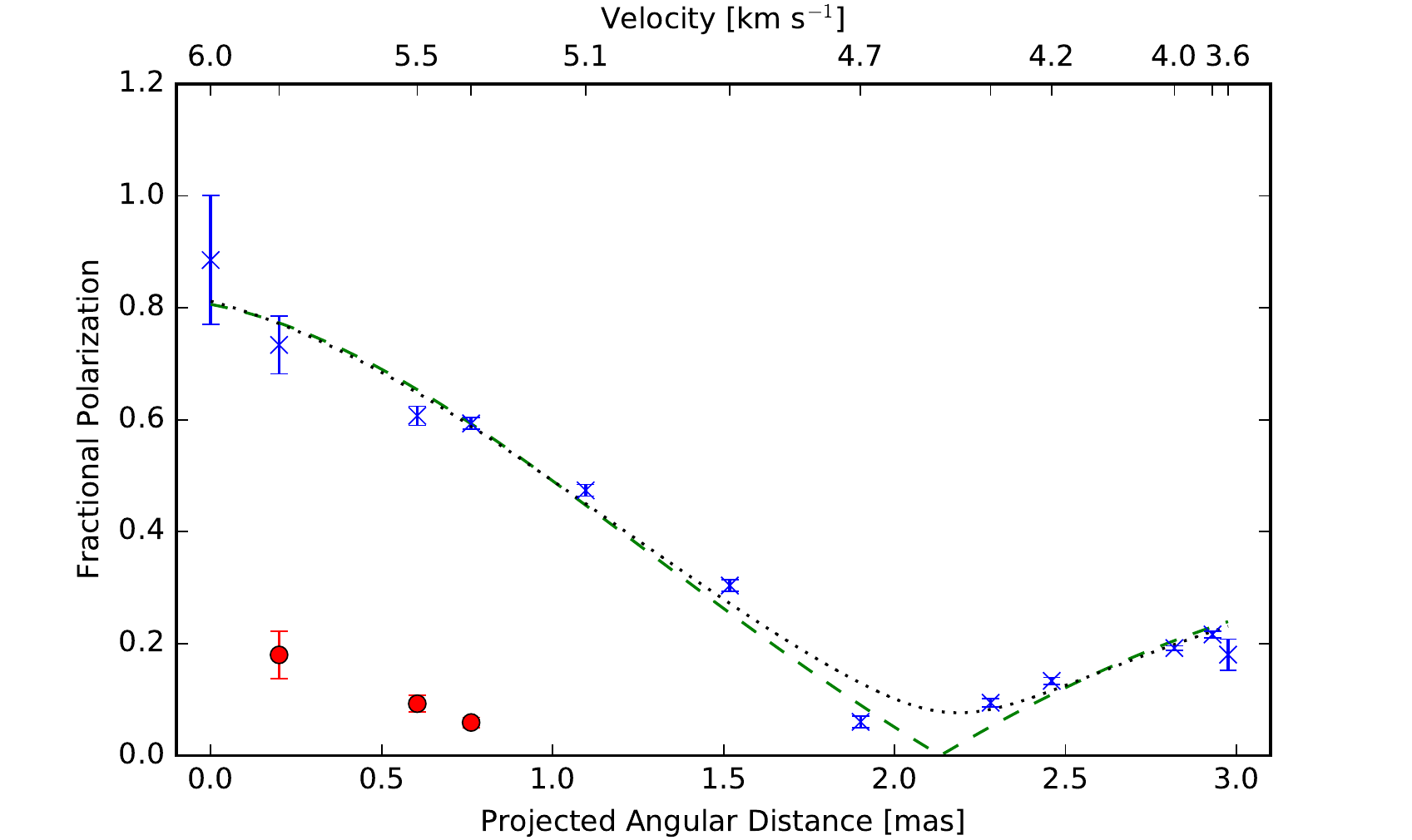}{0.49\textwidth}{(d)}}
\gridline{\fig{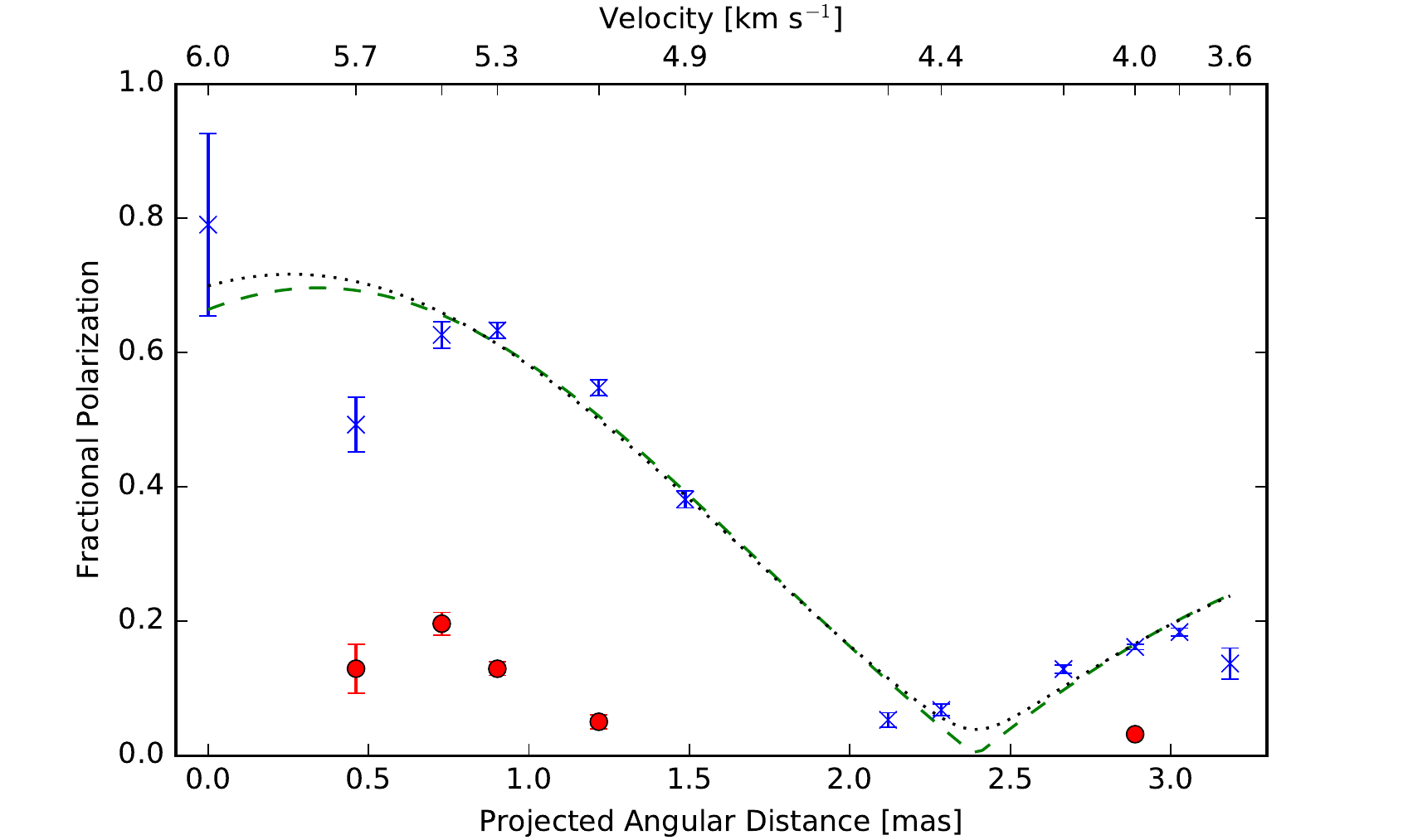}{0.49\textwidth}{(e)}}
\caption{Measured Polarization Fractions as a function of angular distance across the feature for BD46AN (a), BD46AO (b), BD46AP (c), BD46AQ (d), and BD46AR (e). The fraction of linearly polarized light with S/N above 7 (blue 'x') as well as the fraction of circularly polarized light with S/N above 3 (red points) are shown. Angular distance is computed with respect to the south--westernmost point in each epoch. The velocity corresponding to the components is denoted on the upper x-axis. The GKK profile for linear polarization for a quadratic approximation of the magnetic field angle with respect to the line of sight ($\theta$) was fit to each epoch with K=0 (dashed green line) and with non-zero K fitting (dotted black line).  \label{fig-6}}
\end{figure*}

\subsection{Linear Polarization Fraction\label{results_ml}}

Fractional linear polarization above the S/N cutoff was extracted at the peak intensity in each channel across the EVPA reversal maser feature in each epoch. This is plotted in Figure \ref{fig-6} as a function of projected angular distance, $d$, from the feature's southwesternmost data point at each epoch. In general, the fractional linear polarization tends to be highest in the SW at $m_l \sim 0.6-0.8$ and decreases to the NE to $m_l \sim 0.05 - 0.1$ before rising again.

\added{The Stokes $\{Q,U,V\}$ values were sampled at the peak-intensity pixel in Stokes $I$ in each velocity channel as noted here and in Sections \ref{results_evpa} and \ref{results_mc}. This reduces the SNR by approximately $\sqrt{N_{pix}}$, where $N_{pix}$ is the mean number of pixels per component in each velocity channel. However, peak $I$ intensity sampling is a simple robust estimator that avoids bias introduced by any differences in the shape and size of associated Stokes $\{I,Q,U\}$ and $V$ components, any related issues concerning imperfect component deblending, and the statistical correlation introduced by averaging interferometer image pixels. We do expect fine-scale polarization structure across individual components, especially at velocities near the minimum $m_l \sim 0$ where $\frac{d\chi}{d\theta}$ is inherently larger (see the upper x-axis in Figure \ref{fig-6} and Figure \ref{fig-7}); this is evident at these velocities in the polarization contour plots over all epochs, such as Figure \ref{fig-3}. In Appendix A we show that peak-intensity sampling for these features introduces no bias in EVPA.}

The projected angular distances, $d$, used in this analysis are calculated as distances from a reference point, defined above as the feature's southwesternmost data point at each epoch. The sampled peak-intensity pixel positions over velocity across the components at each epoch are substantially linear in projection on the sky as described above in Section \ref{pmspread}. However, to ensure that our method of calculating the projected distance across the feature was not significantly affecting the shape of the measured profiles in $m_l$ and EVPA, we also calculated the distance $d$ by projecting orthogonally onto the best fit line for the peak-intensity pixel positions over velocity at each epoch. The mean offset between these two estimates for $d$ over all five epochs is 0.021 mas with a standard deviation of 0.028 mas and only one data point across all epochs had an offset exceeding 0.058 mas. The method for estimating $d$ is robust and does not significantly affect the shape of either the $m_l$ profile discussed here or the EVPA profile discussed in Section \ref{results_evpa}.

\subsection{Electric Vector Position Angle\label{results_evpa}}

\begin{figure*}[t!]
\figurenum{7}
\gridline{\fig{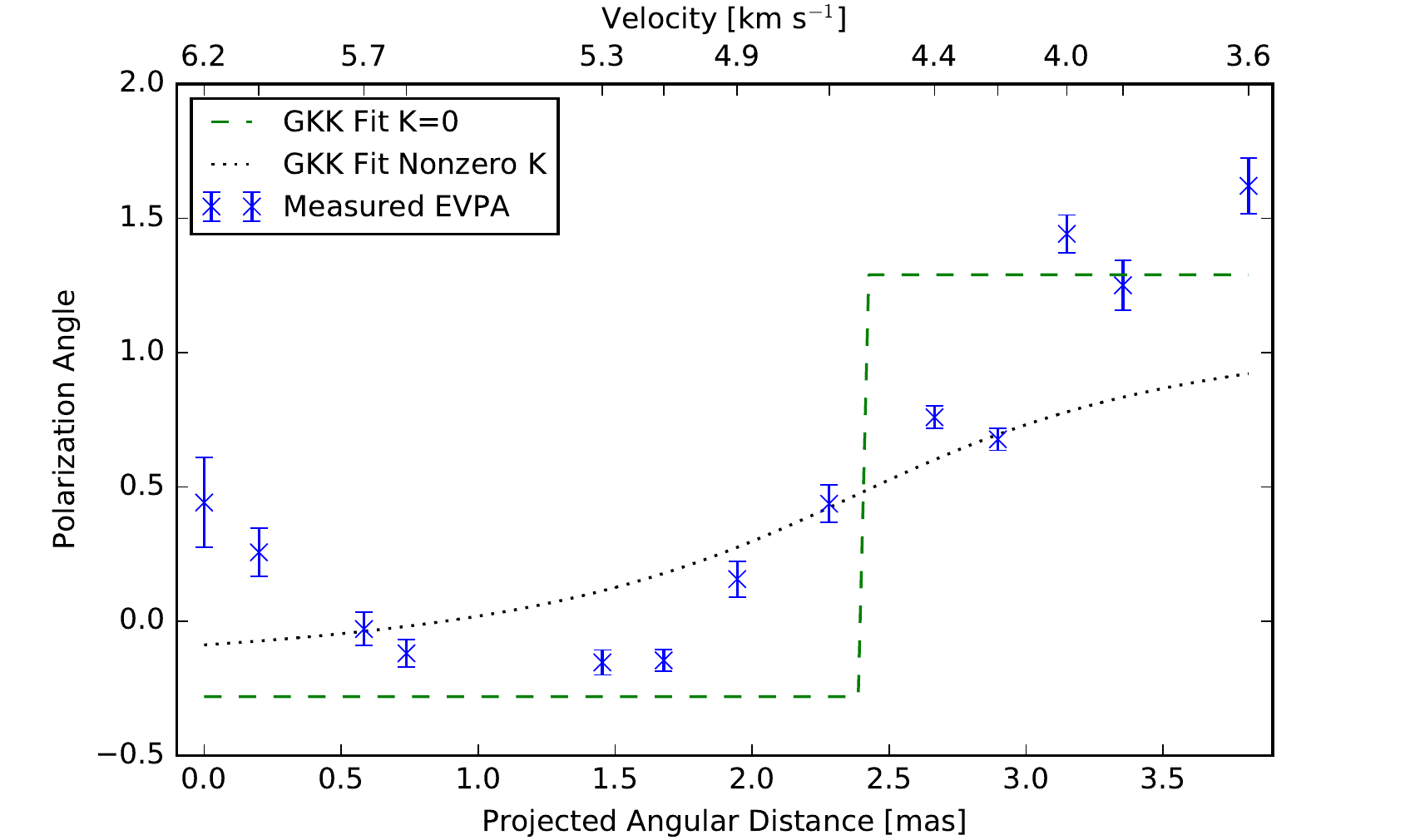}{0.49\textwidth}{(a)}
          \fig{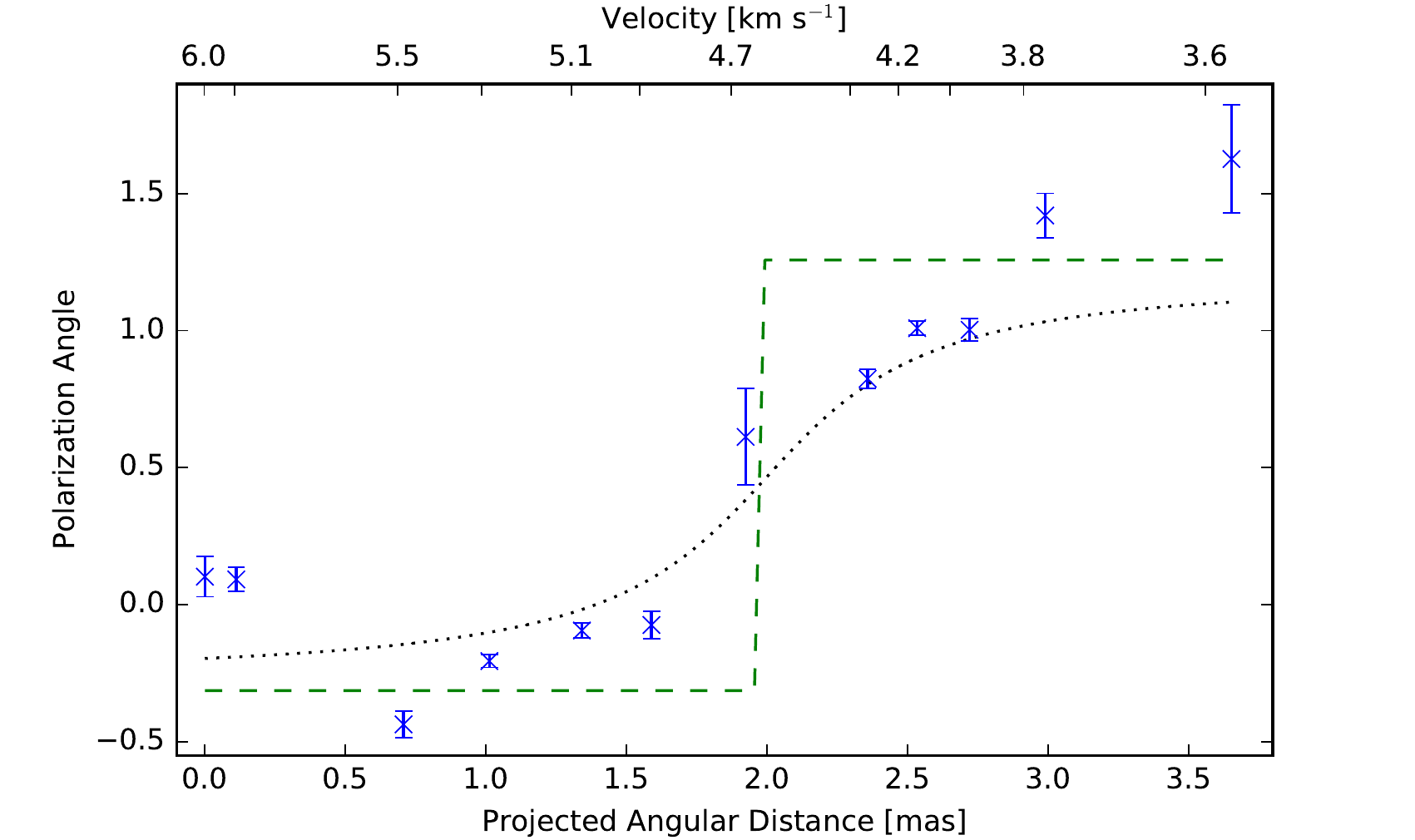}{0.49\textwidth}{(b)}}
\gridline{\fig{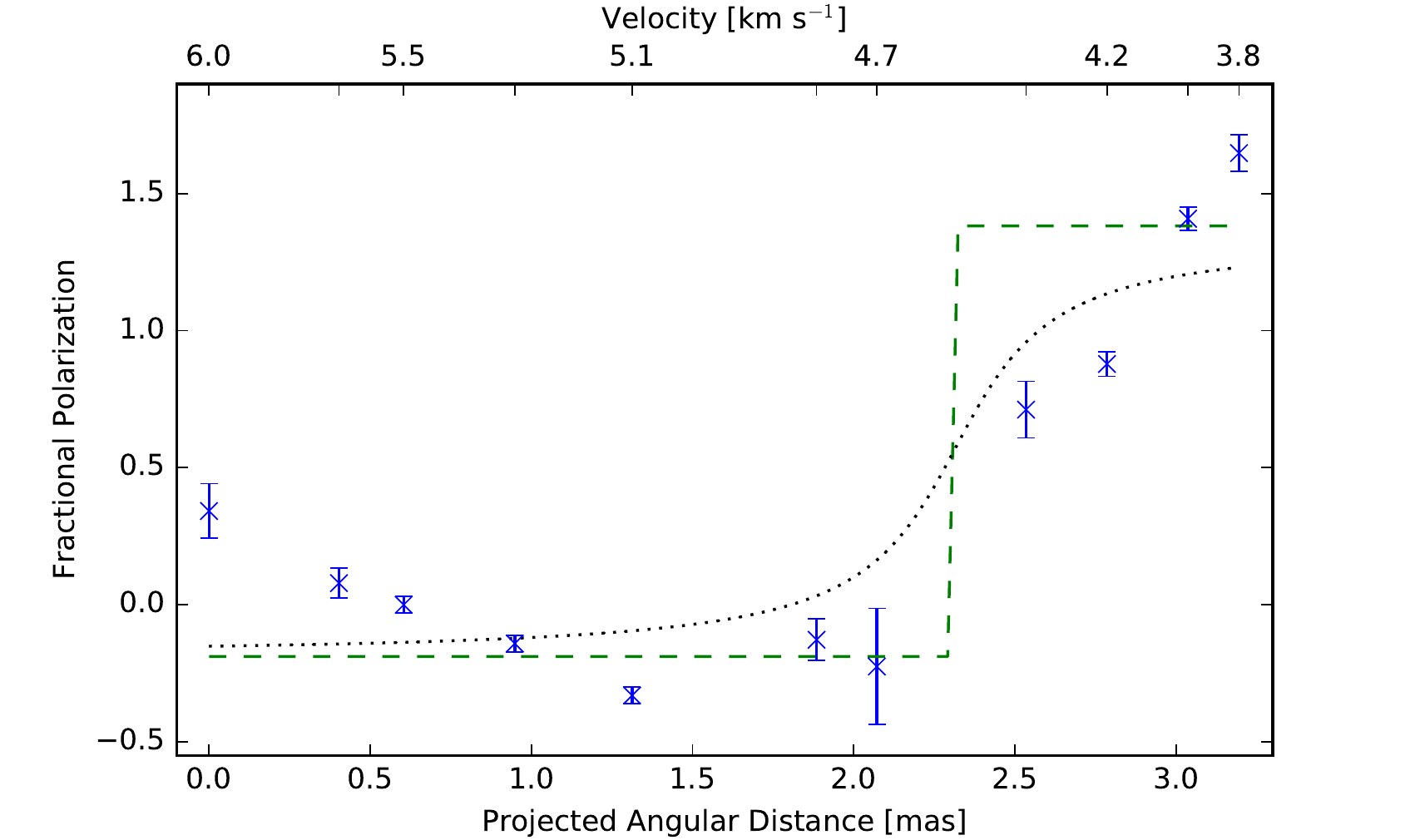}{0.49\textwidth}{(c)}
          \fig{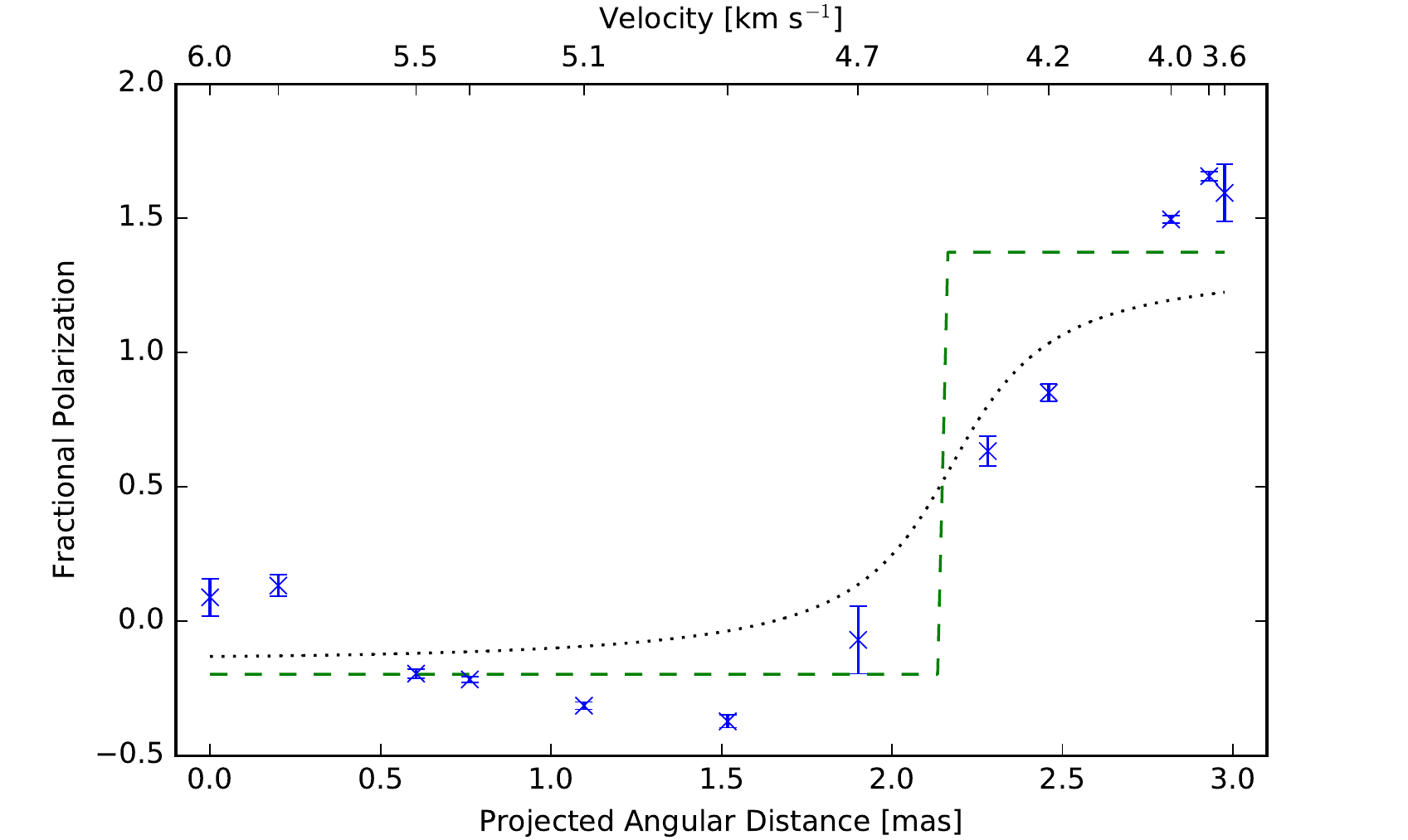}{0.49\textwidth}{(d)}}
\gridline{\fig{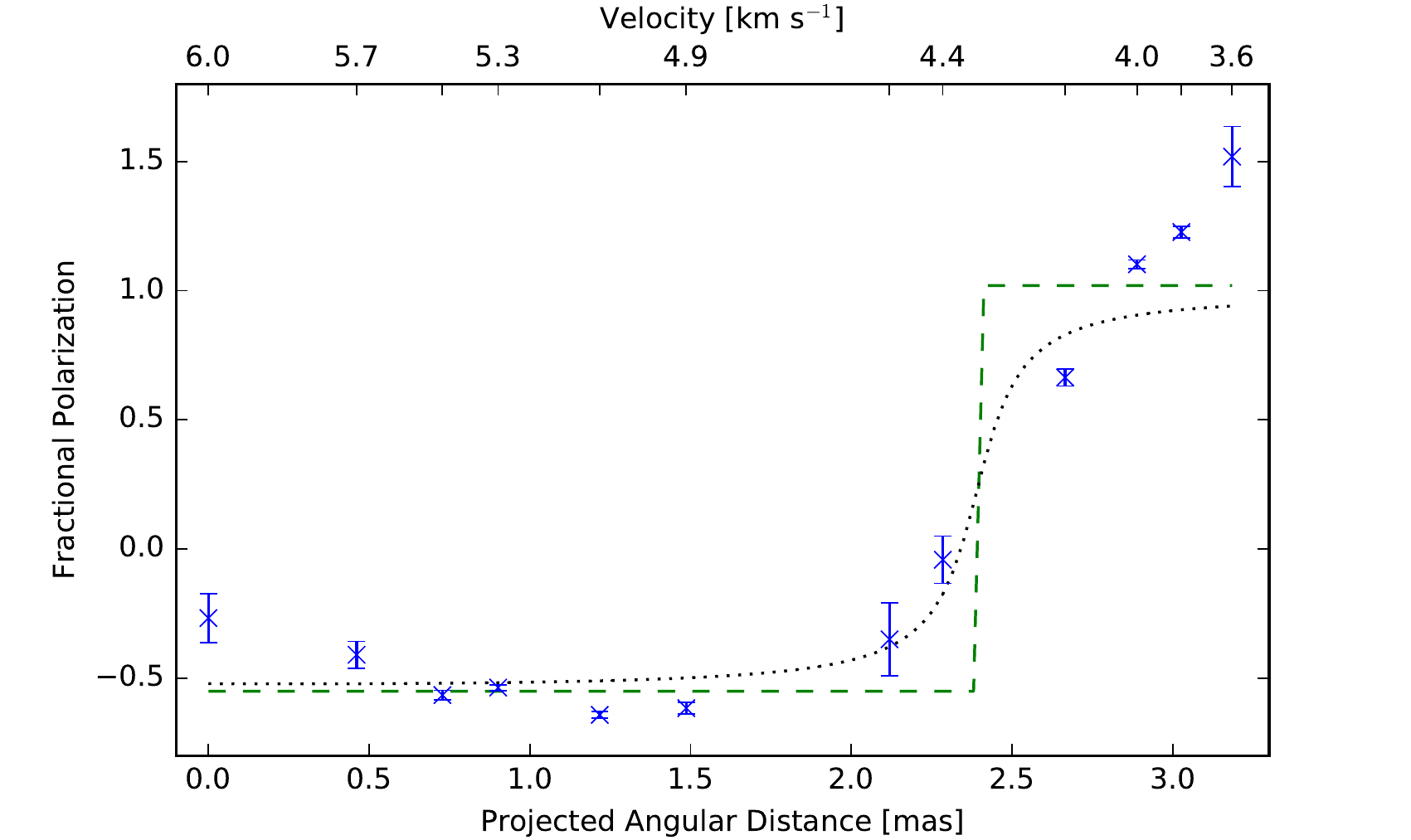}{0.49\textwidth}{(e)}}
\caption{Relative position angle of linear polarization (blue x's with error bars) as a function of angular distance across the feature for (a) BD46AN, (b) BD46AO, (c) BD46AP, (d) BD46AQ, and (e) BD46AR. Again, angular distance is computed with respect to the south- and westernmost coordinates  in each epoch and velocity corresponding to the components is denoted on the upper x-axis. Systemic error in absolute EVPA is not included. Lines show expected EVPA profile from the GKK fits shown in Figure \ref{fig-6}, for $K=0$ (green dashed line) and nonzero $K$ (black dotted line). Fitted nonzero K values for each epoch are given in Table \ref{tbl-2}. \label{fig-7}}
\end{figure*}

The EVPA of the peak-intensity pixel described above in Section \ref{results_ml} was also calculated, and is plotted for points with $m_l$ S/N $\geq 3$ in each channel in Figure \ref{fig-7}. As in Figure \ref{fig-6}, points are plotted across each feature as a function of distance from the southwesternmost data point of the feature at each epoch. Errors shown are determined by propagation of uncertainty in Stokes Q and U derived from mean off-source thermal rms values, and do not include systematic errors in calibrating absolute EVPA, as the following analysis focuses on the relative change in EVPA across the feature.
In all epochs, the EVPA rotation appears to be relatively smooth, with the bulk of the rotation occurring near the minimum in $m_l$ visible in Figure \ref{fig-6}.There is also a slight decrease in EVPA) with distance at lower projected angular distances. 

Previous work (eg. \citealt{cotton08,kemb09}) have noted alignment between the EVPA and the proper motion of maser features, and postulated that either the maser feature is traveling along the magnetic field line or the masing material is dragging the magnetic field. The GKK solution results in the EVPA being parallel to the projected magnetic field for $\theta \leq \theta_F$ and perpendicular to the projected magnetic field for $\theta > \theta_F$. Assuming GKK, the measured linear polarization across the feature $m_l(\theta)$ implies that the magnetic field would be parallel to the EVPA in the southwest and perpendicular to the EVPA in the northeast of the maser feature. The resulting orientation of the projected magnetic field in each epoch is shown in Figure \ref{fig-5}. In epoch BD46AN, this implies a magnetic field that is perpendicular to the fitted proper motion discussed in Section \ref{pmspread}. Subsequent epochs show an offset in alignment that is less than perpendicular, but it does not approach parallel monotonically. The offset between the inferred GKK magnetic field direction and the proper motion direction decreases to $\sim 30^\circ$ for epoch BD46AP and increases again to $\sim 45^\circ$ for BD46AR.

\subsection{Circular Polarization\label{results_mc}}
\begin{figure}[t!]
\figurenum{8}
\plotone{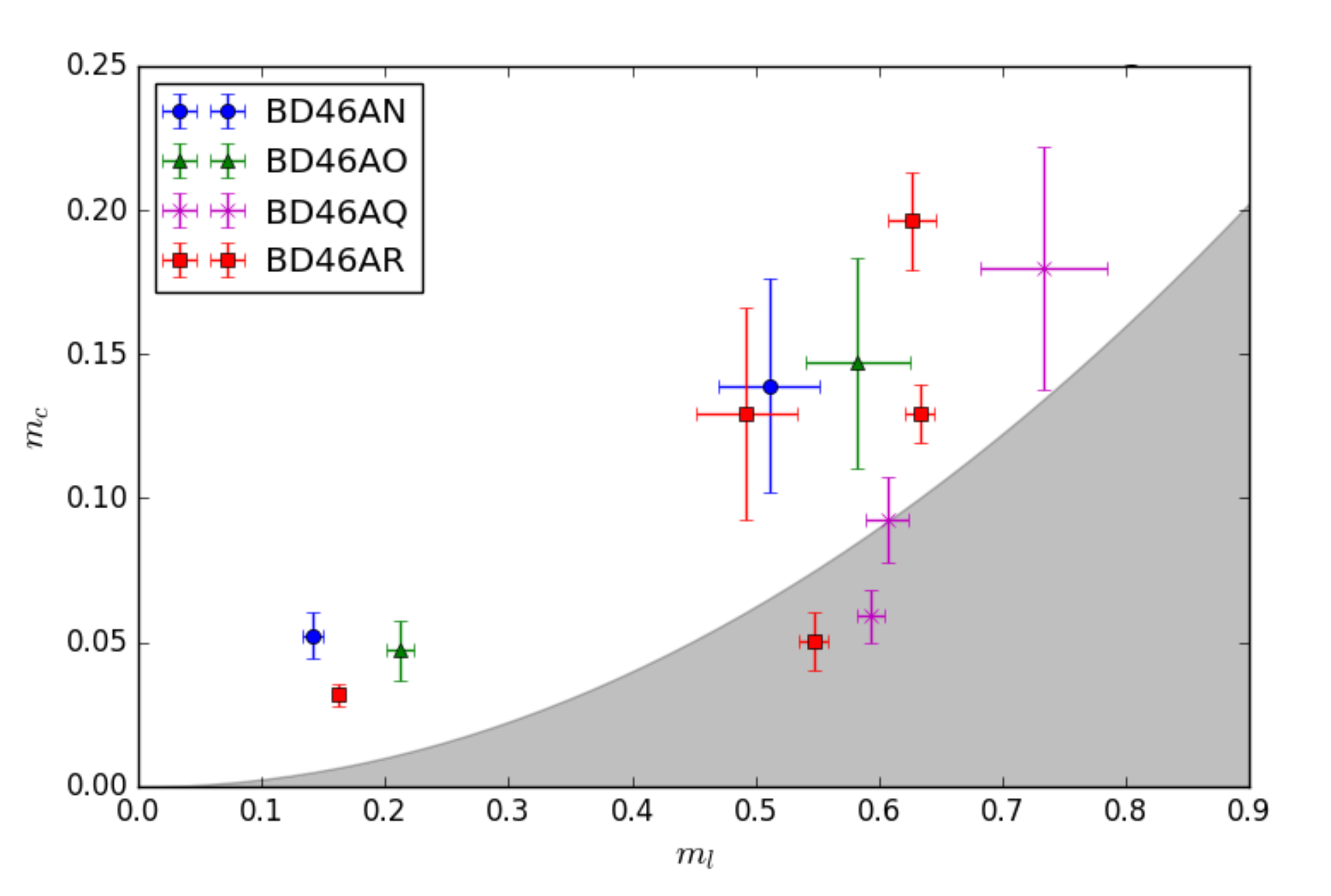}
\caption{Fraction of circularly--polarized light vs. fraction of linearly--polarized light for circular polarization fraction above S/N=3 for all epochs. Region corresponding to $m_c < m_l^2 / 4$ is shaded in each plot. \label{fig-8}}
\end{figure}

The circularly-polarized fraction of maser emission, $m_c$, with $S/N>3$ at the position of peak Stokes $I$ brightness is also shown for each epoch in Figure \ref{fig-6}. This fractional circular polarization, $m_c$, is also plotted as a function of linear polarization fraction, $m_l$ in Figure \ref{fig-8}. Generally, high values of $m_c$ are associated with high values of $m_l$, but not all high values of $m_l$ are accompanied by high $m_c$.

\section{Discussion}\label{disc}

\subsection{Magnetic Fields and the Van Vleck Angle\label{gkk}}

As discussed in Section \ref{Intro}, GKK derived their solutions in a number of asymptotic limits. Recent estimates of the individual parameters \citep{assaf13, kemb09, kdrgx11} suggest that the SiO $\nu=1, J=1-0$ masers studied here fall within the $\Delta \omega \gg g \Omega \gg R \gg \Gamma$ limit, where $\Delta \omega$ is the bandwidth of amplified maser radiation, $g \Omega$ is the Zeeman splitting rate, $R$ is the stimulated emission rate, and $\Gamma$ is the damping frequency. 

To compare, we estimated $R$ for the data presented here using the relation from \citet{plamb03},

\begin{equation}
R = 23 \left( \frac{T_B}{2 \times 10^{10} \textrm{ K}}\right)\left( \frac{d\Omega}{10^{-2} \textrm{ sr}^{-1}} \right)\textrm{ s}^{-1},
\end{equation}

where $T_B$ is the brightness temperature and $g\Omega$ is the beaming angle. We followed the method of \citet{assaf13} to estimate beaming angle, $d\Omega = \left( \frac{d_F}{L} \right)^2$, where $d_F$ is the angular FWHM of the component in Stokes I and $L$ is the actual angular size of the component. Since our target feature was double-peaked in Stokes I in many channels, we measured only the FWHM of the brighter peak, which was consistently being sampled for the profile. Our resulting calculated stimulated emission rates range from $R\sim 20 - 60$ s$^{-1}$.  \citet{kemb09} estimates $B \sim 580-720$ mG for TX Cam resulting in $g \Omega \sim 750$ s$^{-1}$.

In these environments, radiative de-excitation dominates over collisional de-excitation, resulting in $\Gamma \sim 1-5$ s$^{-1}$ \citep{kwan74, kemb09}. While $R > \Gamma$, the  required minimum $R/\Gamma$ to qualify as sufficiently saturated to achieve a GKK form is uncertain, as discussed in Section \ref{Intro}. \citet{watwyld01} require $R/\Gamma \gtrsim 30$ to obtain $m_l$ as high as 0.7. However, \citet{eli96} argues that the GKK form will be applicable at $R/\Gamma > x_B$, where $x_B$ is the ratio Zeeman splitting to the Doppler linewidth. From \citet{eli96,kdmd97}, $x_B = \left( 8.2 \times 10^{-4} \right) B / \Delta \nu_D$, where $B$ is in Gauss and $\Delta \nu_D$ is the Doppler linewidth in km s$^{-1}$. Using $B \sim 0.6$ G, derived from equating the bulk kinetic energy density and thermal energy density \citep{kemb09}, and our Doppler linewidth of 0.6 km s$^{-1}$, the requirement for saturation from \citet{eli96} becomes $R/\Gamma > 8 \times 10^{-4}$. \replaced{As our estimated values for these observations are $R/\Gamma \sim 4 - 60$, the degree of saturation deduced from estimates of $R,\ \Gamma$, and $g\Omega$ is highly dependent on the theory applied.}{According to our estimated values for these observations above, $R/\Gamma \sim 4 - 60$. However, the degree of saturation that is thereby implied is highly dependent on the value of $R/\Gamma$ prescribed by any individual theory for the onset of saturation.}

The observational tests for GKK that are performed here are two--fold: the profile of the linear polarization fraction across the EVPA rotation and the functional form of the EVPA rotation itself. As noted above, the GKK solution for this regime assumes zero circular polarization and negligible Faraday rotation.

In this limit, GKK gives the solution
\begin{eqnarray}
Y=-1, \qquad Z=0, \qquad \textrm{for } \sin^2 \theta \leq \frac{1}{3} \nonumber \\
Y=\frac{ 3 \sin^2 \theta - 2}{3 \sin^2 \theta}, \qquad Z=K, \qquad \textrm{for } \sin^2 \theta \geq \frac{1}{3},\label{eq1}
\end{eqnarray}
where $Y=Q_B/I$, $Z=U_B/I$, and $\theta$ is the angle between the magnetic field and the line of sight. $K$ is not formally constrained by GKK, and is technically only subject to the general constraints that linear polarization not exceed total stokes I - that is, $K^2 \leq 1-Y^2$. However, \citet{eli91} argues that the lack of constraint on $K$ makes both directions of resulting $U_B$ equally likely, causing $U_B \rightarrow 0$ on average. Unlike the observed $Q$ and $U$, $Q_B$ and $U_B$ are oriented such that $Q_B>0$ is perpendicular to the projected magnetic field. Therefore, we fit the magnitude of the linear polarization fraction to the data, as in \citet{kdrgx11} and \citet{assaf13}. To this end, $\theta$ was approximated as a quadratic function of the projected angular distance, $d$, that passes through the Van Vleck Angle, $\theta_F = \arcsin \left( \sqrt{ 2/3 } \right)$, at some flip distance, $d_f$:
\begin{equation}
\theta \textrm{ [rad]} = p_0 \left( d^2 - d_f^2 \right) + p_1 \left( d - d_f \right) + \theta_F \label{eq2}
\end{equation}

\begin{figure*}[t!]
\figurenum{9}
\plotone{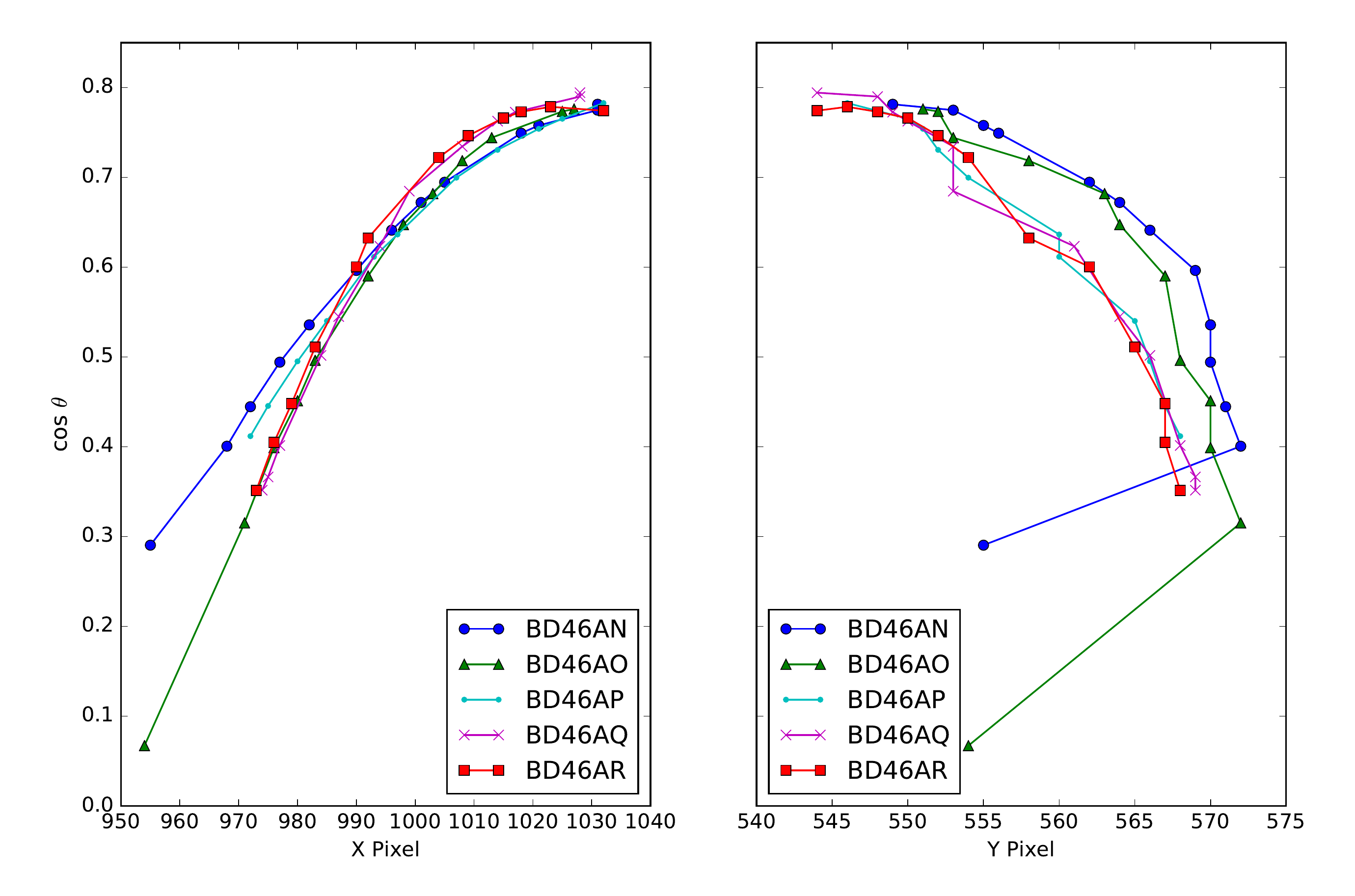}
\caption{$\cos \theta$ as a function of position along the X (left) and Y (right) axes for each epoch, assuming GKK fit. \label{fig-9}}
\end{figure*}

Figure \ref{fig-6} shows the best fit GKK profile for each epoch with $K=0$. The final best--fit parameters and the $\chi^2$ value are shown in Table \ref{tbl-2}, and the resulting $\cos \theta$ values as a function of both X and Y location is shown in Figure \ref{fig-9}.

\begin{deluxetable*}{lrrrrrrr}
\tabletypesize{\scriptsize}
\tablecaption{GKK Best Fit for $m_l$.\label{tbl-2}}
\tablehead{\colhead{Epoch Code} & \colhead{$p_0$} & \colhead{$p_1$} & \colhead{$d_f$} & \colhead{$v_f \left[\textrm{km s}^{-1}\right]$} & \colhead{$|K|$} & \colhead{$\chi^2$} & \colhead{$w_i$} }
\startdata
BD46AN & 0.02933 &  0.04618 & 2.409 & 4.4 & 0 & 18.73 & 0.9418 \\
BD46AO & 0.05273 &  0.03219 & 1.990 & 4.4 & 0 & 27.17 & 0.9567 \\
BD46AP & 0.02862 &  0.05708 & 2.305 & 4.3 & 0 & 4.308 & 0.9769 \\
BD46AQ & 0.05477 &  0.02478 & 2.134 & 4.3 & 0 & 9.551 & 0.9573 \\
BD46AR & 0.06618 & -0.04546 & 2.392 & 4.2 & 0 & 11.57 & 0.9581 \\
\hline
BD46AN & -0.01184 &  0.1456  & 2.701 & 4.2 & 0.1350  & 5.924 & 0.05817 \\
BD46AO &  0.01866 &  0.1035  & 2.042 & 4.4 & 0.1145  & 24.59 & 0.04327 \\
BD46AP &  0.02232 &  0.07233 & 2.327 & 4.3 & 0.05635 & 3.911 & 0.02310 \\
BD46AQ &  0.04782 &  0.03386 & 2.187 & 4.3 & 0.07620 & 6.671 & 0.04268 \\
BD46AR &  0.06192 & -0.03193 & 2.393 & 4.2 & 0.03848 & 11.18 & 0.04187 \\
\enddata
\tablecomments{Best fit parameters and $\chi^2$ for $m_l$ fitting as seen in Figure \ref{fig-6} for both $K=0$ and fitted $K$, as well as the Akaike Weight \citep{akaike73, hurtsai89}, $w_i$, comparing the $K=0$ and nonzero $K$ fits in each epoch. The corresponding velocity of $d_f$, $v_f$, is estimated assuming that, below the velocity resolution of the observations, the velocity between consecutive channels is approximately proportional to the projected angular distance of the sampled data.}
\end{deluxetable*} 

For completeness, we repeated this process, allowing $|K|$ to be fit as an additional free parameter. However, the additional parameter did not provide a sufficient improvement in fit as the relative likelihood of the $K=0$ model is $>0.94$ compared to a relative likelihood of the non-zero $K$ model of $<0.06$, as shown by the Akaike weights in Table \ref{tbl-2}.  Furthermore, most of the best fit $K$ values are quite small, having only a slight effect on the resulting profile (see Figure \ref{fig-6}).

The second test of the GKK theory preformed here is an analysis of the EVPA profile across the maser feature. The expected EVPA profile for the GKK model can be determined by applying the predictions for $Y$ and $Z$ to the definition of EVPA:

\begin{eqnarray}
\textrm{EVPA}_B &=& \frac{1}{2} \arctan{\left( Z, Y \right)} + \gamma \nonumber \\
&=& \frac{1}{2} \arctan \left( K, \frac{3 \sin^2 \theta - 2}{3 \sin^2 \theta} \right) + \gamma,\label{eq3}
\end{eqnarray}

where $\gamma$ is the rotational offset between the EVPA in the magnetic field coordinate frame used in GKK and the observed coordinate frame. If $K=0$, the EVPA profile across the flip manifests itself as a step function, where the direction of polarization changes by $\pi/2$ instantaneously. A non-zero $K$ serves to smooth the step function somewhat, with larger $|K|$ resulting in a smoother rotation. However, for $K \neq 0$, the extremal values are approached much more slowly. Figure \ref{fig-7} shows the EVPA as a function of distance in each epoch (without the added systemic error in absolute EVPA) along with expected profiles from the GKK fits to the fractional linear polarization $m_l(\theta)$. 

Although the transition region where the EVPA changes by $\pi/2$ is more compact at some epochs than others, none show the instantaneous flip predicted under GKK with $K=0$. In some cases, the inclusion of a nonzero $Z=K$ is a good fit to the rate of EVPA rotation near the Van Vleck angle (eg. BD46AO and BD46AR in Figure \ref{fig-7} (b) and (e)), while in BD46AN, it predicts a more gradual and smaller rotation than is seen in the data.  The inclusion of a nonzero $K$ also leads to an underestimate of the total amount of rotation in all cases. Even assuming GKK with $K=0$ may slighty underestimate the total angular change in some epochs, such as BD46AQ and BD46AR. We note that none of these profiles account for the slight EVPA counter-rotation that is seen at $d \sim 0.0 - 1.0$ mas in every epoch. 

Beam averaging was investigated for its possible smoothing effects on the EVPA rotation. The restoring beam, with a major axis of $\sim 0.5$ mas, exceeds the typical projected angular displacement of the maser feature at successive frequencies. A trial profile of EVPA as a function of projected angular distance as described in Equations \ref{eq2}--\ref{eq3} was convolved with a 1-D Gaussian beam, assuming a concurrent $m_l$ profile as described in Equation \ref{eq1}. In these circumstances, beam convolution actually sharpens the EVPA rotation. This is because $m_l$ is highest at the extremal position angles and minimized when the rate of EVPA rotation is at maximum. The linear polarization serves to weight the EVPA toward the extremal angles, causing the EVPA transition to appear more abrupt. This, then, would not explain the observed smoother EVPA rotation than expected by the GKK model with $K=0$.


There are several other factors that could cause a smoother EVPA rotation than predicted by GKK in the $K=0$ limit. Each of our peak-intensity pixel data points sample a single line of sight at a specific LSR channel velocity of width $\sim 0.22$ km/s. To first-order we assume that we are sampling a single effective $\theta$ at each data point, but it is instructive to relax this assumption as a possible explanation for the smoother EVPA profile that is observed than expected from GKK. In the observed features in the current data, the bulk of the EVPA rotation occurs over $\sim 6$ frequency channels near the inferred Van Vleck angle. We construct a model where $\theta_{k_v}$ along the line of sight in each of these channels $k_v$ is distributed discretely in four bins $m$ as: $\theta_{k_v} \in  \{\theta_0+(1+m-k_v)\Delta_\theta,\ \forall\ m=0,..3\}$. The gradient in $\theta$, namely $\Delta\theta$, is uniform by bin number $m$ and channel number $k_v$. Using this model we calculated intensity-weighted mean Stokes Q, U, EVPA, and $m_l$ profiles for the transition channels $k_v$ near the Van Vleck angle, and found that the the observed smooth EVPA rotation in the data would require only a net change $\Delta\theta \sim 9^{\circ}$ in $\theta$. Along the line-of-sight, this must occur within the coherent velocity path length of the maser feature, i.e. the length along which velocity dispersion is less than the internal velocity dispersion of the gas \citep{jiyune05}, and may be of the order $10^{13} - 10^{14}$ cm \citep{moran79, alross86, barv87}.

Another possible, although arguably unlikely, explanation is that there is an abrupt fine-scale change in the orientation of the magnetic field below our sampling resolution that happens to occur both near the inferred Van Vleck transition region and with an opposite direction so that it smooths out the net rotation in EVPA from the expected abrupt $\pi/2$
change.

We also consider the possibility that the smoother EVPA transition may be a result of Faraday rotation. The original derivation in GKK assumed negligible Faraday rotation. No analytic solution currently exists for the GKK model with non--zero Faraday rotation. Neither estimated electron density nor magnetic field strength are well-constrained in the NCSE. Estimates adopted by \citet{assaf13} suggest that Faraday rotation would only affect the polarization angle of the SiO ($\nu=1$, $J=1-0$) maser by $\sim 15^\circ$ for R Cas. An independent analysis by \citet{richter16} determined that Faraday rotation was unlikely to be a significant effect for SiO masers toward the supergiant VY CMa. Prior work has also noted that the EVPA of the linearly-polarized SiO maser emission in the TX Cam imaging campaign analyzed here has been noted to be predominantly oriented tangentially to the maser ring \citep{kemb09, kdmd97}. Significant and pervasive Faraday rotation would disrupt this trend; however this does not rule out localized regions of higher Faraday rotation.

Faraday rotation causes depolarization, and GKK argue that large amounts of Faraday rotation will lead to no observed linear polarization. A zeroth-order estimate of the amount of surviving polarization can be determined as $\sin \left( 2 \Delta \Psi_{obs} \right) / \left( 2 \Delta \Psi_{obs} \right)$ \citep{burn65}, where in this instance $\Delta \Psi_{obs}$ is the difference between the observed EVPA and the expected position angle for GKK. From an examination of $\Delta \Psi_{obs}$ for most data points here, this gives a surviving polarization fraction of $\gtrsim 0.9$ times the original polarization fraction. However, it decreases to $\sim 0.5 - 0.8$ around the location of the EVPA flip and  $\gtrsim 0.7-0.8$ at very low projected angular distance from the origin. Given these estimates, it is possible that Faraday rotation could play a part, since no observed linear polarization fractions are above the limit set by Faraday depolarization. However, a rigorous test of Faraday polarization within the GKK model is beyond the scope of this paper as it requires a closer integration with maser radiation transport. 

\begin{figure}
\figurenum{10}
\plotone{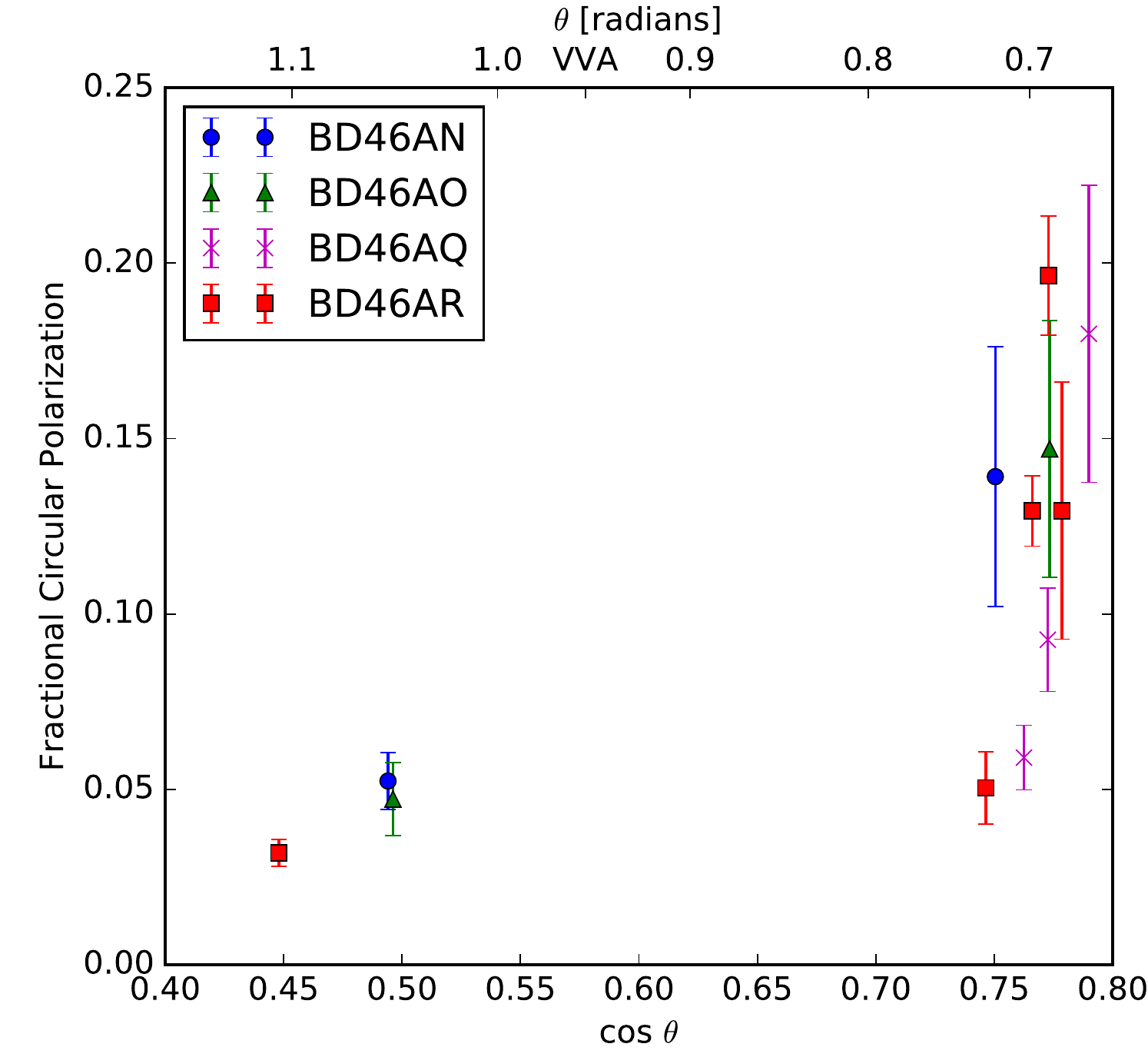}
\caption{Fractional circular polarization as a function of $\cos \theta$ (lower axis) or theta (upper axis) in each epoch for S/N $>3$. The Van Vleck Angle is denoted on the upper axis as VVA.\label{fig-10}}
\end{figure}

\subsubsection{Zeeman Circular Polarization\label{zeemanv}}

As noted above GKK assumed no circular polarization in deriving their asymptotic solutions. Subsequent work including non-zero circular polarization produced divergent predictions. \citet{eli96} found $m_c \propto 1/ \cos \theta$ while work by \citet{gray12} predicts $m_c \approxprop \cos \theta$. Finally, \citet{watwyld01} found that $V \propto \cos \theta$ only in the highly unsaturated limit. With increased saturation, the behavior of V with $B \frac{\partial I}{\partial \nu} \cos \theta$ is only solvable numerically in their work, and produces a peaked function with a maximum  near $\cos \theta \sim 0.2 - 0.3$, increasing only to $\sim 1$ at $\cos \theta = 1$ (see Figure 1 in \citet{watwyld01}). This final relation is explored below.

Our measured fractional circular polarization as a function of $\cos \theta$ as derived from the $K=0$ GKK fit is shown in Figure \ref{fig-10} for all points with S/N $>3$. The results show a general trend of increased $m_c$ for larger $\cos\theta$, which is most consistent with \citet{gray12}. The relation seen in our analysis is not purely linear, and there is some significant scatter at large $\cos \theta$. However, the function defined in \citet{gray12} for $m_c$ is not perfectly linear with $\cos\theta$, even though the $\cos \theta$ dependence dominates. This may be the cause of any non-linearity. 


The work of \citet{eli96} predicts that the GKK solutions for masers such as SiO, where the Zeeman splitting is much less than the Doppler linewidth, are valid even at moderate saturation. This analysis is accompanied by two auxiliary predictions: first, polarization is only possible for $\sin^2 \theta > 1/3$ ($\cos\theta \lesssim 0.817)$. We note that the values of $\theta$ from the GKK $K=0$ fit to the current data all have $ \sin^2 \theta \geq 0.369$ across the extent of the feature. The data are therefore consistent with the theory in this regard. However, the second auxiliary prediction of this theoretical work is that $m_c \propto 1/\cos \theta$. As discussed above, our observations support an approximate proportionality to $\cos \theta$, rather than the inverse. 

\begin{figure}
\figurenum{11}
\plotone{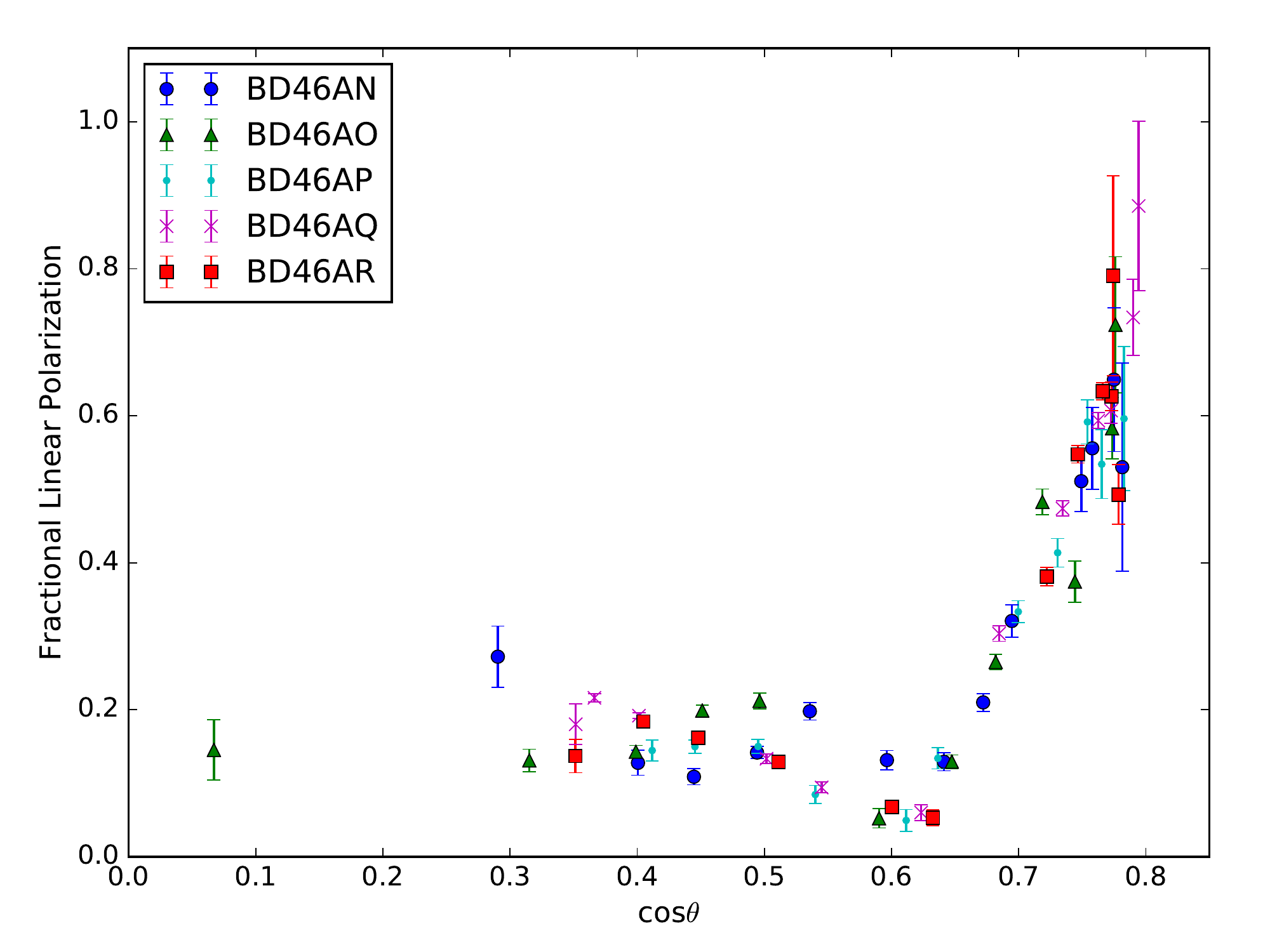}
\caption{Fractional linear polarization as a function of $\cos \theta$ in each epoch, assuming GKK.\label{fig-11}}
\end{figure}

Contrary to \citet{eli96}, \citet{watwyld01} predict that the GKK profile for fractional linear polarization is approached much more slowly as saturation increases, and then are only reached for large $\cos \theta$. The dependence of $m_l$ on $\cos \theta$ (as derived from the fit to GKK) is shown in Figure \ref{fig-11}. The linear polarization fraction reaches $m_l \sim 0.5 - 0.9$ around $\cos \theta \sim 0.8$. These values could only be reached by some of the most saturated masers in the model of \citet{watwyld01}. 

As noted above, \citet{watwyld01} also consider $V/\left( p B \partial I / \partial \nu \right)$ as a function of $\cos \theta$ and degree of saturation for the case of weak Zeeman splitting. Here $p$ is the Zeeman coefficient for the particular transition. \citet{watwyld01} find that this quantity should be directly proportional to $\cos \theta$ only for highly unsaturated masers, increasing to a singly--peaked function as saturation increases. Unfortunately, due to the low frequency resolution for our observations (due to correlator constraints), we were unable to  estimate $\delta I / \delta \nu$ via finite differencing with sufficient accuracy to compare our results to this model in a meaningful manner.

If the circular polarization is, in fact, Zeeman in origin, this indicates a stronger magnetic field. The Zeeman splitting is given by $x_B = \frac{3 \sqrt{2}}{16} v_{peak} \cos \theta$ \citep{eli96}, where $v_{peak}$ is the ratio of Stokes V to Stokes I at maximum Stokes V. When combined with the Zeeman splitting equation cited above \citep{eli96,kdmd97}, one may estimate $B / \cos \theta = 320 v_{peak} \Delta v_D $, where again $B$ is in Gauss and $\Delta v_D$ is in km s$^{-1}$. Using again $\Delta v_D = 0.6$ km s$^{-1}$, the average value for this feature across all five epochs is $B/ \cos \theta \sim 12$ G.

\citet{wiewat98} considered non-Zeeman origins for $m_c$, such as circular polarization arising from linear polarization conversion. They predict that this process should generally produce $m_c \lesssim m_l^2 / 4$. Figure \ref{fig-8} shows how the measured $m_c$ for the current data compare to this limit (shaded region) as a function of $m_l$. Most of the points lie above this limit. \citet{wiewat98} do note that velocity variations could produce scatter in this relation, so a more rigorous check would be that $\left< m_c \right> < \left< m_l^2 / 4 \right>$. Average values for each epoch are shown in Table \ref{tbl-3}. While some individual epochs, such as BD46AQ, have similar $\left<  m_c \right>$ and $\left< m_l^2/4 \right>$, most epochs have $\left<m_c \right> \sim 2 \times \left<m_l^2/4\right>$, including the averages across all epochs. Therefore, our data are not consistent with this form of non--Zeeman circular polarization, agreeing with the findings by \citet{herp06} for the SiO $\nu=1$, $J=2-1$ maser transition and with \citet{cotton08,richter16} for other SiO maser data.

\begin{deluxetable}{lrrrrr}
\tabletypesize{\scriptsize}
\tablewidth{8pt}
\tablecaption{Average $m_c$ and $m_l^2/4$ for Each Epoch for $S/N \geq 3$.\label{tbl-3}}
\tablehead{\colhead{Epoch Code} & \colhead{N} & \colhead{$\left<m_c\right>$} & \colhead{$\sigma_{\left<m_c\right>}$} & \colhead{$\left<m_l^2/4\right>$} & \colhead{$\sigma_{\left<m_l^2/4\right>}$} }
\startdata
BD46AN     &  2 &  0.0958 & 0.0268 & 0.0351 & 0.0148 \\
BD46AO     &  2 &  0.0971 & 0.0269 & 0.0481 & 0.0153 \\
BD46AP     &  0 &     --- &    --- &    --- &    --- \\
BD46AQ     &  3 &  0.1105 & 0.0264 & 0.1049 & 0.0160 \\
BD46AR     &  5 &  0.1075 & 0.0192 & 0.0681 & 0.0108 \\
All Epochs & 12 &  0.1046 & 0.0238 & 0.0685 & 0.0137 \\
\enddata
\end{deluxetable} 

\subsection{Local Magnetic Field Curvature\label{localBcurve}}

A\deleted{s discussed in Section \ref{Intro}, a}brupt curvature in the magnetic field could in principle cause an EVPA rotation of this scale independently, without invoking the GKK model. This could account for the smoother change in the observed EVPA across the feature, as the direction of the polarization would simply be tracing the projected magnetic field. The major $\sim \pi/2$ rotation occurs over an average span of $\sim 1.5$ mas or $0.6$ AU in the image plane, using a distance of $\sim 390$ pc to TX Cam \citep{olivier01}. Assuming that the radius of the photosphere is $R_*\sim 2$ AU \citep{cahneli79,pegourie87}, the rotation occurs over a span of $\sim 0.3 R_*$ at a distance of $\sim 2 R_*$ from the photosphere. 
  However, there is no physical reason for a magnetic field that is rotating $\sim 90^\circ$ in the observational plane to also induce the observed functional form in $m_l(\theta)$ observed here. 

\subsection{Anisotropy \& Non-Zeeman Polarization}

\subsubsection{Anisotropic Pumping}

\citet{aram05} investigated the effect of radiative anisotropic pumping of dichroic masers on the resulting linear polarization. According to this model, the direction of linear polarization is determined by the anisotropy factor, 

\begin{equation}
    w = \frac{1}{2} \frac{\left(\tau^3 - 2 \tau \right) E_1\left(\tau\right) - \left( \tau^2 - \tau - 2 I_0 / S_0 \right) e^{-\tau}}{2 + 2 \tau E_1\left( \tau \right) + \left( I_0/S_0 - 2 \right) e^{-\tau}},
\end{equation}

where $\tau$ is the total optical depth of the slab at that frequency, $E_1\left(\tau\right)$ is the first exponential integral function, $I_0$ is the unpolarized incident radiation field, and $S_0$ is the source function of the masing slab. Under this model, linear polarization would be primarily oriented radially to the maser ring when $w>0$ or tangentially when $w<0$.  \citet{aram05} describe the polarization on either side of $w=0$ only as either "mainly radial" or "mainly tangential", without describing the specific EVPA profile expected for such a transition. However, this would be consistent with the general decrease in $m_l$ that is observed around the location of the rotation. \citet{aram05} do not derive a specific profile for this linear polarization fraction, and it is beyond the scope of the current work to do so here. At $w=0$, the emitted radiation is unpolarized, consistent with the decrease in linear polarization fraction that is observed around the location of the rotation in the current data. \citet{kemb09} argued that anisotropic pumping is unlikely to be a dominant effect globally, as the strength of linear polarization in these masers decreases with distance from the star and the opposite is observed in this $J=1-0$ observing campaign for TX Cam. However, the fractional linear polarization observed in $J=5-4$ SiO masers around supergiant stars by \citet{shin04} and \citet{vlem11} can reach $\sim 0.6 - 0.8$. Following the work of \citet{west84}, they argue that higher rotational transitions would require a much larger magnetic field strength to reach these high levels of linear polarization, and that a more plausible explanation is some contribution from anisotropic pumping \citep{shin04, vlem11}. In addition, according to \citet{aram05}, the regime in which a large EVPA rotation of $\sim \pi/2$ could occur coincides with a suppression in masing; this argues against this mechanism. 

\subsubsection{Circular Polarization by Anisotropic Resonant Scattering\label{houde}}

\citet{houde14} analyzed the production of circular polarization by maser radiation scattering off foreground material. The primary observational predictions for this theory are most visible in the Stokes $\left\lbrace I,Q,U,V \right\rbrace$ spectra. Unfortunately, our frequency resolution is poor and we do not have more than $\sim3-4$ data points with reasonable $S/N$ across individual emission features along any particular line of sight. Therefore, a rigorous fitting to the expected profile is not possible. 

However, comparisons to some of the more general predictions can still be performed. In this theory, Stokes V is determined by the amount of scattering and the angle between the initial linear polarization and the magnetic field in the scattering material. For typical conditions in SiO maser regions around AGB stars and for most angles, the produced Stokes V will be comparable to either Stokes Q or U, depending on the orientation of the foreground magnetic field. Low levels of circular polarization such as those measured here ($\sim0.1$) will only be produced if the linear polarization is almost perpendicular to the foreground magnetic field. 

BD46AN has low Stokes V compared to total linear polarization across all 9 analyzed lines-of-sight. In some, either Stokes U or Q is low enough to be $\sim V$. However, the general trend is that $V\sim U$ at extremal frequencies, while $V \sim Q$ at intermediate frequencies. This only occurs when Stokes $U$ or $Q$, respectively, is on the order of $0.1$, while the other is the dominant source of linear polarization. Under this interpretation, the foreground magnetic field would be nearly perpendicular to our Stokes $Q$ for material that is scattering slightly lower and higher frequencies, but intermediate frequencies are scattered by material with a magnetic field that is nearly perpendicular to Stokes $U$. Furthermore, the lack of higher Stokes $V$ in between these regions implies that the foreground magnetic field is not rotating slowly, but rather flips abruptly twice. 

\section{Conclusion}\label{concl}

In this paper, we have compared observations of a maser feature with a $\pi/2$ EVPA rotation persisting across five epochs in an image series against multiple theories of maser polarization. The fraction of linear polarization across the EVPA reversal $m_l(\theta)$ is remarkably consistent with the asymptotic solution derived by \citet{GKK}, and arising from the angle between the projected magnetic field and the line of sight passing through the critical Van Vleck angle $\theta_F$. However, the smooth rotation of EVPA observed $\chi(\theta)$ is not consistent with this theory without inclusion of Faraday rotation or a net variation of $\sim 9^{\circ}$ along the line of sight. Our data do not, however, conclusively rule out other theoretical interpretations of maser polarization in these environments.

We examined other possible explanations for the smooth EVPA rotation $\chi(\theta)$, including local curvature in the projected magnetic field \citep{sokerclayton} or a change in anisotropy conditions \citep{aram05}. The former would not intrinsically explain the decrease in linear polarization fraction during the rotation. The latter would be consistent with the decrease in linear polarization as $w\sim 0$, though the exact behavior expected for $m_c \left( w \right)$ has yet to be derived explicitly and is beyond the scope of this work. 

Circular polarization, on the other hand, is the most consistent with the prediction from \citet{gray12} that $m_c \approxprop \cos \theta$. This relation shows moderate scatter at large $\cos \theta$ however, and additional high-sensitivity data are needed to provide more complete sampling of this relation over a wider range of $\cos \theta$. 
We were unable to analyze the prediction by \citet{watwyld01} for $V/\left(p B\delta I/\delta \nu \right)$ due to insufficient velocity resolution in the current data, an artifact of correlator limitations at the time these data were taken. If the circular polarization is Zeeman in origin (Section \ref{zeemanv}), it implies a strong magnetic field of $B/ \cos \theta \sim 12$ G on average across all five epochs.

\acknowledgements 
We thank Robert Harris for reviewing a draft of this paper before submission. Our sincere thanks are also due to the anonymous referee for their helpful comments and recommended revisions. This material is based upon work supported by the National Science Foundation Graduate Research Fellowship Program under Grant No. DGE - 1144245. MDG acknowledges funding from the UK Science and Technology Facilities Council (STFC) as part of the consolidated grant ST/P000649/1 to the Jodrell Bank Centre for Astrophysics at the University of Manchester.

\software{AIPS \citep{aips}, Scipy \citep{scipy}, LSTSQ \citep{lstsq}}

\appendix\section{EVPA estimation \label{evpa_appendix}}
As discussed in Section \ref{results_evpa}, the EVPA profiles used in the analysis were extracted from the pixel with peak Stokes I in each channel across the target feature at each epoch. To ensure that this pixel-based sampling is not biased in the presence of EVPA rotation substructure across individual components, we performed both an analytical and empirical evaluation of the peak extraction method versus the mean EVPA averaged over the component in each channel. 

A first-order analytic model assume Stokes I is Gaussian in one dimension ($x$) along some axis of symmetry with a full-width half-max of $\sigma_x$, where the interferometric observing beam, $B(x)$ is represented by a Gaussian profile with FWHM $\sigma_b$.
At a sub-pixel level, we assume that EVPA gradient is linear with $x$ and that $m_l=p$ is constant across the maser component:

\begin{eqnarray}
    Q(x) = p I(x) \cos \left( 2 \alpha x \right) \\
    U(x) = p I(x) \sin \left( 2 \alpha x \right).
\end{eqnarray}
As a function of $x$, the EVPA is:
\begin{equation}
    \chi (x) = 1/2 \arctan \left( U (x) / Q(x) \right) = \alpha x \textrm{ mod } \pi.
\end{equation}
At pixel center the true EVPA is $\chi (0) = 0$. The interferometrically measured values of Stokes I, Q, U, and V include beam convolution and averaging over the center pixel:
\begin{equation}
    I_0 = \frac{1}{\Delta} \int _{-\Delta/2} ^{\Delta/2} B(x) * I(x) dx,
\end{equation}
where $\Delta$ is the pixel size. Due to its symmetric integrand, $U_0 = 0$. As a result, $\chi _0 = \chi(0)=0$ independent of $\Delta$, with no bias. Also, if we assume an offset in peak I of $\Delta/2$ then, to first-order, $\chi_{\Delta/2} = 1/2 \arctan \left( \alpha \Delta \right)$. Along the fitted direction of the GKK magnetic field in the components in the current data, typically $\alpha \lesssim 0.05$ rad pix$^{-1}$, resulting in $\chi_{\Delta/2} \sim 0.02$ radians $\sim 1^{\circ}$ for single-pixel averaging.

As an empirical test to investigate if the peak I pixel sampling is biasing measured EVPA ($\chi_{peak}$) due to resolved substructure in the component, we calculated intensity-weighted average Stokes Q and U for the components in each channel and epoch using a limiting $7 \sigma$ Stokes I contour cutoff to define the feature. The EVPA for each channel component was then calculated from the mean Q and U. The resulting mean EVPA ($\chi_{mean}$) profile showed the same functional form as that based on the peak-pixel samples but with lower scatter as expected. We found also $\chi_{peak}-\chi_{mean} =-0.0089 \pm 0.015$ radians. Using both empirical and first-order analytic methods we therefore conclude that the peak-I data sampling method imposes no significant statistical bias on the estimated EVPA.

\newpage 
\section{Figure Sets \label{figsets}}

\subsection{Figure Set 1: Frequency--Averaged Stokes I Contours of $\nu=1,J=1-0$ SiO Maser Ring Towards TX Cam with Linear Polarization Vectors}

\begin{figure}[htbp]
\figurenum{1.1}
\epsscale{1.2}
\plotone{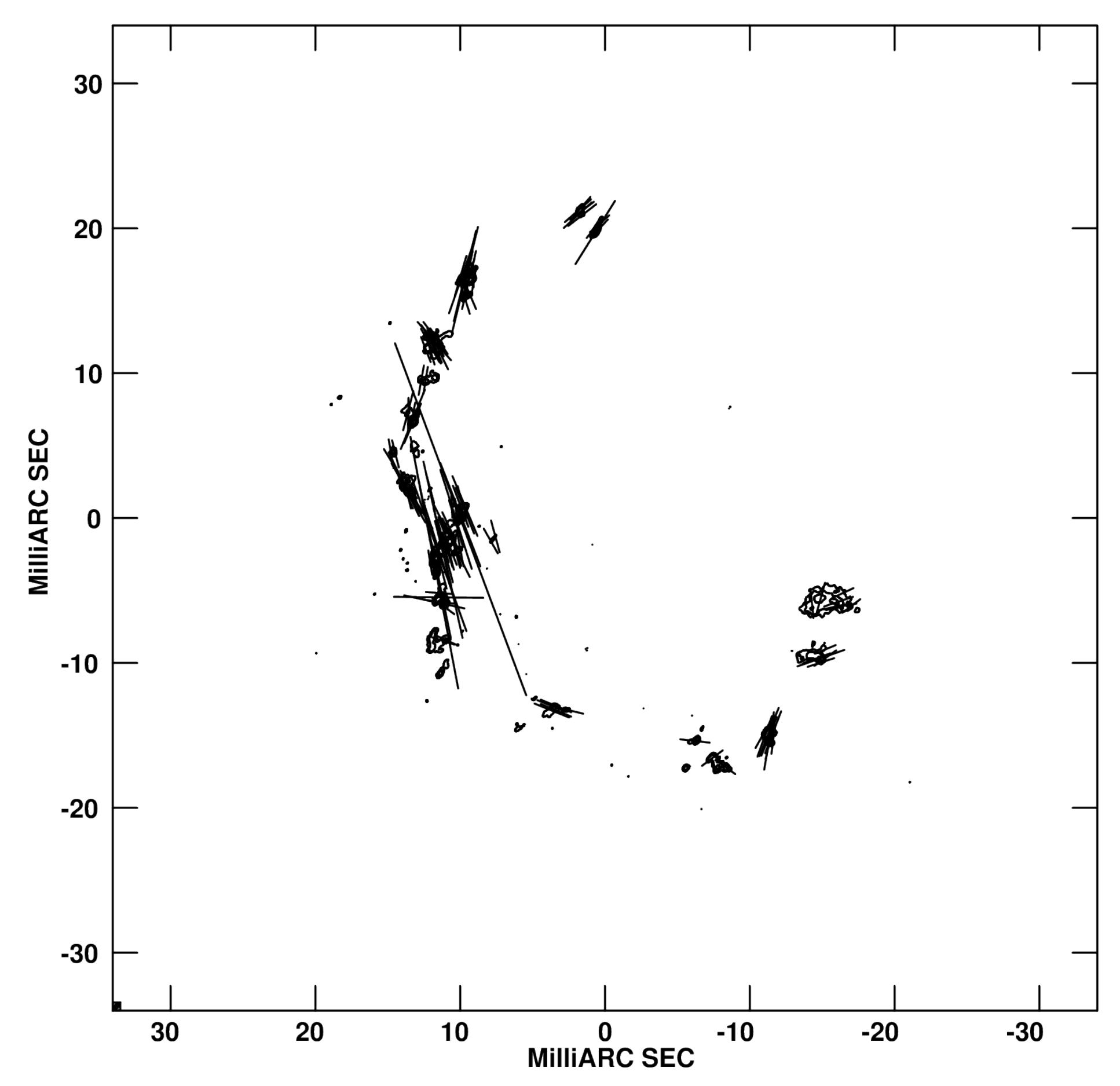}
\caption{Stokes I contours and linear polarization vectors for the full maser ring in epoch BD46AN. Contour levels are $\{-10,-5,5,10,20,40,80,160,320\} \times \sigma$, where $\sigma_{AN} = 1.5209$ mJy beam$^{-1}$. Vectors are at the angle of the EVPA and with a length proportional to the zeroth moment linearly polarized intensity such that 1 mas in length indicates $P=4$ mJy beam$^{-1}$. Spatial coordinates are with reference to the center of the aligned subimage.}
\end{figure}

\begin{figure}[htbp]
\figurenum{1.2}
\epsscale{.68}
\plotone{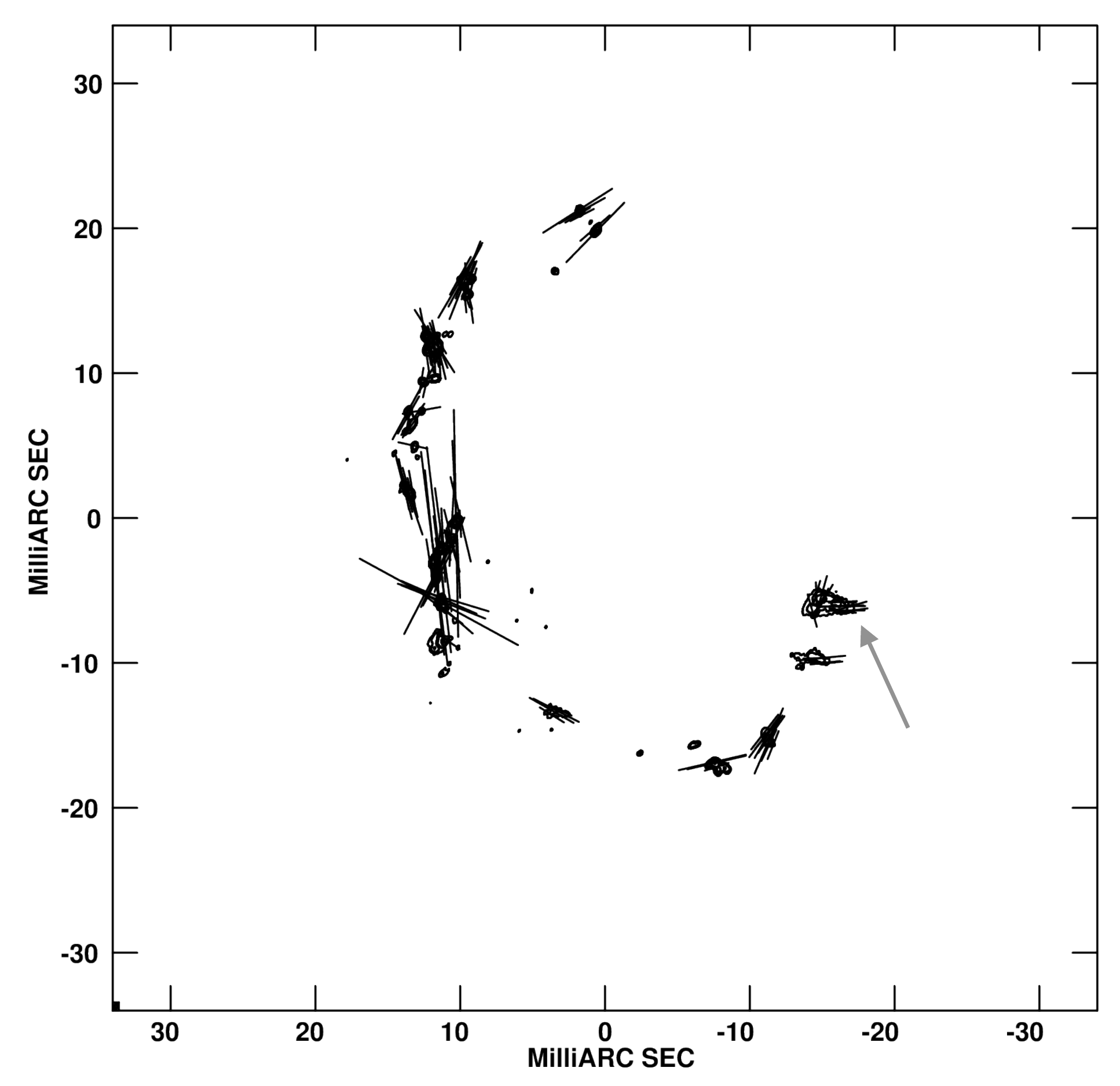}
\caption{Stokes I contours and linear polarization vectors for the full maser ring in epoch BD46AO. Contour levels and spatial scale are as in 1.1, with instead $\sigma_{AO} = 1.6430$ mJy beam$^{-1}$.}
\end{figure}

\begin{figure}[htbp]
\figurenum{1.3}
\epsscale{.68}
\plotone{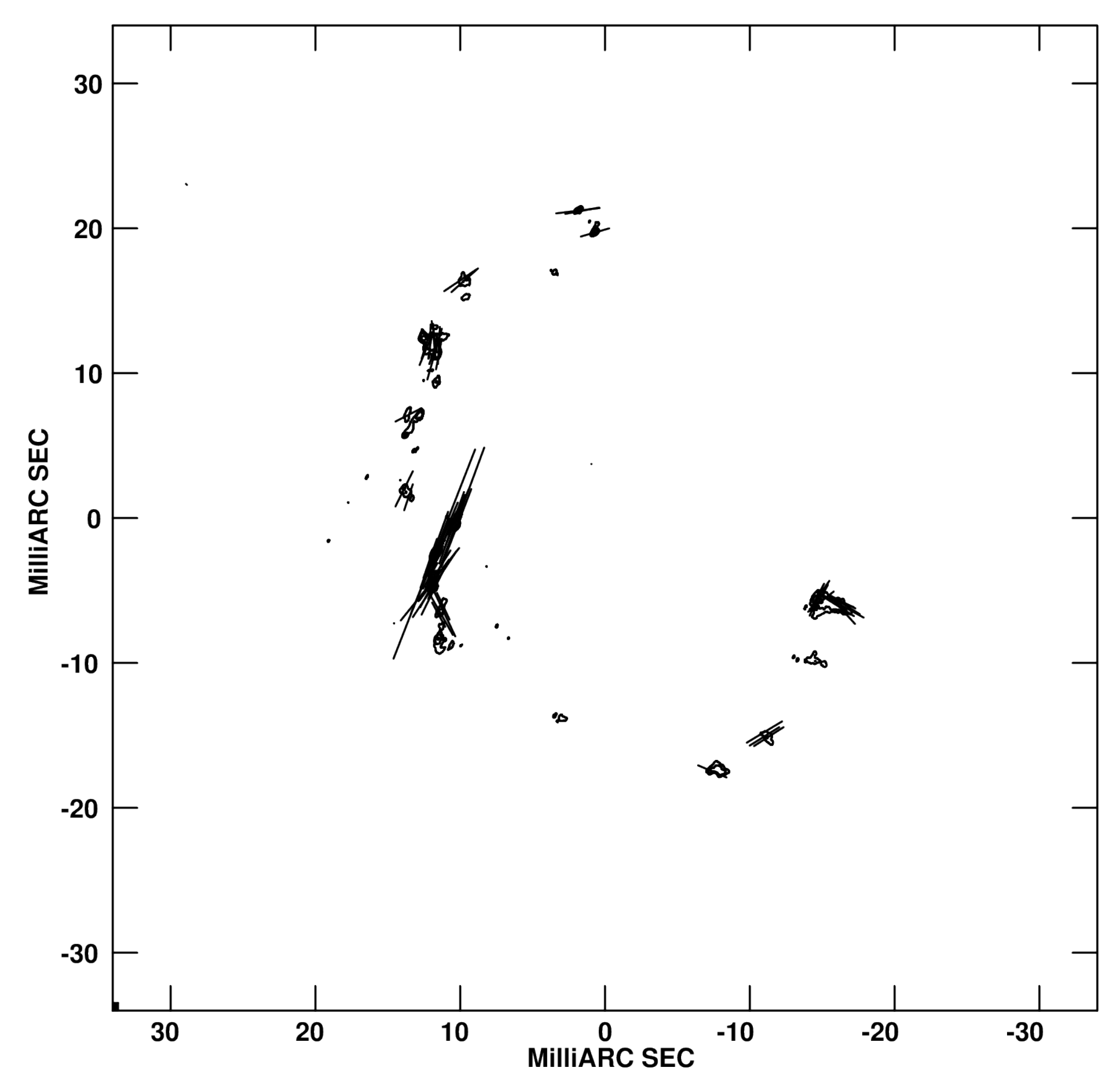}
\caption{Stokes I contours and linear polarization vectors for the full maser ring in epoch BD46AP. Contour levels and spatial scale are as in 1.1, with instead $\sigma_{AP} = 1.8594$ mJy beam$^{-1}$.}
\end{figure}

\begin{figure}[htbp]
\figurenum{1.4}
\epsscale{.68}
\plotone{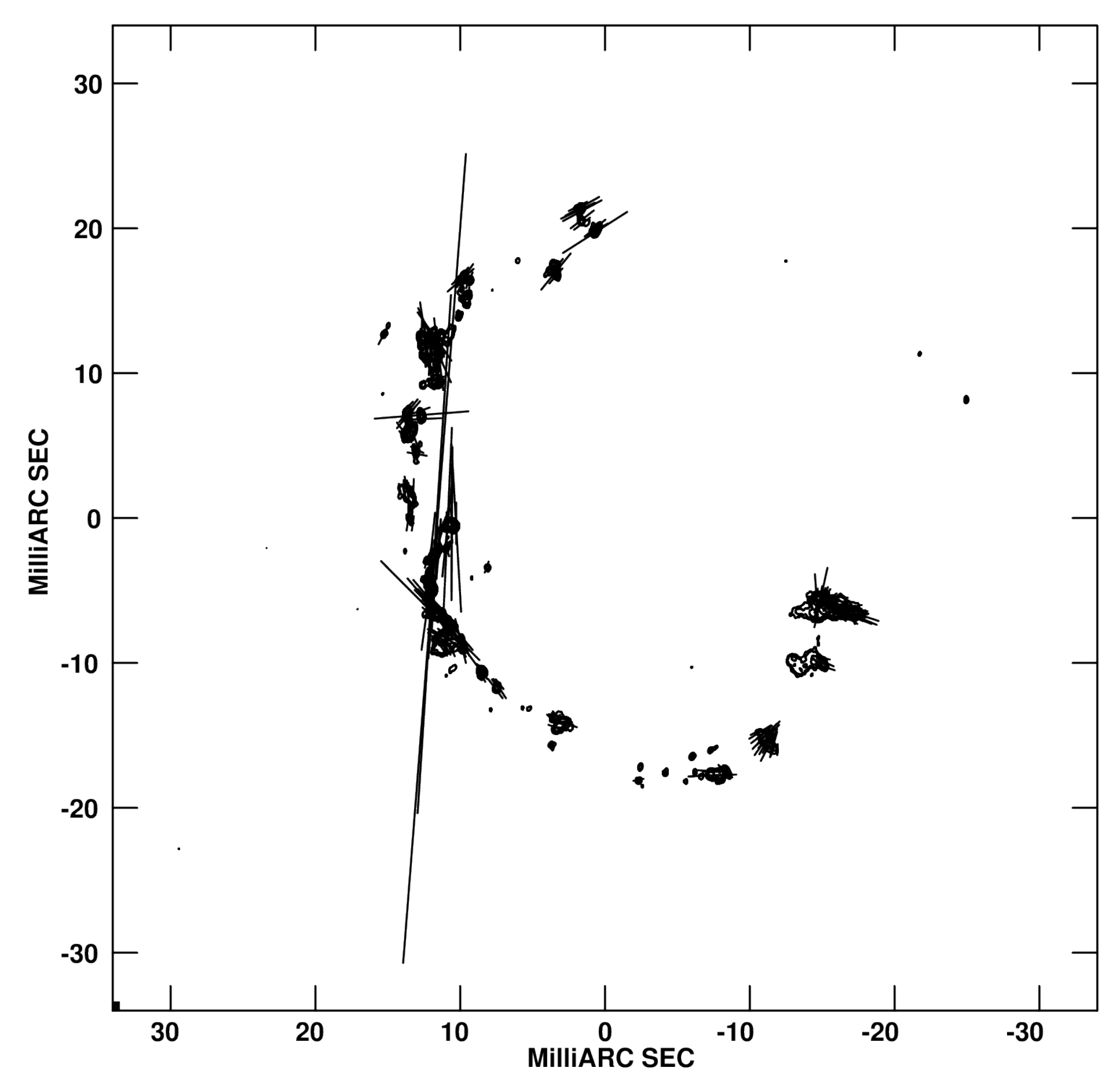}
\caption{Stokes I contours and linear polarization vectors for the full maser ring in epoch BD46AQ. Contour levels and spatial scale are as in 1.1, with instead $\sigma_{AQ} = 0.64265$ mJy beam$^{-1}$.}
\end{figure}

\begin{figure}[htbp]
\figurenum{1.5}
\epsscale{.68}
\plotone{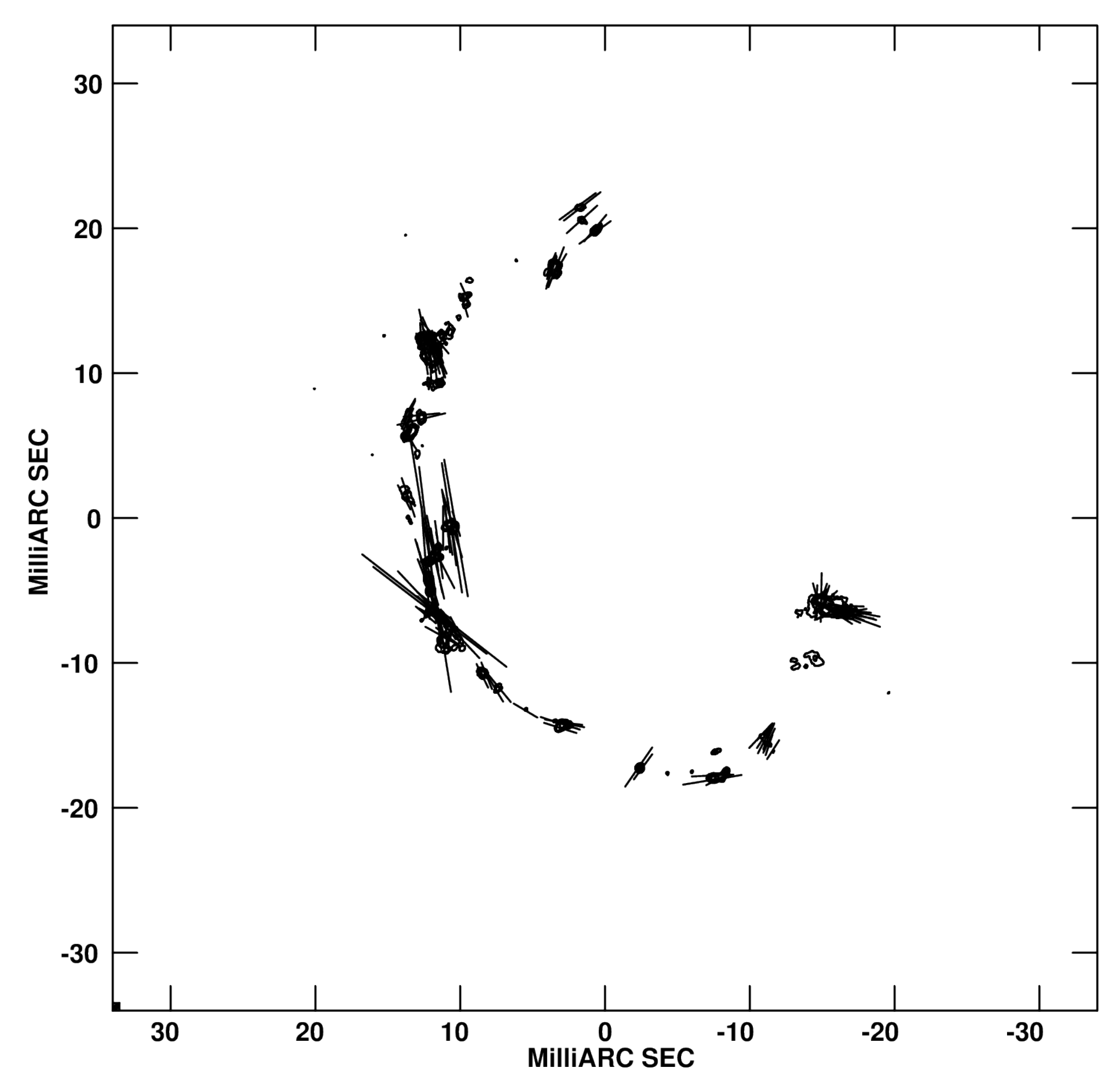}
\caption{Stokes I contours and linear polarization vectors for the full maser ring in epoch BD46AR. Contour levels and spatial scale are as in 1.1, with instead $\sigma_{AR} = 1.4597$ mJy beam$^{-1}$.}
\end{figure}

\newpage

\subsection{Figure Set 2: Frequency--Averaged Stokes I Contours with Linear Polarization Vectors of $\nu=1,J=1-0$ SiO Maser Feature with $\pi/2$ EVPA Rotation}

\begin{figure*}[htbp]
\figurenum{Set 2}
\gridline{\fig{AO_ICON_PVEC_FEAT.pdf}{0.45\textwidth}{(2.1)}
             \fig{AO_ICON_PVEC_FEAT.pdf}{0.45\textwidth}{(2.2)}}
\gridline{\fig{AP_ICON_PVEC_FEAT.pdf}{0.45\textwidth}{(2.3)}
             \fig{AQ_ICON_PVEC_FEAT.pdf}{0.45\textwidth}{(2.4)}}
\gridline{\fig{AR_ICON_PVEC_FEAT.pdf}{0.45\textwidth}{(2.5)}}
\caption{Stokes I contours and linear polarization vectors for the target maser feature in epoch (2.1) BD46AN, (2.2) BD46AO, (2.3) BD46AP, (2.4) BD46AQ, and (2.5) BD46AR. Contour levels are$\{-10,-5,5,10,20,40,80\} \times \sigma$, where (2.1) $\sigma_{AN} = 1.5209$ mJy beam$^{-1}$, (2.2) $\sigma_{AO} = 1.6430$ mJy beam$^{-1}$, (2.3)  $\sigma_{AP} = 1.8594$ mJy beam$^{-1}$, (2.4) $\sigma_{AQ} = 0.64265$ mJy beam$^{-1}$, and (2.5) $\sigma_{AR} = 1.4597$ mJy beam$^{-1}$. The lowest contour visible here is $5\sigma$. Vectors are at the angle of the EVPA and with a length proportional to the zeroth moment linearly polarized intensity such that 1 mas in length indicates $P=10$ mJy beam$^{-1}$. Spatial coordinates are with reference to the center of the aligned subimage.}
\end{figure*}

\newpage

\subsection{Figure Set 3: Channel-Level Stokes I Contours with Linear Polarization Vectors of $\nu=1,J=1-0$ SiO Maser Feature with $\pi/2$ EVPA Rotation}

\begin{figure}[htbp]
\figurenum{3.1}
\epsscale{.98}
\plotone{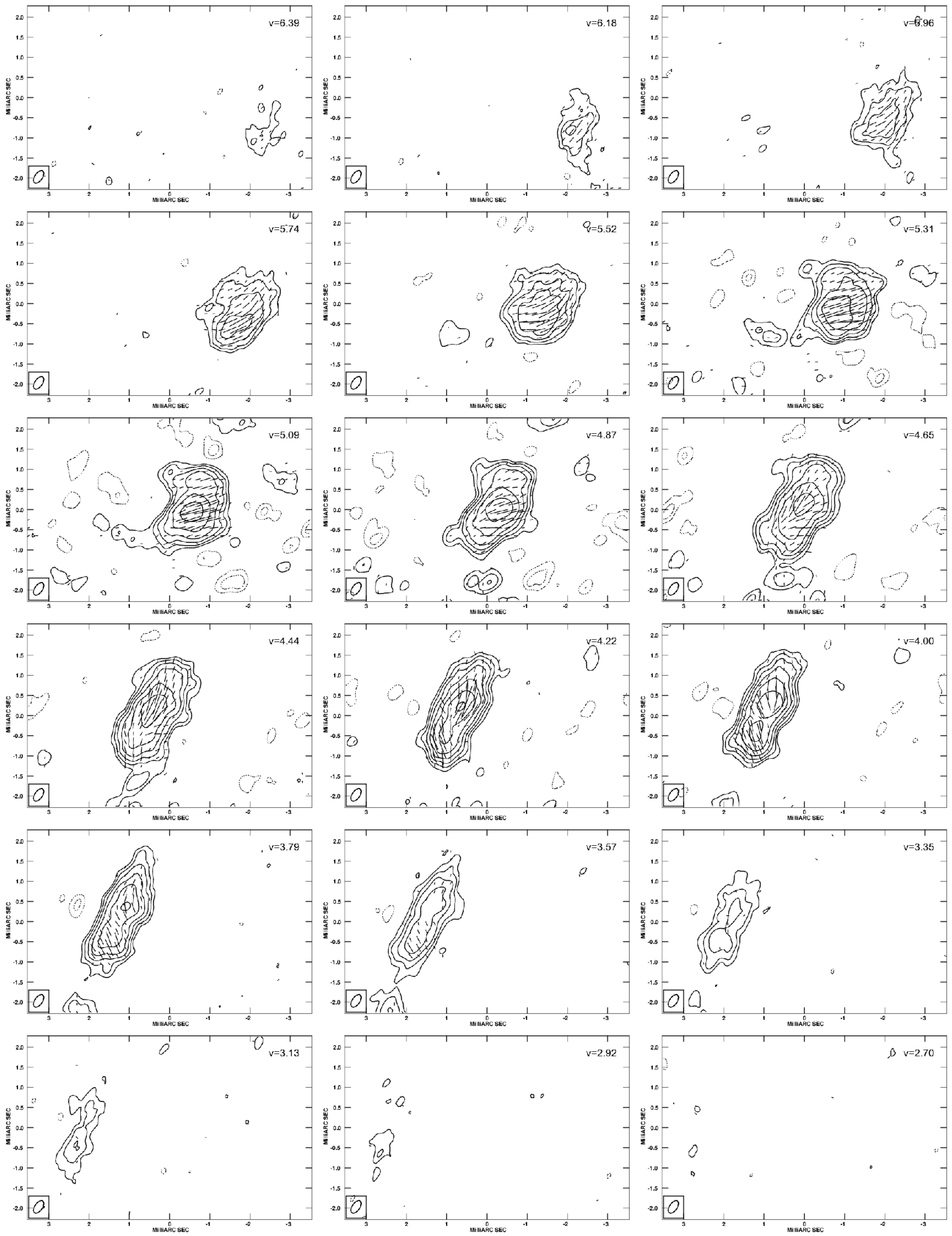}
\caption{Stokes I channel images, labeled with the LSR velocity (km s$^{-1}$), with linear polarization vectors for epoch BD46AN. Contour levels are $\{ -12, -6, -3, 3, 6, 12, 24, 48, 96, 192, 384, 768\} \times \sigma$,, where $\sigma_{AN} = 11.349$ mJy beam$^{-1}$. Vectors are at the angle of the EVPA and with a length proportional to the linearly polarized intensity such that 1 mas in length indicates $P=0.5$ Jy beam$^{-1}$. Spatial coordinates are with reference to the center of the aligned subimage.}
\end{figure}

\begin{figure}[htbp]
\figurenum{3.2}
\epsscale{1.15}
\plotone{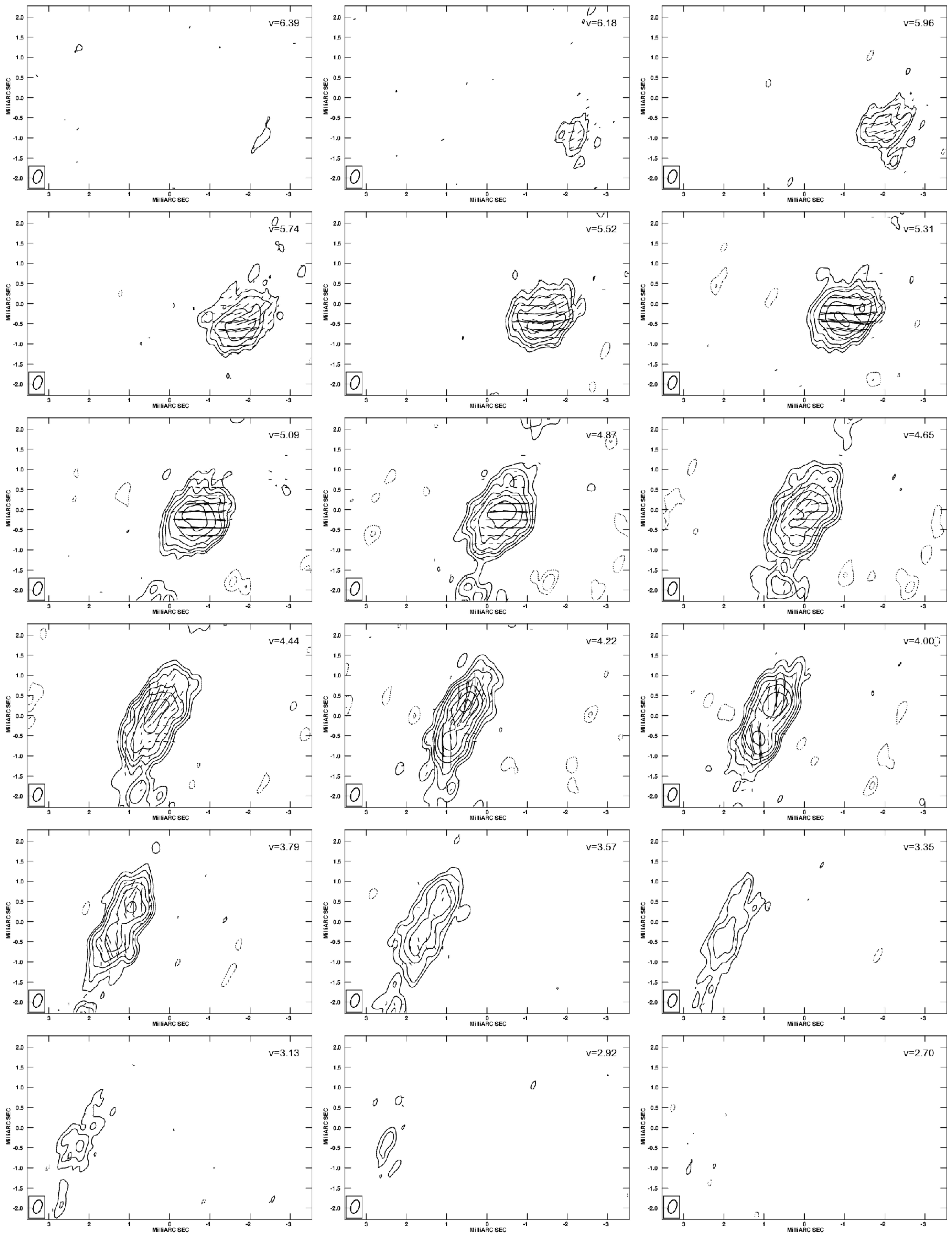}
\caption{Stokes I channel images with linear polarization vectors for epoch BD46AO. Contour levels and spatial scale are as in 3.1, with instead $\sigma_{AO} = 12.403$ mJy beam$^{-1}$.}
\end{figure}

\begin{figure}[htbp]
\figurenum{3.3}
\epsscale{1.15}
\plotone{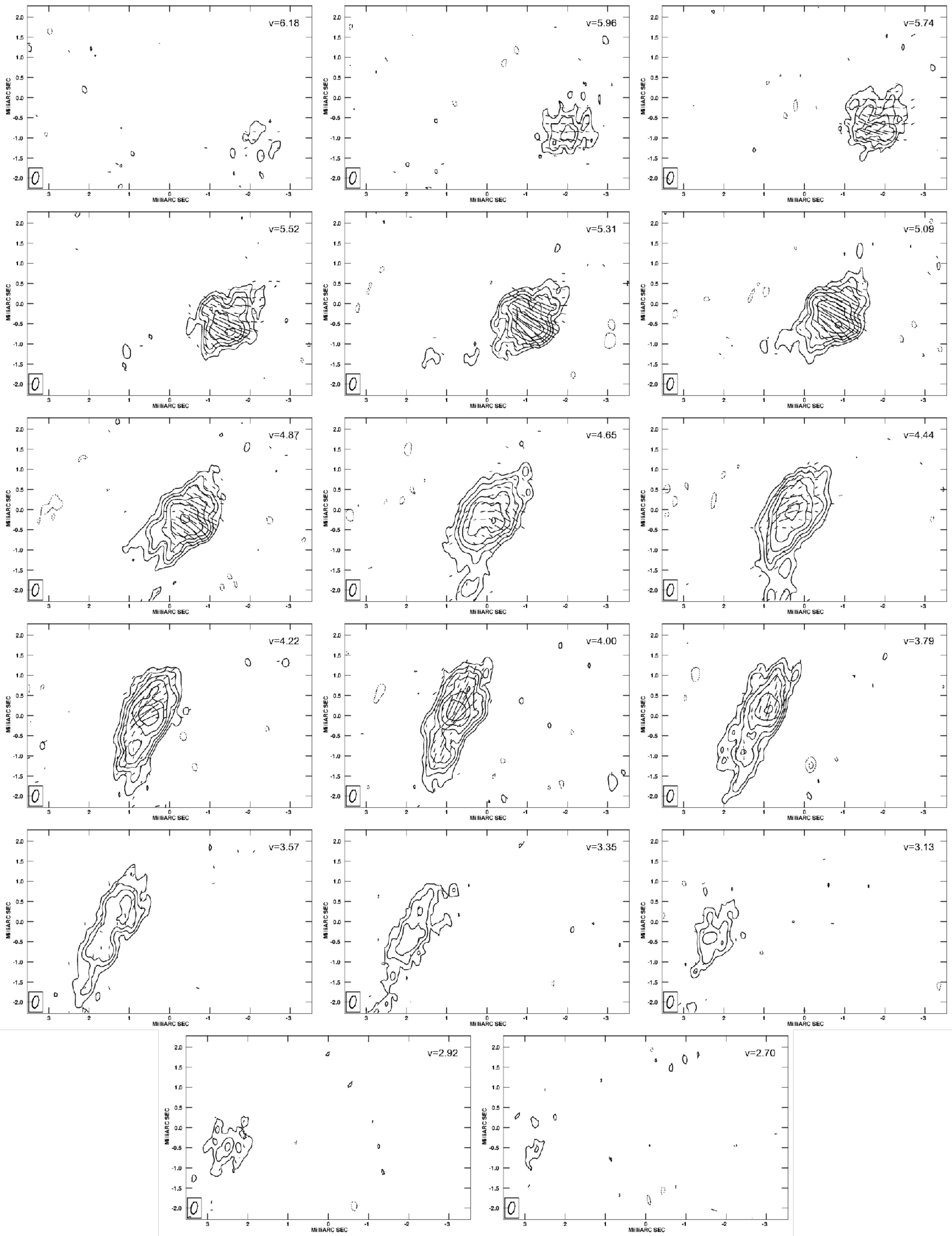}
\caption{Stokes I channel images with linear polarization vectors for epoch BD46AP. Contour levels and spatial scale are as in 3.1, with instead $\sigma_{AP} = 13.293$ mJy beam$^{-1}$.}
\end{figure}

\begin{figure}[htbp]
\figurenum{3.4}
\epsscale{1}
\plotone{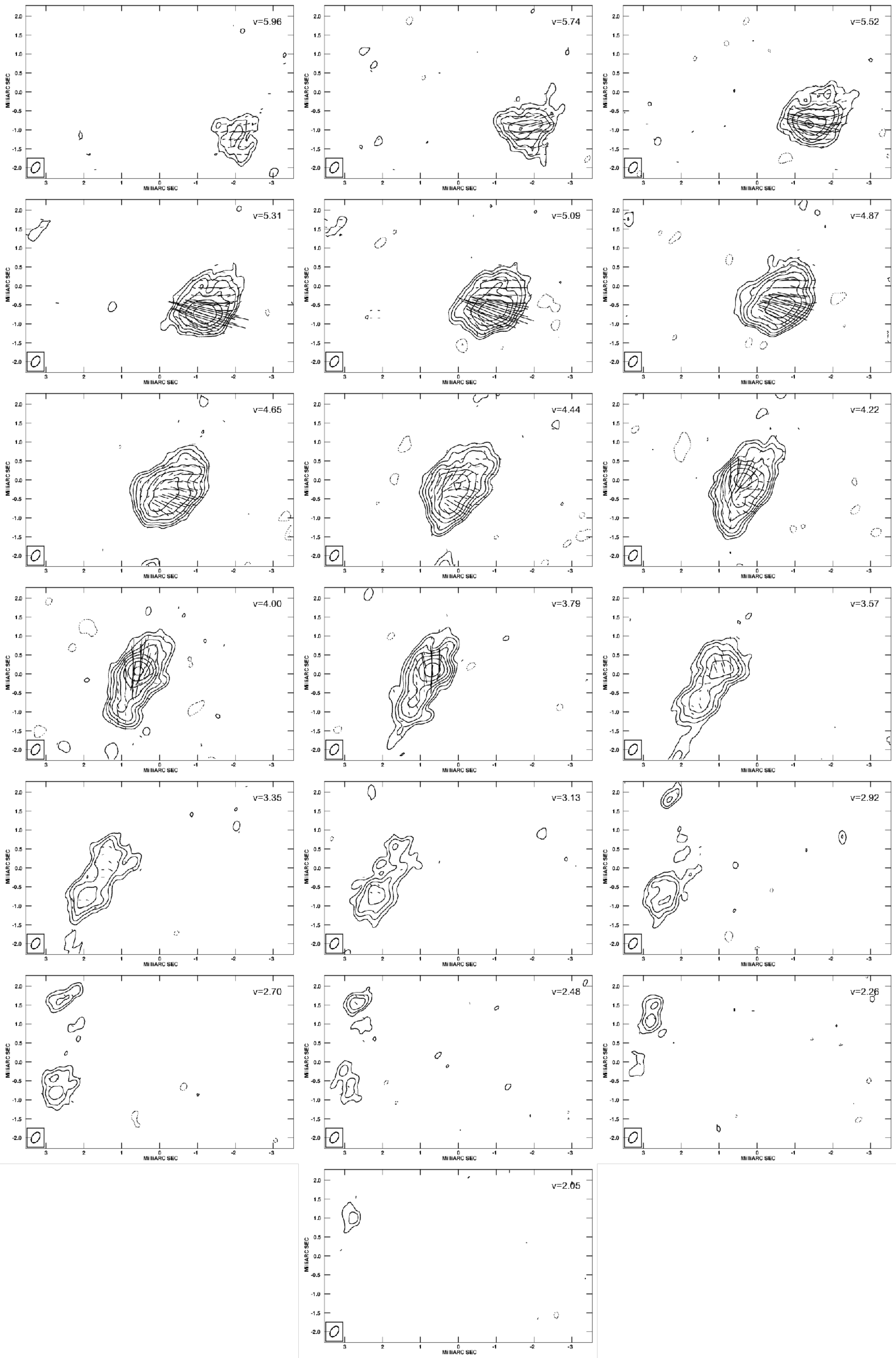}
\caption{Stokes I channel images with linear polarization vectors for epoch BD46AQ. Contour levels and spatial scale are as in 3.1, with instead $\sigma_{AQ} = 5.5904$ mJy beam$^{-1}$.}
\end{figure}

\begin{figure}[htbp]
\figurenum{3.5}
\epsscale{1}
\plotone{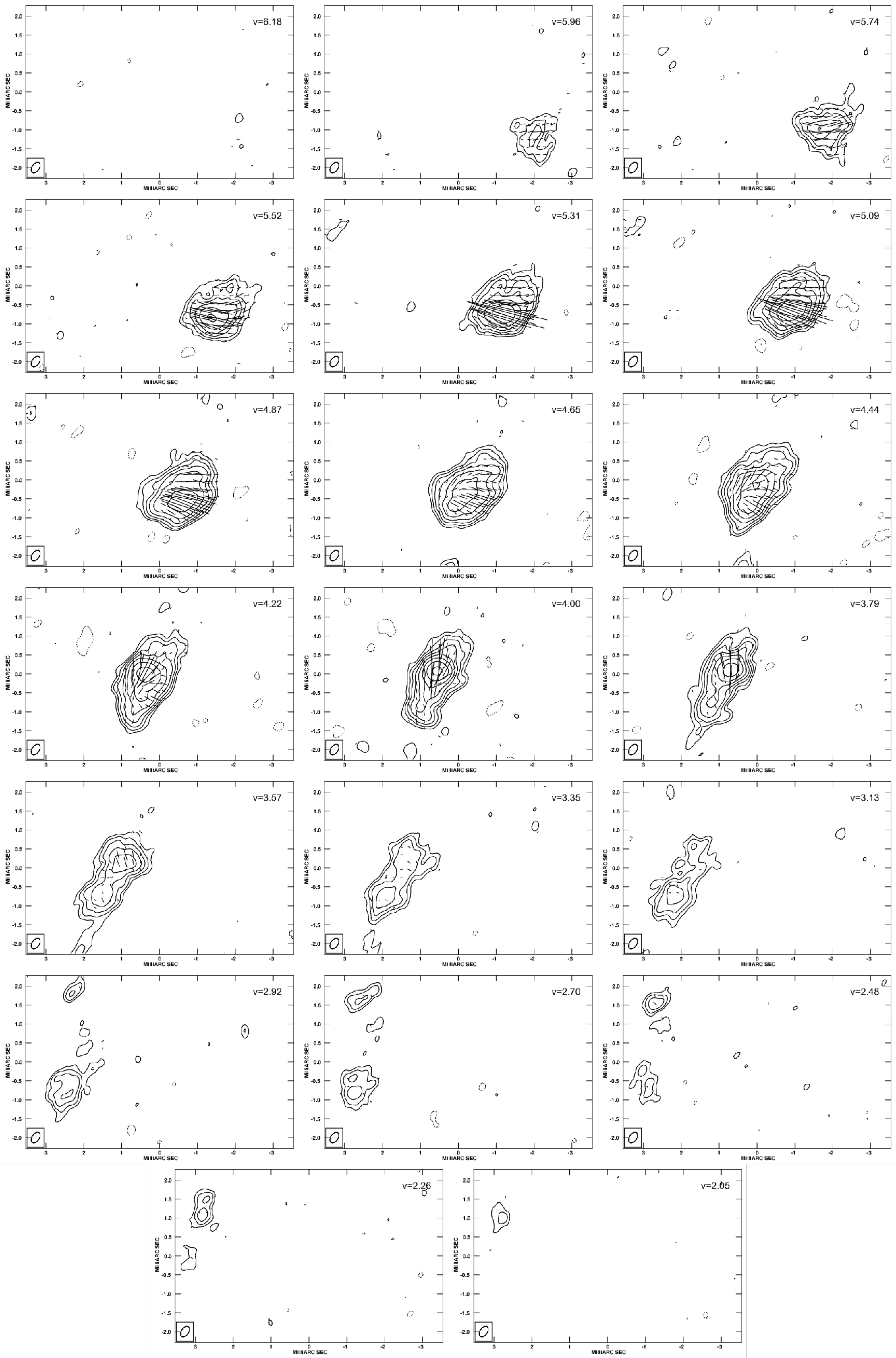}
\caption{ I channel images with linear polarization vectors for epoch BD46AR. Contour levels and spatial scale are as in 3.1, with instead $\sigma_{AR} = 11.334$ mJy beam$^{-1}$.}
\end{figure}

\clearpage
\thispagestyle{empty}
\hfill

\subsection{Figure Set 4: Channel-Level Stokes V Contours of $\nu=1,J=1-0$ SiO Maser Feature with $\pi/2$ EVPA Rotation}

\begin{figure}[h!]
\figurenum{4.1}
\epsscale{.88}
\plotone{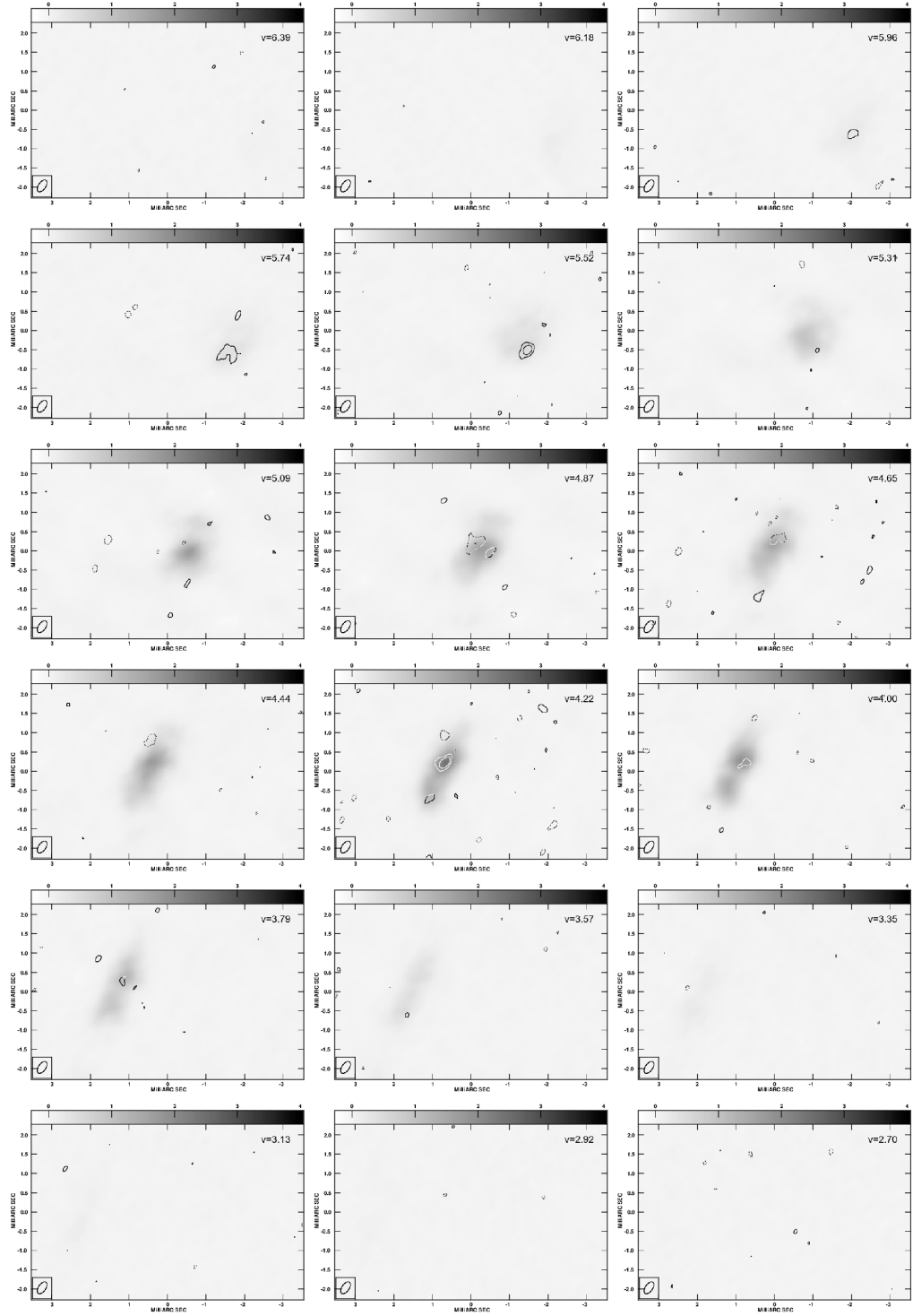}
\caption{Stokes V contours over Stokes I (greyscale) in each channel for epoch BD46AN, labeled with the LSR velocity (km s$^{-1}$) Contour levels are $\{-24, -12, -6, -3, 3, 6, 12, 24\} \times \sigma$, where $\sigma_{AN} = 11.349$ mJy beam$^{-1}$. Spatial coordinates are with reference to the center of the aligned subimage.}
\end{figure}

\begin{figure}[h!]
\figurenum{4.2}
\epsscale{1}
\plotone{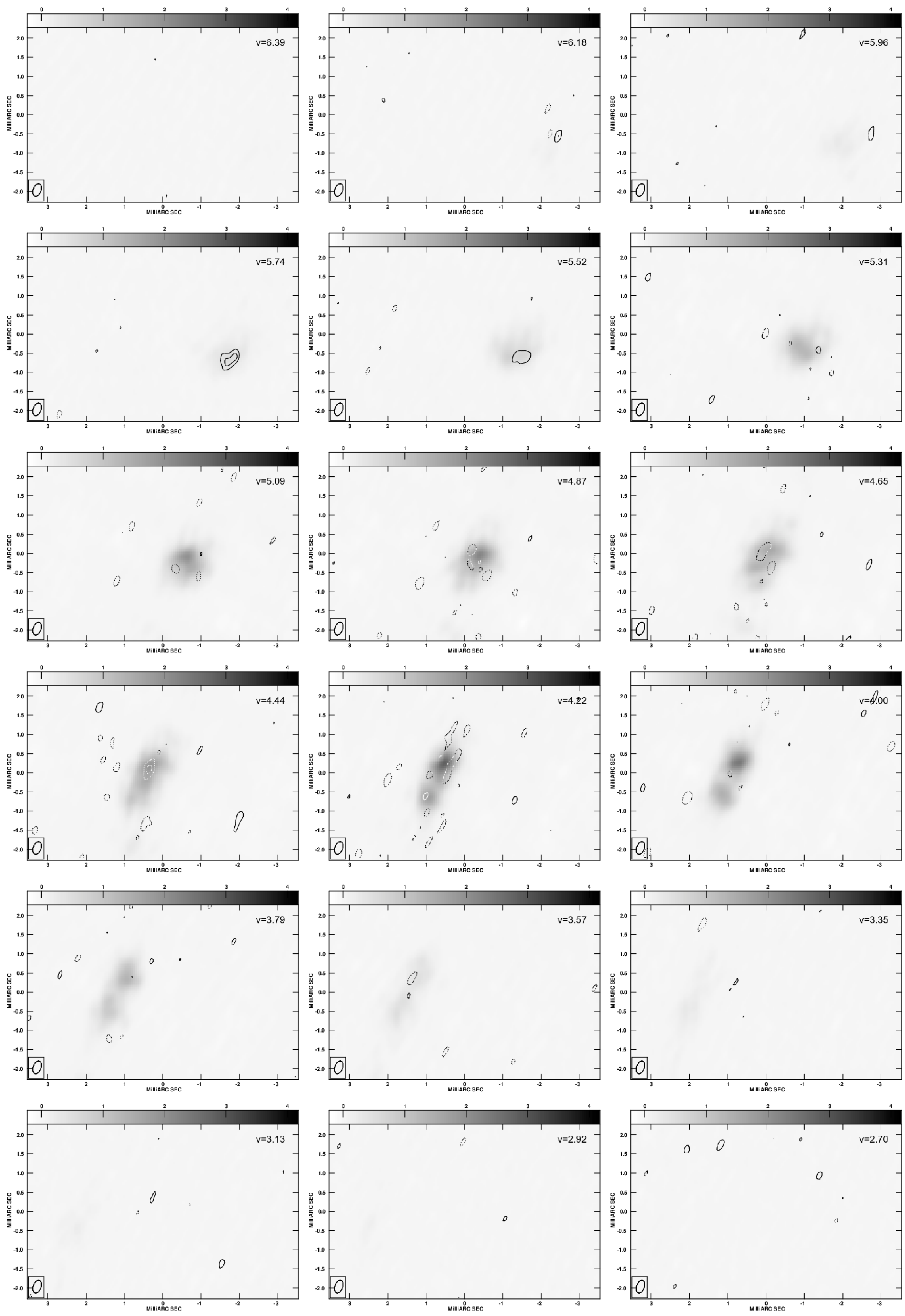}
\caption{Stokes V contours over Stokes I (greyscale) in each channel for epoch BD46AO. Contour levels and spatial scale are as in 4.1, with instead $\sigma_{AO} = 12.403$ mJy beam$^{-1}$.}
\end{figure}

\begin{figure}[htbp]
\figurenum{4.3}
\epsscale{1}
\plotone{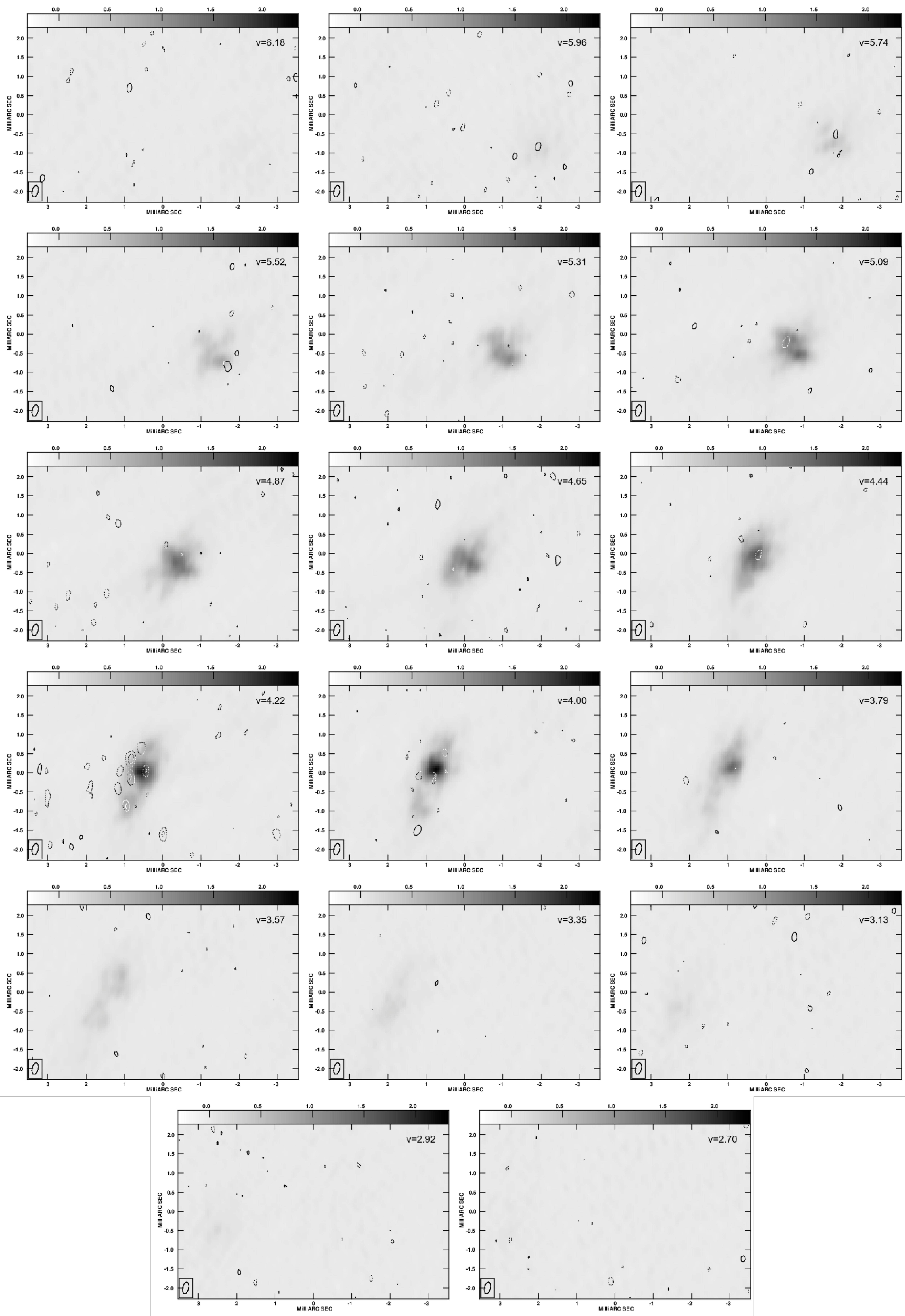}
\caption{Stokes V contours over Stokes I (greyscale) in each channel for epoch BD46AP. Contour levels and spatial scale are as in 4.1, with instead $\sigma_{AP} = 13.293$ mJy beam$^{-1}$.}
\end{figure}

\begin{figure}[htbp]
\figurenum{4.4}
\epsscale{.88}
\plotone{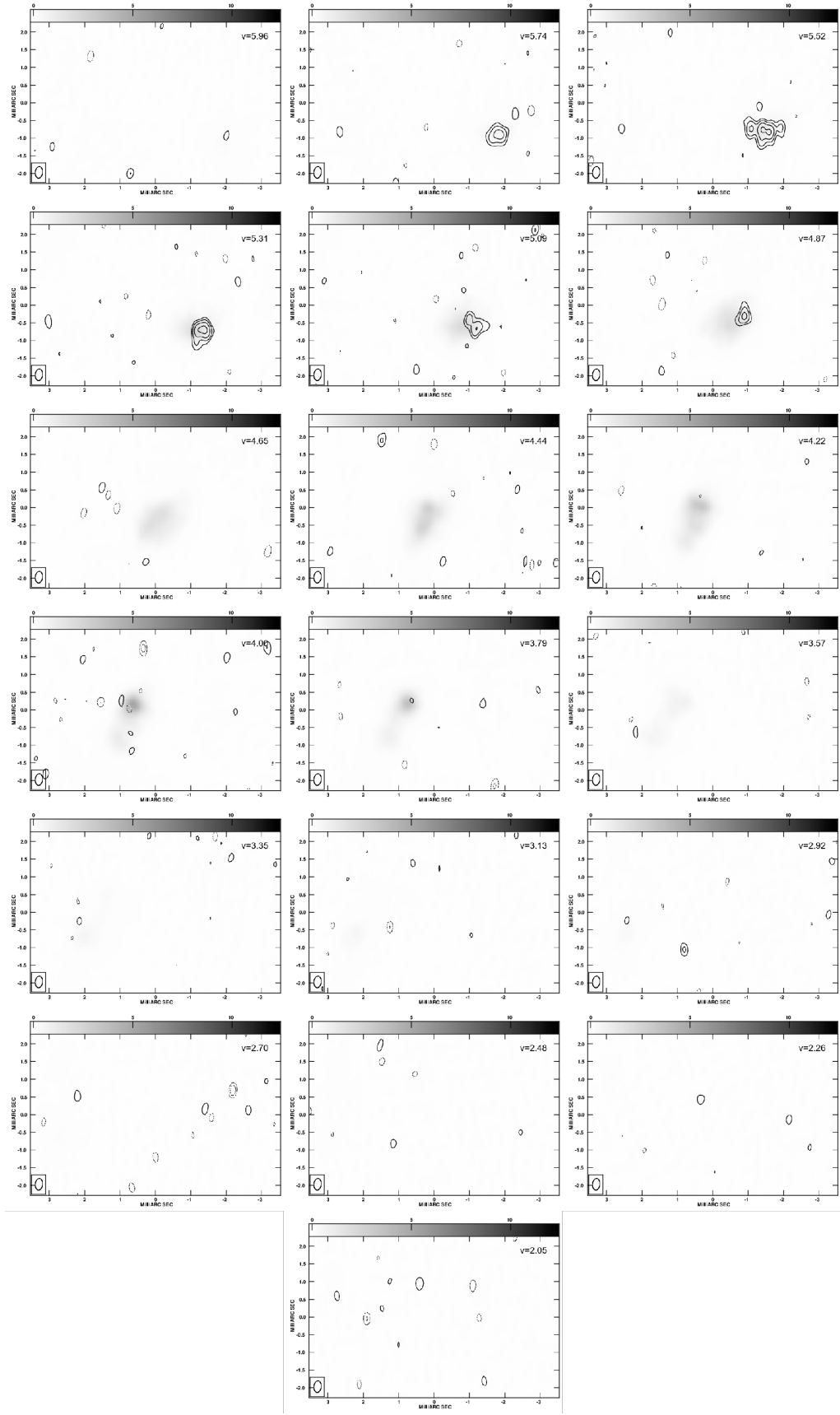}
\caption{Stokes V contours over Stokes I (greyscale) in each channel for epoch BD46AQ. Contour levels and spatial scale are as in 4.1, with instead $\sigma_{AQ} = 5.5904$ mJy beam$^{-1}$.}
\end{figure}

\begin{figure}[htbp]
\figurenum{4.5}
\epsscale{.88}
\plotone{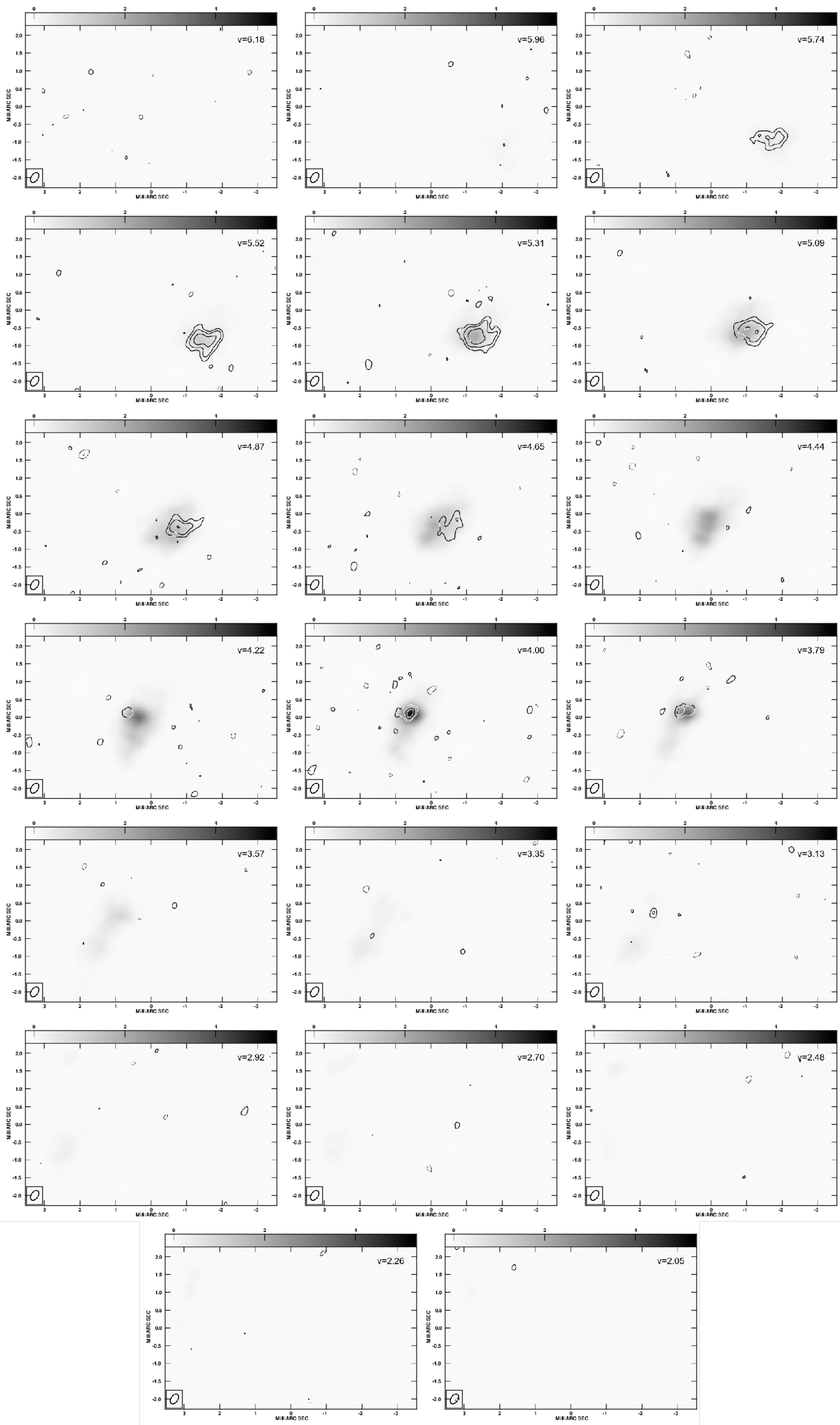}
\caption{Stokes V contours over Stokes I (greyscale) in each channel for epoch BD46AR. Contour levels and spatial scale are as in 4.1, with instead $\sigma_{AR} = 11.334$ mJy beam$^{-1}$.}
\end{figure}

\end{document}